\documentclass[preprint2]{proto}
\usepackage{times}
\usepackage{xcolor}
\usepackage{graphicx}
\usepackage{wasysym}
\usepackage{float}
\usepackage{dblfloatfix}
\usepackage[T1]{fontenc}

\graphicspath{{./}{Figures/}}

\providecommand\aap{Astron.\ Astrophys.} % Astron and Astrophys
\providecommand\apjl{Astrophys.\ J.} % Astrophysical Journal, Letters
\providecommand\apjs{Astrophys.\ J.\ Suppl.} % Astrophysical Journal, Supplement

\begin{document}

\title{\textbf{\LARGE The Asteroid-Comet Continuum\footnote{Chapter in press for the book \textbf{Comets III}, edited by K. Meech and M. Combi, University of Arizona Press}}}

\author {\textbf{\large David Jewitt}}
\affil{\small\em Department of Earth, Planetary and Space Sciences, University of California at Los Angeles}

\author {\textbf{\large Henry H.\ Hsieh}}
\affil{\small\em Planetary Science Institute, Honolulu}

\begin{abstract}

\begin{list}{ } {\rightmargin 1in}
\baselineskip = 11pt
\parindent=1pc
{\small 
The practical distinctions between asteroids and comets, viewed as products of accretion on either side of the snow line, are less clear-cut than previously understood.  In this chapter, we discuss the numerous solar system populations which have physical and dynamical properties that conflict with any simple diagnosis of their nature and origin.  Studies of these so-called ``continuum'' or ``transition objects'', which include many of the most intriguing bodies in the solar system, have implications for a broad range of scientific topics from the demise of comets and the activation of asteroids to the production of interplanetary debris and the origin of the terrestrial planet volatiles.  We present an overview of the current state of knowledge concerning the asteroid-comet continuum and discuss the numerous physical processes behind the activity shown by small bodies in the solar system.
\\~\\~\\~}%leave this in to get the correct vertical space after the abstract
\end{list}
\end{abstract}  

%-----------------------------------------------------------------------------------------------------------
\section{\textbf{INTRODUCTION}}
\label{section:intro}

%\subsection{Background\label{section:background}}

%Fundamentally, scientific interest in small solar system bodies lies in their potential to be used as compositional and dynamical probes of the origin and evolution of our solar system.  This of course requires work to achieve a good understanding of the compositional and dynamical properties of small bodies themselves, part of which has naturally involved the classification of small bodies into various groups that share common properties.  In recent years, however, some of these traditional classifications of small solar system bodies have given way to a more complex picture.
%\\

Traditionally, small solar system bodies have been classified as either asteroids (complex assemblages of refractory minerals) or comets (ice rich bodies containing a mix of volatile and refractory solids).  
Observationally, asteroids and comets are differentiated by the absence or presence of detectable mass loss.  Comets, in essence, appear fuzzy while asteroids do not. Cometary activity typically takes the form of comae (unbound ejected material surrounding the central body) and tails.
%Traditionally, small solar system bodies in the inner solar system (i.e., excluding trans-Neptunian objects for our purposes here) are classified as either asteroids (complex assemblages of refractory minerals) or comets (ice rich bodies containing a mix of volatile and refractory materials).  
%Observationally, the distinction between asteroids and comets are based empirically upon observations of mass loss.  Comets, in essence, appear fuzzy while asteroids do not,
%where cometary activity typically takes the form of comae (unbound ejected material surrounding the central body, or nucleus) and tails (ejected material swept away from the nucleus in a particular direction, often the anti-solar direction or along an object's orbit).
%Asteroids possess a  fractured ``rubble pile'' internal structure due to past impacts and some preserve volatiles like comets. Comets are probably more pristine, both collisionally and thermally, but they are also much less volatile-rich than originally envisioned. 
\\

%As mentioned in Section~\ref{section:intro}, the defining characteristic between asteroids and comets has historically simply been that comets exhibit activity (i.e., mass loss) in the form of comae (unbound ejected material surrounding the central body, or nucleus) and tails (ejected material swept away from the nucleus in a particular direction, often the anti-solar direction or along an object's orbit), while asteroids do not.

Meanwhile, the classical dynamical
%(but imperfect)
distinction between comets and asteroids is based on the Tisserand parameter (with respect to Jupiter),
\begin{equation}
T_J = \frac{a_J}{a} + 2 \left[(1 - e^2)\frac{a}{a_J}\right]^{1/2} \cos(i)
\end{equation}
where $a$, $e$ and $i$ are the orbital semi-major axis, eccentricity and inclination, respectively, and $a_J = 5.2$~au is the semi-major axis of Jupiter. $T_J$ parameterizes the relative velocity between an object and Jupiter at their closest approach, and is approximately conserved in the circular restricted three-body dynamics approximation \citep{tisserand1896_tj,murray2000_solarsystemdynamics}, even in the event of close encounters \citep{carusi1995_tisserand}. Jupiter itself has $T_J$ = 3. Asteroids typically have $T_J > 3$ and comets have $T_J < 3$ \citep{vaghi73}.    \\

\begin{figure*}[htb!]
\begin{center}
\includegraphics[width=12.0cm]{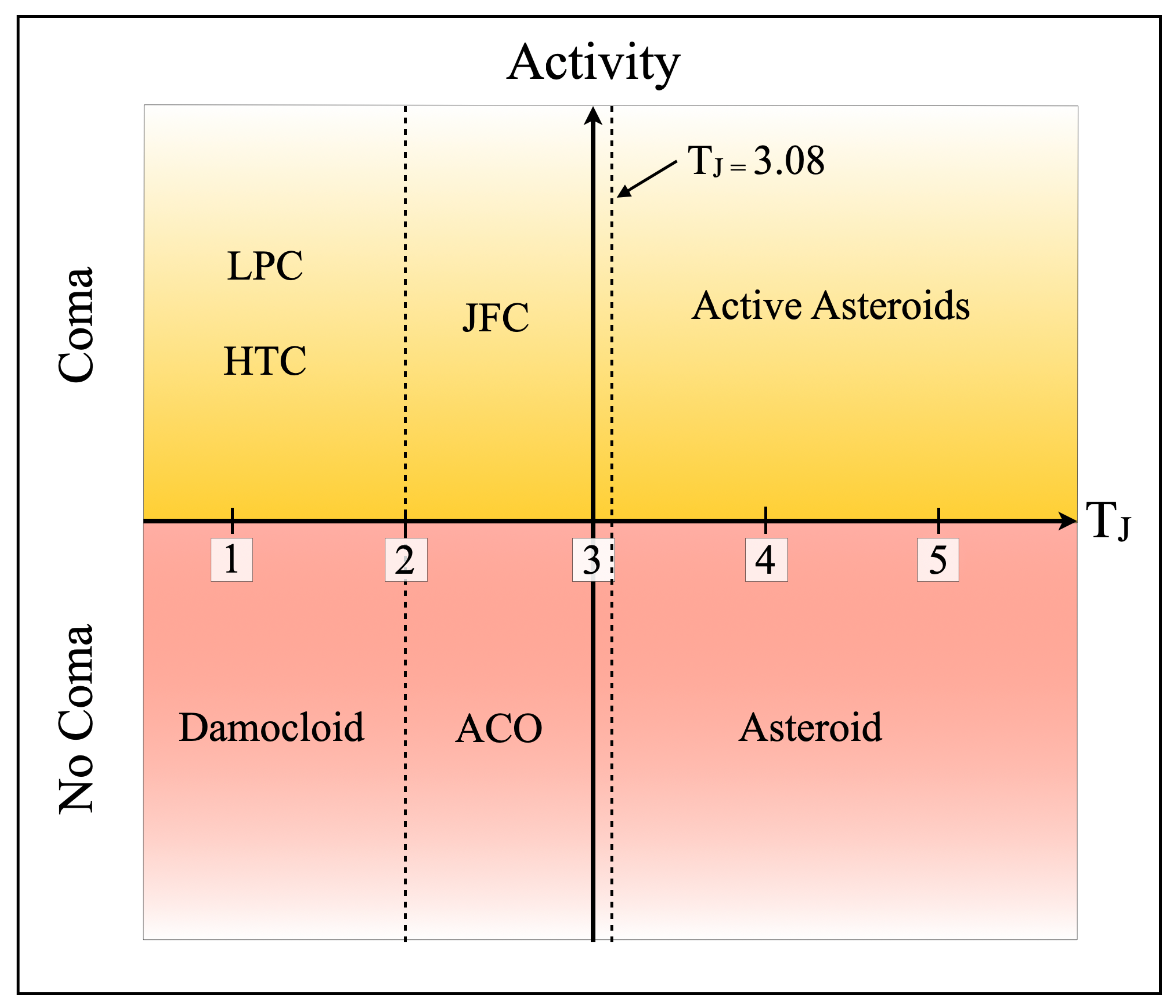}
\end{center}
%\caption{Tisserand parameter ($T_J$) vs.~mass loss rate ($dM/dt$) for a selection of comets and active asteroids, showing the wide diversity of properties.  Almost six orders of magnitude separate the most active comets from the low acivity active asteroids.  The sensitivity limits on the detection of dust coma, currently near 0.1 kg s$^{-1}$, are marked with a horizontal dashed line. 
%LPC, HTC and JFC denote long-period comets, Halley-type comets, and Jupiter-family comets, respectively. Object HB = C/Hale-Bopp (1995 O1). $T_J$ = 3.08 marks the nominal JFC vs.~active asteroid dividing line. The sensitivity limit near 0.1 kg s$^{-1}$ represents the detection of weak continuum coma \citep{Sonnett2011}. Spectroscopic limits to the detection of gas are several times higher.}
\caption{Classification diagram for small bodies discussed in this chapter, showing the Tisserand parameter with respect to Jupiter ($T_J$) vs.~whether or not a coma has ever been detected.  Acronyms are explained in Section~\ref{section:intro}.  Centaurs (not shown) can appear anywhere in this plot.  The vertical dashed line marked $T_J$ = 3.08 denotes the nominal lower bound for active asteroids.}
\label{figure:tiss_vs_massloss}
%\end{center}
\end{figure*}

To make sense of the alphabet soup used to describe small solar system populations, we show in Figure \ref{figure:tiss_vs_massloss} a simple classification scheme based on $T_J$ and whether or not observable mass loss exists.  Within the comet population, long period comets (LPCs) have $T_J < 2$ and orbital periods $P_{\rm orb}\ge200$~yr, Halley-type comets (HTCs) have $T_J < 2$ and $P_{\rm orb}<200$~yr, and Jupiter-family comets (JFCs) have $2 \le T_J \le 3$ \citep{levison1996_comettaxonomy}. The different orbital properties of these populations reflect their origins, with LPCs supplied from the Oort Cloud, JFCs from the Kuiper belt, and HTCs from a thus far ambiguous source, with likely contributions from both the Oort cloud and the scattered disk component of the Kuiper belt \citep[][also see chapter by Kaib and Volk]{dones2015_cometreservoirs}.   Damocloids and Asteroids on Cometary Orbits (ACOs) are inactive bodies with $T_J <$ 2 and 2 $\le T_J \le$ 3, respectively, paralleling the distinction between LPCs and JFCs.
\\

In the classical view of the solar system described above, the asteroids and comets formed at different temperatures in locations interior and exterior to the snow line, respectively, and have been preserved more or less in their formation locations for the past 4.5 Gyr.  In principle, these different formation conditions and source regions should produce objects with distinct physical and dynamical properties.  As such, small solar system objects should be uniquely classifiable into one of these categories or the other.
\\

Recently, however, a more nuanced picture has emerged, with classical asteroids and comets now understood as simply end-members of observational, physical, and dynamical continua.  A key practical problem with the observational classification of objects based only on observations of activity is that it is dependent on instrumental sensitivity (Section~\ref{section:activity_detection}).
%For illustration, Figure~\ref{figure:tiss_vs_massloss} shows the wide range of mass loss properties of various active small solar system bodies with different dynamical properties, including active asteroids (discussed further in Section~\ref{section:AA}), LPCs, HTCs, and JFCs.
%, including a few examples of well-known comets as well as active asteroids
%including a few well-known comet examples and objects Filacchione et al.~in this book)which have $T_J > 3$ and so are dynamically asteroid-like but which show measurable mass loss indicating a comet-like nature.
%The schematic Figure~\ref{figure:tiss_vs_massloss} shows the wide range of dynamical and mass loss properties of comets and asteroids, including a few examples of well-known comets as well as active asteroids
%including a few well-known comet examples and objects which have $T_J > 3$ and so are dynamically asteroid-like but which show measurable mass loss indicating a comet-like nature.
%These are the active asteroids to be discussed in Section \ref{section:AA}, some of which 
%\citep[known as main-belt comets, or MBCs;][]{hsieh2006_mbcs} 
%are indeed ice-rich like comets from the Kuiper belt and Oort cloud, while others of which eject mass for reasons unrelated to sublimation.
Neither does activity necessarily imply the presence of sublimating volatiles; we now know that many other processes besides ice sublimation can lead to visible mass loss (Section~\ref{section:activity_mechanisms}). Activity is possible for a much broader range of objects beyond just those containing ice.
\\

We also now recognize that rocky asteroids and icy comets did not originate in wholly distinct regions of the solar system, as was formerly believed (Section~\ref{section:volatile_distribution}).  Comets may be  less volatile-rich than the half rock, half ice mixture envisioned by \cite{whipple1950_cometmodel1}, and they contain a curious combination of the most volatile ices from the outermost regions of the protoplanetary disk and high-temperature, crystalline silicates from the inner edge of the protoplanetary disk \citep[][also see chapters by Bergin et al.\ and Filacchione et al.]{Westphal09}.  Isotopic evidence from meteorites shows two types of material which accreted separately (perhaps interior and exterior to Jupiter's orbit, and perhaps not simultaneously), but which are now intermingled in the asteroid belt \citep{Warren11,Lichtenberg21}.  As we discuss later, there is clear evidence for near-surface ice in the main asteroid belt where its survival was previously thought to be implausible.  Some of this mixing across the protoplanetary disk might have been forced by the radial migration of the planets, the latter identified first from the unexpectedly dense resonant populations in the Kuiper belt.   The snow line itself is now understood to be a dynamic entity, which moved in response to time-dependent heating sources in the protoplanetary disk \citep{Harsono15}.
\\

Finally, dynamically, $T_J$ is an imperfect  discriminant due to its neglect of planetary perturbers other than the Sun and Jupiter and lack of consideration of nongravitational perturbations due to asymmetric outgassing.
%Gravitational perturbations from other planets and forces due to asymmetric outgassing can act to alter $T_J$, 
%limiting its diagnostic value.  
We now know that not only is the dynamical boundary between asteroids and comets better characterized as a range of $T_J$ values rather than a sharp threshold at a single $T_J$ value (i.e., $T_J=3$), the boundary is also porous, with objects capable of passing through to the other side in both directions, so obscuring their true dynamical origins (Section~\ref{section:dynamics}).
\\

%For all these reasons, we are in the middle of a dramatic unveiling of the nature of the comets and asteroids, and a reassessment of the relationships between them.
%In this chapter we discuss so-called ``transition objects'', whose properties lie in the middle between the traditional asteroid and comet end-members.
In this chapter, which builds on earlier accounts by \cite{Jewitt12}  and \cite{JHA15}, we review the current state of knowledge concerning objects that sit in the ambiguous space between classical rocky asteroids and classical icy comets.  
We will consider objects that  originate as either classical comets or classical asteroids and evolve observationally, physically, or dynamically into the other type of object, as well as objects for which having characteristics of both asteroids and comets is simply part of their intrinsic nature.  While these objects have often been termed ``transition objects'' in earlier literature, the title of this chapter reflects the fact that not all of these objects are actually in the midst of transitioning from one type of small body to another.  Instead, there is simply a broad diversity of objects that span the range of observational, physical, and dynamical properties commonly attributed to either asteroids or comets in a continuous, rather than discrete, fashion \citep[e.g.,][]{hsieh2017_continuumobjects}.
\\

In addition to the topics above, we will discuss  key examples of objects that have properties of both classical asteroids and comets: active asteroids (Section~\ref{section:AA}), inactive comets (Section~\ref{section:inactive_comets}), and Centaurs (Section~\ref{section:centaurs}).
%We will specifically discuss active asteroids (Section~\ref{section:AA}), defunct comets (Section~\ref{section:deadcomets}), and Centaurs (Section~\ref{section:centaurs}).  
While these do not comprise an exhaustive list, they provide instructive examples relevant to the entire asteroid-comet continuum for small solar system bodies.  We will then conclude with a discussion of future research prospects (Section~\ref{section:future}).  
\\

%We will first discuss the various processes by which observable activity can arise on small solar system bodies, which include, but are not limited to, the sublimation of volatile material to which cometary activity is classically attributed (Section~\ref{section:activity_mechanisms}).  We will then provide a brief summary of relevant issues in solar system dynamics (Section~\ref{section:dynamics_general}).
%and dynamical topics relevant to this subject.  We will then discuss key groups of small solar system bodies that

%\subsection{Distinctions\label{section:distinctions}}

\section{ACTIVITY DETECTION}
\label{section:activity_detection}

\subsection{Overview}
\label{section:activity_detection_overview}

One factor in the growing recognition of asteroid-comet continuum objects is a steady improvement in the ability of  astronomers to  detect weak activity, enabled by the use of larger, more sensitive telescopes, and by increasing numbers of wide-field, time-resolved surveys.  A second factor is the recognition that weak activity exists to be detected, which adds an equally important, albeit psychological, dimension to the growing rate with which these objects are perceived.   The most direct method of activity detection is resolved imaging, which typically reveals the presence of comet-like features like comae or tails.  
Other methods,  including photometric analysis, spectroscopic detection of gas, detection of debris streams associated with small solar system bodies, and detection of non-gravitational perturbations,  can also reveal the presence of activity.
\\

\subsection{Resolved Imaging}
\label{section:activity_visual_detection}
Resolved imaging of mass loss is  the gold standard and most common means by which objects are determined to be active. 
The optical scattering efficiency (measured as cross-section per unit mass) of dust is much larger than that of resonance fluorescence from common gaseous molecules. Therefore, direct imaging typically reveals only ejected dust, whether or not gas is present. Even in classical comets, for which we are sure that gas drag drives mass loss, gas is commonly undetectable against the continuum of sunlight scattered from dust, particularly in observations taken at heliocentric distances  $r_H\gtrsim2$~au.   
\\

For weakly active objects, direct imaging of mass loss requires sensitivity to faint near-nucleus surface brightness features and is thus a function of telescope size, angular resolution, sky brightness, and more.  Most active asteroids and Centaurs have been discovered in data from wide-field surveys, for which activity detection is typically not prioritized. Unfortunately, the surface brightness detection limits of most of those surveys are not well documented, making it difficult to use their data to quantitatively characterize populations of active bodies.  Two exceptions are the wide but shallow survey of \citet{Waszczak13} and the deep but narrow-field ``Hawaii Trails Project'', where the latter survey resulted in the discovery of a fan-like dust tail associated with 176P/LINEAR, then known as asteroid (118401) 1999 RE$_{70}$, with an $R$-band surface brightness of $\Sigma_R=25.3$~mag~arcsec$^{-2}$ \citep{hsieh2009_htp}. The relatively shallow depths of most wide field surveys, though, mean that low-level activity is not well constrained for the vast majority of asteroids. Many asteroids that are currently considered inactive could in fact be weakly active but have thus far escaped detection \citep{Sonnett2011}.
\\

\begin{figure}[htb!]
\begin{center}
\includegraphics[width=5.5cm]{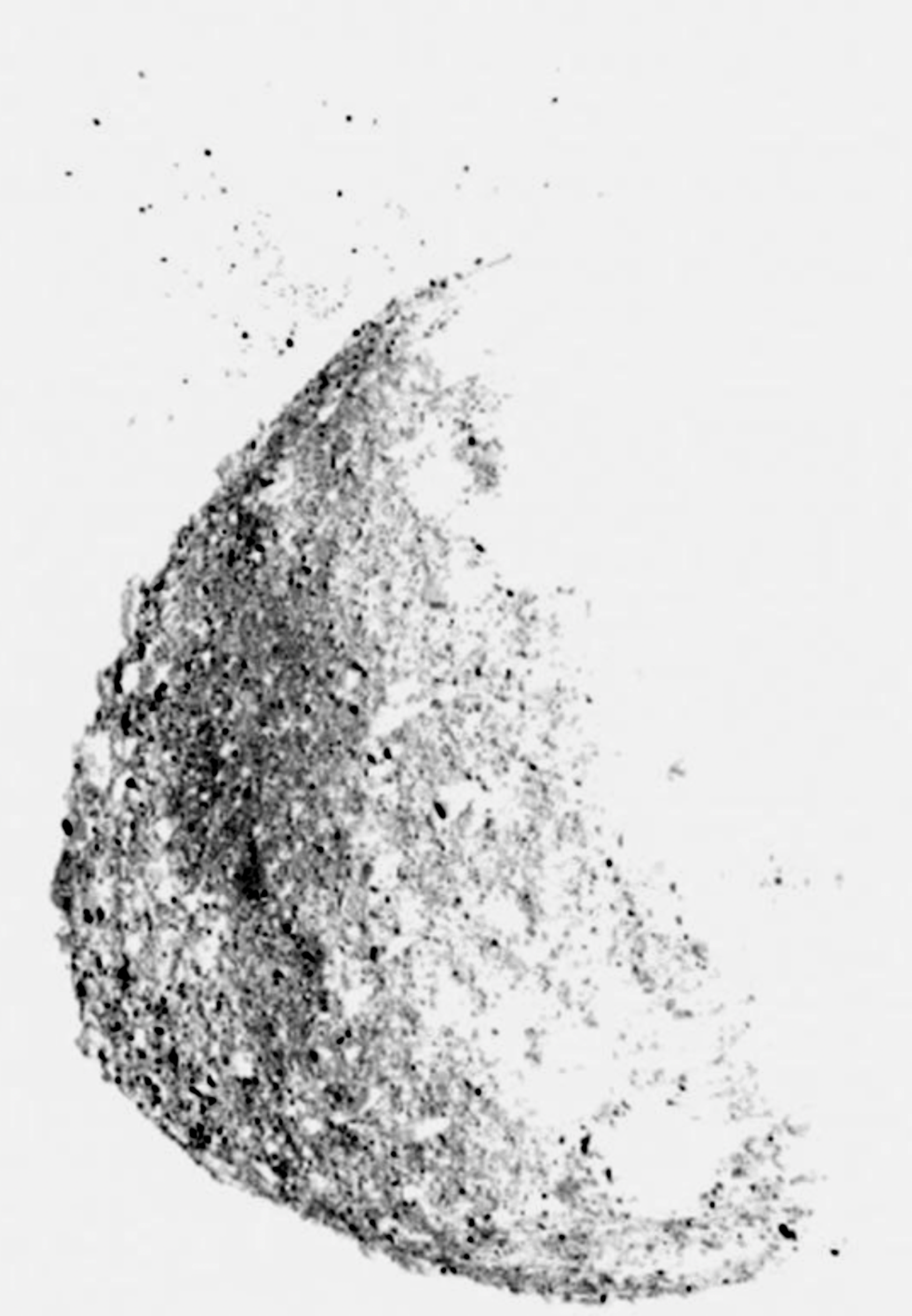}
\end{center}
\caption{Reverse polarity image of 0.5 km diameter asteroid (101955) Bennu from OSIRIS-REx showing a  swarm of centimeter-sized particles in the upper left of the image. Image credit: NASA/Univ.\ of Arizona.}
\label{figure:bennu}
%\end{center}
\end{figure}

This view is anecdotally supported by the discovery of activity in (3200) Phaethon, (3552) Don Quixote,
and (101955) Bennu. Phaethon (Section~\ref{section:phaethon}) and Don Quixote \citep{mommert2014_donquixote} were only found to be active after intensive targeted observational efforts motivated by independent indicators that the objects could be active (i.e., Phaethon's association with the Geminid meteor stream and Don Quixote's obviously cometary orbit with $T_J=2.3$),
%; see Sections~\ref{section:debris_streams} and \ref{section:inactive_comets}).
while activity was observed unexpectedly in situ for Bennu by the visiting OSIRIS-REx spacecraft (Figure~\ref{figure:bennu}; Section~\ref{section:bennu}).  %The possibility that large numbers of asteroids may exhibit low-level activity that is not directly observable for individual objects has also been proposed by \citet{Sonnett2011}.
\\

%Often considered the defining characteristic of cometary bodies, sublimation-driven activity 

%Although sublimation-driven activity is a defining characteristic of cometary bodies, the fact that detection of that activity can depend on when an object is observed presents a key complication. 

Identification of active objects is also complicated by the fact that activity is usually transient. Active icy objects can become dormant when their surface volatiles are depleted or buried by a non-volatile crust \citep[][]{jewitt1996_dormantcomets}.  Additionally, even modest orbital eccentricities can strongly affect water ice sublimation rates (because sublimation is an exponential function of temperature and hence of heliocentric distance) and therefore dust production rates.  Thus, even very deep observations of main-belt objects away from perihelion do not exclude the possibility of sublimation-driven activity when closer to the Sun \citep[e.g.,][]{hsieh2015_ps1mbcs}. Meanwhile, visible mass loss caused by other processes, e.g., impacts and rotational destabilization, can be very short-lived and can occur at any time (not just near perihelion), meaning that it is almost always discovered by chance in all-sky surveys \citep[e.g.,][]{birtwhistle2010_p2010a2,larson2010_scheila,smith2019_gault}.  As such, the vast majority of such events have undoubtedly gone undetected due to the fact that we are, and will continue to be for the foreseeable future, unable to observe all small solar system bodies at all times.
\\

\subsection{Photometric Detection}
\label{section:activity_detection_photometry}

Potentially very sensitive searches for activity  can be conducted by comparing measurements of the apparent magnitude, $m_{\lambda}(r_H,\Delta,\alpha)$, of an object in a filter with effective wavelength, $\lambda$, with predictions based on the absolute magnitude in the same filter, $H_{\lambda}$.  The absolute magnitude is defined by
\begin{equation}
    H_{\lambda} = m_{\lambda}(r_H,\Delta,\alpha) -5\log_{10}(r_H \Delta) + 2.5\log_{10}(\Phi(\alpha))
\end{equation}
where $r_H$ and $\Delta$ are the object's instantaneous heliocentric and geocentric distances, respectively, in au, and $0 < \Phi(\alpha) < 1$ is the phase function\footnote{The phase function is equal to the ratio of the scattered light at phase angle $\alpha$ to that at $\alpha$ = 0\degr. It is affected both by the apparent illuminated fraction of the surface  and by microscopic scattering effects in the regolith.}.  Photometric measurements that are brighter than would be expected from an object's absolute magnitude and rotational lightcurve amplitude would then imply the presence of unresolved ejected dust within the seeing disk of that object.
\\

Anomalous brightening was first used to identify the activity of (2060) Chiron \citep{Hartmann90}, with resolved coma directly imaged soon afterwards \citep{Luu90, Meech90}.  Since then, photometry has been used to search for and identify activity in other objects \citep[e.g.,][]{cikota2014_activeasts,hsieh2015_324p}.  Unfortunately, absolute magnitudes for most asteroids are computed from imprecise photometric data collected by the IAU's Minor Planet Center, limiting the accuracy with which their brightnesses can be predicted at any given time.  An additional limitation is imposed by the unknown rotational variability of many objects.  As a result, while photometric analysis provides very clear detections of activity for bright comets already showing resolved comae and tails \citep[e.g., Figure~7 of][]{Ferrin21}, results are  ambiguous for objects with poorly characterized nucleus properties, as is true of most asteroids.  We expect that photometric activity detection will become much more effective when data from the Vera C.\ Rubin Observatory (see Section~\ref{section:future}) become available. Those data will enable both the computation of precise absolute magnitudes for large numbers of small solar system bodies and precision photometry in individual observations for activity detection.
\\

\subsection{Spectroscopic Detection}
\label{section:activity_detection_spectroscopy}

Spectroscopy  is essentially the only means for acquiring unequivocal evidence of volatile sublimation and can also provide additional detail about the types of volatiles present.  In practice, however, spectroscopy has so far played only a small role in the study of asteroid-comet continuum objects.  
\\

At optical wavelengths, two major limitations of spectroscopy are lack of sensitivity to the dominant volatile (water) and the strong heliocentric dependence of resonance fluorescence emission.  Water has no observable bands in the optical, forcing the use of trace species (e.g., CN, which has an emission band around 3883\AA)  as proxies.  For example, one hour of spectroscopic observations targeting CN with the Keck telescope reaches production rates $Q_{CN}\sim10^{22}$~s$^{-1}$ at $r_H=1$~au, falling to $Q_{CN}\sim10^{24}$~s$^{-1}$ at $r_H=3$~au \citep{Jewitt12}.
The water production rate in JFCs is 400 times that of CN \citep{Ahearn95}, implying spectroscopic detection limits of $Q_{H_2O}\sim4\times10^{24}$~s$^{-1}$ ($\sim0.1$~kg~s$^{-1}$) at $r_H=1$~au but only $Q_{H_2O}\sim4\times10^{26}$~s$^{-1}$ ($\sim10$~kg~s$^{-1}$) at $r_H=3$~au.
%The water production rate in JFCs is $10^{2.6}$ times that of CN \citep{Ahearn95}, giving spectroscopic limits to $Q_{H_2O} \sim 10^{24.6}$ s$^{-1}$ ($\sim 0.1$ kg s$^{-1}$) at $r_H$ = 1 au but only $Q_{H_2O} \sim 10^{26.6}$~s$^{-1}$ ($\sim10$~kg~s$^{-1}$) at $r_H=3$~au.
This is why JFCs and LPCs that are visibly active at $\geq$3 au and active Centaurs at much largely distances commonly show no optical evidence for gas.   Worse, there is no fundamental reason why the ratio of CN to water in classical comets should be applicable to other objects like active asteroids, and in fact, reason to believe that it may be far lower \citep[e.g.,][]{prialnik2009_mbaice}, meaning that water production rate limits inferred from CN observations are of dubious value. %For all these reasons, optical spectroscopy has not played a leading role in the investigation of the asteroid-comet continuum objects.
\\

Spectroscopic observations at non-optical wavelengths are better suited to studying volatile sublimation (see chapter by Bodewits et al.) but have not been used extensively for the subjects of this chapter. One exception is the successful detection of outgassing by (1) Ceres in the form of multiple detections of the $1_{10}-1_{01}$ ground-state transition line of ortho-water at 556.939 GHz using the {\it Herschel Space Observatory} \citep{kuppers2014_ceres}. Detections of CO rotational transitions at submillimeter and millimeter wavelengths have also been secured in comets and Centaurs at large heliocentric distances (see chapter by Fraser et al.).
\\

\subsection{Debris Stream Detection}
\label{section:activity_detection_debris}

The association between meteoroid streams and active comets has been long recognized \citep[e.g.,][]{whipple1951_cometmodel2}. There is also  a growing number of identified associations between meteoroid streams and asteroids \citep[see][]{jenniskens2008_meteorshowerparents,jenniskens2015_meteoroidstreams_ast4}.  The latter associations have been interpreted as indicating unseen mass loss from those otherwise apparently inert bodies.
Famously, the association of the Geminid meteoroid stream with Phaethon \citep{Gustafson89,Williams93} led to the first suggestions that Phaethon was losing mass (see Section~\ref{section:phaethon}).  
Other suspected asteroidal meteor stream parents \citep[e.g.,][]{ryabova2002_geographos_meteors,babadzhanov2003_adonis,babadzhanov2015_extinctcomets_meteors} could also be experiencing unseen mass loss (see chapter by Ye and Jenniskens).  Divergence between the orbital elements of debris streams and their parents results from gravitational perturbations by the planets and, potentially, from unknown non-gravitational effects \citep[e.g.,][]{kanuchova2007_quadrantidparents}.
\\

Finally, several asteroids, notably the near-Earth asteroids (NEAs) 2201 Oljato and 138175 (2000 EE104), are associated with distinct and repeated disturbances in the magnetic field carried by the solar wind, called Interplanetary Field Enhancements (IFEs; Section~\ref{section:IFEs}), tentatively interpreted as drag from charged nanodust particles left behind along the orbits of these bodies.
\\

\subsection{Non-Gravitational Acceleration}
\label{section:activity_detection_NGA}
Anisotropic mass loss from a small body creates a recoil force that can measurably perturb the orbit relative to purely gravitational motion, where non-gravitational accelerations are well-known to exist for comets \citep{whipple1950_cometmodel1}. Sublimation-driven mass loss from a spherical nucleus of radius $r_n$ and density $\rho$ at rate $\dot{M}$ leads to an acceleration
\begin{equation}
    \alpha_{ng} = \frac{3 k_R V_g \dot{M}}{4\pi \rho r_n^3}
\end{equation}
in which $V_g$ is the gas velocity and $k_R \sim 0.5$ is a dimensionless constant to account for the fact that the flow is neither perfectly isotropic ($k_R$ = 0) nor perfectly collimated ($k_R$ = 1).  For illustration, for plausible values $V_g$ = 500 m s$^{-1}$, $\rho$ = 500 kg m$^{-3}$, and $r_n = 10^3$ m, we find $\alpha_{ng} = 10^{-10} \dot{M}$ m s$^{-2}$.  Expressed as a fraction of the local gravitational attraction to the Sun at 3 au ($g_{\odot} = 7\times10^{-4}$ m s$^{-2}$), this is $\alpha_{ng}/g_{\odot} \sim 1.7\times10^{-7} \dot{M}$, which should be detectable for the best astrometrically observed asteroids for production rates of $\dot{M} \sim 1$ to 10 kg $s^{-1}$ \citep{Hui17}.  This method is limited in its application mainly by the reliability of astrometry, particularly when plate-era data are used to obtain long astrometric arcs. % Yarkovsky acceleration \citep{Greenberg20}.  
\\

%Chiron: and even found to exhibit CN emission \citep{bus1991_chiron}.

%Another method for activity detection is the direct detection of sublimation products via spectroscopic observations.  While visible gas or dust emission activity commonly accompanies spectroscopic detections of sublimation products in classical comets, leaving no ambiguity

%\subsection{Debris Streams}
\label{section:debris_streams}

%Near-Earth Asteroids 2201 Oljato \citep{Russell84} and 138175 (2000 EE104) \citep{Lai17} are associated with distinct and repeated disturbances in the magnetic field carried by the solar wind, called Interplanetary Field Enhancements (IFEs).  These disturbances, detected by spacecraft crossing the orbits of the asteroids, typically last from minutes to hours and have profiles distinct from magnetic structures emanating from active regions in the Sun. 

%Numerous sub-populations exist in this space, some, like the active asteroids, only recently discovered.  Previous reviews relevant to this subject include \cite{Jew04}, \cite{Bertini11}, \cite{Jew12}, \cite{JHA15}, \cite{Snodgrass17} and \cite{Kasuga19}.

\section{\textbf{ACTIVITY MECHANISMS}}
\label{section:activity_mechanisms}

\subsection{Overview}
\label{section:activity_overview}

A variety of mechanisms in addition to sublimation can lead to observable mass loss, expanding the range of objects that we must consider to have the potential for activity. We briefly discuss the major mass loss mechanisms, which may operate alone or in conjunction.
\\

\subsection{Ice Sublimation}
\label{section:ice_sublimation}

%\begin{figure*}[htb!]
%\begin{center}
%\includegraphics[width=5.0in]{figures/fig_asteroid-family-mbc-formation2b.pdf}
%\end{center}
%\caption{Illustration of processes that could produce small bodies with near-surface ice that are more reasonably susceptible to impact-triggered sublimation-driven activity than larger and older parent bodies for which activation of sublimation-driven activity would be expected to be more difficult given the greater depth of buried ice.
%From \citet{hsieh2018_activeastfamilies}}
%\label{figure:asteroid_family}
%%\end{center}
%\end{figure*}

There is now substantial evidence that water ice is currently present (and possibly widespread) in objects with dynamically stable orbits in the main asteroid belt, and not just in objects from the outer solar system.  This revelation raises the possibility of being able to conduct present-day studies of ice possibly from a part of the protosolar disk not sampled by the classical comet population,
%has implications for constraining the regions of the protosolar disk in which planetesimals were able to accrete ice \citep[e.g.,][]{encrenaz2008_water,martin2012_snowline}, 
and also presents new opportunities for studies of the thermal and dynamical evolution of small bodies both in the early and present-day solar system \citep[e.g.,][]{schorghofer2008_mbaice,demeo2014_astbeltmapping,hsieh2016_tisserand}.
\\

Exposed, dirty (low albedo) water ice at asteroid belt distances is quickly lost by sublimation. However, \citet{fanale89} noted that water ice can persist at relatively shallow depths (e.g., a few centimeters to meters)
%, depending on latitude and obliquity) 
when protected by a nonvolatile surface layer. Its persistence is more likely at high latitudes on objects with larger semimajor axes, slow rotation rates, low thermal conductivity, and low obliquity.
%Shallow ice is susceptible to exposure by modest impacts \citep[see][]{Haghighipour16} or other disturbances of the surface.
\citet{schorghofer2008_mbaice} confirmed these findings, introducing the concept of a ``buried snowline'', i.e., the distance from the Sun beyond which subsurface ice can persist over the age of the solar system.  That said, near-surface asteroid ice is likely to be dominated by water ice, given the greater thermal instability of other volatiles in the asteroid belt \citep{prialnik2009_mbaice}.  \citet{schorghofer2020_icepreservation} combined models of the thermal and dynamical evolution of  Themis asteroid family members, finding that most of their subsurface ice would be retained even during transitions to near-Earth orbits, assuming stable spin pole orientations.  This raises the intriguing possibility that asteroid ice could one day be sampled by a near-Earth mission, and reminds us that the outer asteroid belt could be a source of at least some terrestrial planet volatiles.
\\

%\subsubsection{Sublimation-Driven Activity\label{section:sublimation_driven_activity}}

\begin{figure*}[htb!]
\begin{center}
\includegraphics[width=5.0in]{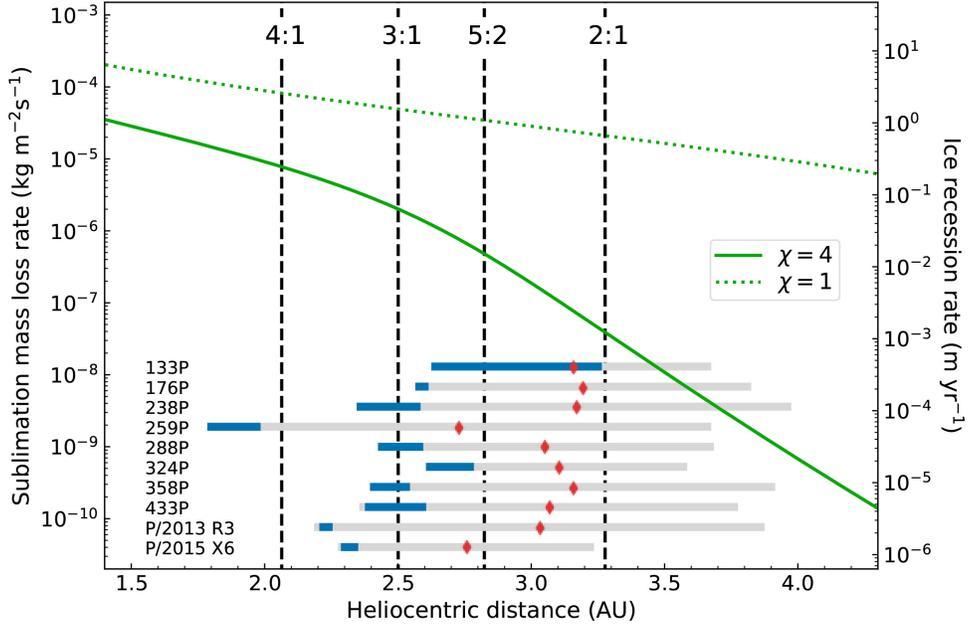}
\caption{Plot of mass loss rates due to water sublimation from a sublimating gray body and ice recession rates per year as functions of heliocentric distance over the range of the main asteroid belt using the isothermal approximation ($\chi=4$; solid curved green line) and subsolar approximation ($\chi=1$; dotted green line).  The positions of the 4:1, 3:1, 5:2, and 2:1 mean-motion resonances with Jupiter are marked with vertical dashed black lines. Also plotted are horizontal light gray bars showing the full heliocentric ranges of a selection of main belt comet (MBC) orbits (see Section~\ref{section:mbcs}) with overlaid blue bars showing the heliocentric range over which activity was detected for those MBCs, and red diamonds indicating each MBC's semimajor axis distance.
After \citet{hsieh2015_ps1mbcs}}.
\label{figure:thermal_plot}
\end{center}
\end{figure*}

Buried ice can be exposed and activated by  triggering events like impacts \citep[e.g.,][]{Haghighipour16}, rotationally-induced landslides \citep[e.g.,][]{scheeres2015_spinmassloss,steckloff2016_hartley2avalanches}, and inward drift of the perihelion distance \citep[e.g.,][]{Rickman90,Fernandez18}.
%revealing the true natures of these previously inert asteroid-like objects as active comets.  
This is only possible, however, if the ice depth is shallow, because deep disturbance of the surface is unlikely.   Shallow ice could exist on fragments from the catastrophic disruption of larger parent bodies, as in those that form asteroid families \citep[see][]{nesvorny2008_beagle,nesvorny2015_astfam_ast4,novakovic2012_288p,novakovic2022_asteroidfamilies}.  Ice-containing fragments would be more easily triggered by, for example, small impacts excavating just a few meters of the surface 
%\citep[Figure~\ref{figure:asteroid_family};][]{hsieh2018_activeastfamilies}.
\citep{hsieh2018_activeastfamilies}.
\\

Ice sublimation rates are very sensitive to temperature, such that even modest temperature differences between perihelion and aphelion give rise to substantial variations in activity strength, and therefore detectability.  Neglecting a small term due to conduction, the energy balance equation at the surface of a sublimating gray body may be written
\begin{equation}
{F_{\odot}\over r_H^2}(1-A) = \chi\left[{\varepsilon\sigma T^4 + L f_D\dot m_{w}(T)}\right]
\label{equation:sublim1}
\end{equation}
where $T$ is the equilibrium surface temperature of the body, $F_{\odot}=1360$~W~m$^{-2}$ is the solar constant, heliocentric distance $r_H$ is in au, $A=0.05$ is the assumed Bond albedo, the distribution of solar heating is parameterized by $\chi$, $\sigma$ is the Stefan-Boltzmann constant, $\varepsilon=0.9$ is the assumed effective infrared emissivity, $L=2.83$~MJ~kg$^{-1}$ is the (nearly temperature-independent) latent heat of sublimation of water ice, $f_D$ represents the reduction in sublimation efficiency caused by a growing rubble mantle, where $f_D=1$ in the absence of a mantle, and $\dot m_w$ is the water mass loss rate due to sublimation.
In this equation, $\chi=1$ applies to a flat slab facing the Sun at all times, known as the subsolar approximation, and produces the maximum attainable temperature, while $\chi=4$ applies to an isothermal surface (e.g., for a fast-rotating body or low thermal inertia), and corresponds to the minimum expected temperature.
\\

The sublimation rate of ice into a vacuum is given by
\begin{equation}
\dot m_{w} = P_v(T) \left(\frac{\mu m_H}{2\pi k T}\right)^{1/2}
\label{equation:sublim3}
\end{equation}
where $m_H = 1.67\times10^{-27}$ kg is the mass of the hydrogen atom, $\mu=18$ is the molecular weight of water, and $k$ is the Boltzmann constant.  The   corresponding ice recession rate, $\dot \ell_{i}$,  is given by 
$\dot \ell_{i} = \dot m_{w}/ \rho$, for an object with a bulk density $\rho$.  
The temperature-dependent sublimation pressure, $P(T)$, is obtained from the Clausius-Clapeyron relation, or from experimental data.  
Solving Equations~\ref{equation:sublim1} and \ref{equation:sublim3} iteratively gives the equilibrium temperature and the sublimation rate of a sublimating gray-body at a given heliocentric distance.
\\

Expected water sublimation rates computed using the equations above, assuming $f_D=1$, for the extreme subsolar and isothermal approximations are plotted in Figure~\ref{figure:thermal_plot}.  We see from the figure that even objects whose orbits keep them entirely confined to the main asteroid belt have expected water sublimation rates that can vary by several orders of magnitude from aphelion to perihelion.  For main-belt objects that are only weakly active near perihelion, despite their modest eccentricities, it is therefore entirely plausible for sublimation in these objects to be detectably strong near perihelion and undetectably weak near aphelion, just like classical comets with much larger perihelion-to-aphelion heliocentric distance excursions.  
\\

\begin{figure}[htb!]
\begin{center}
\includegraphics[width=3.0in]{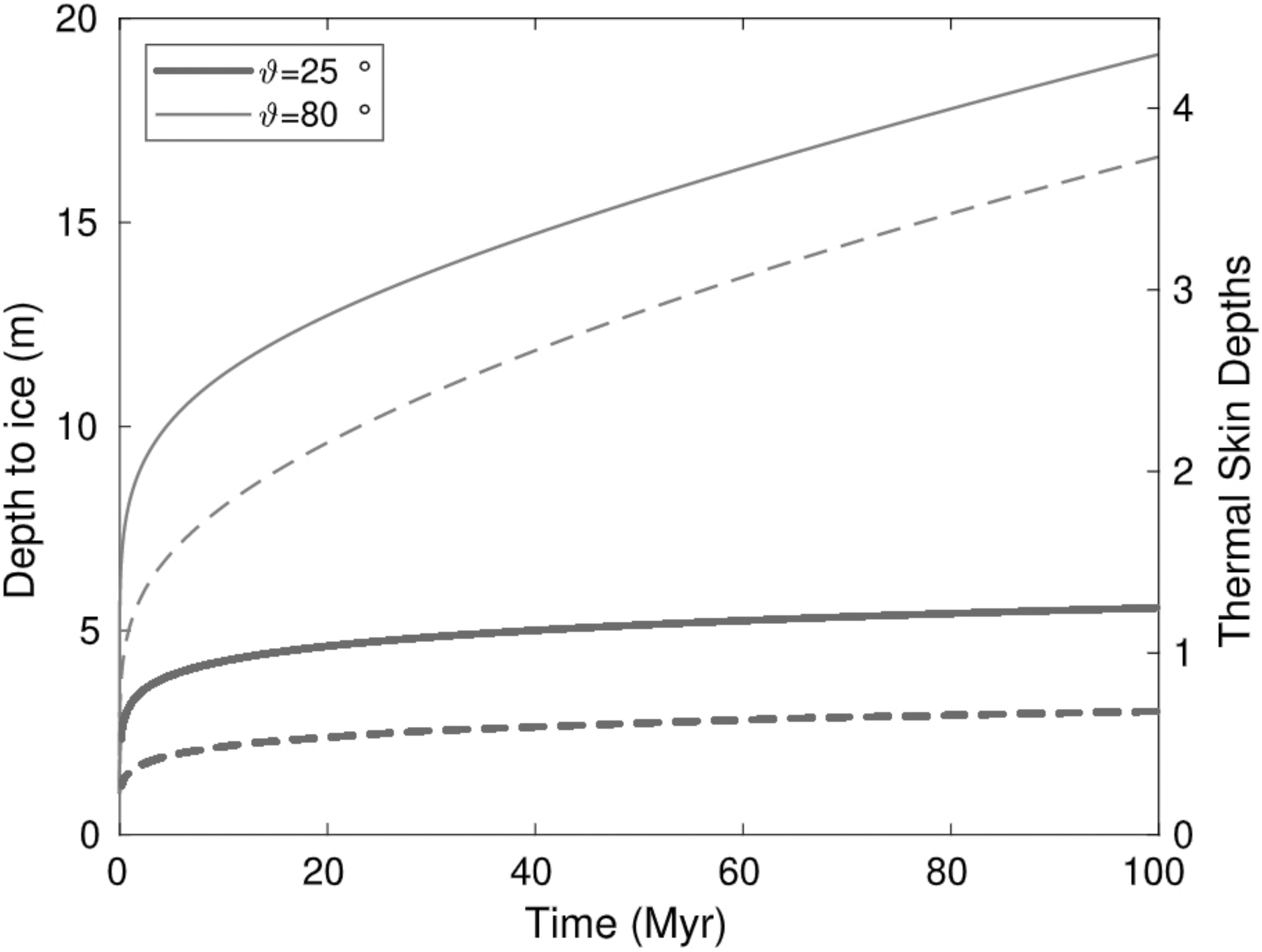}
\caption{Plot of depth to ice in a thermal model as a function of time for a main-belt asteroid in a stationary orbit with $a=3$~au and $e=0$ for obliquities of $\vartheta=25^{\circ}$ (lower set of curves) and $\vartheta=80^{\circ}$ (upper set of curves). The assumed thermal diffusivity is $\kappa = 3.8\times10^{-7}$ m$^2$ s$^{-1}$. Within the first thermal skin depth, ice retreat is rapid, but total retreat is small for $\vartheta=25^{\circ}$, even after 100~Myr.  Solid lines mark ice evolution using uniform thermal properties for all regolith, while dashed lines mark ice evolution using different thermal properties for ice-free and ice-rich regolith.
From \citet{schorghofer2020_icepreservation}}
\label{figure:ice_preservation}
\end{center}
\end{figure}

Observationally, the expected characteristics of sublimation-driven activity include  long-duration mass loss and recurrent activity \citep[e.g.,][]{hsieh2012_scheila}.  Unlike mass loss produced by impulsive events like impacts (Section~\ref{section:impacts_general}), sublimation-driven activity is expected to persist  as long as an object is warm enough for sublimation to occur and volatile material remains.  Determining the duration of emission events often requires the use of numerical dust models, however, which are typically under-constrained by observations.  Recurrent activity correlated with perihelion strongly suggests thermal modulation of the mass loss process, which is a natural characteristic of volatile sublimation.  Other mechanisms require more contrived conditions to produce similar behavior \citep[e.g.,][]{hsieh2004_133p}, leading us to consider recurrent mass loss near perihelion to be the most dependable indicator of sublimation-driven activity currently available using remote observations.
\\

\subsection{Impacts}
\label{section:impacts_general}

Images of asteroids from spacecraft reveal cratered landscapes that record a violent collisional past.  Observations of the consequences of impacts can place important constraints on the material properties and internal structure of the bodies that experienced those impacts, motivating decades of crater studies \citep{marchi2015_cratering_ast4}, theoretical modeling \citep{jutzi2015_impacts_ast4} and impact experiments in laboratory settings \citep[e.g.,][]{housen2018_impactsporousasteroids}, while impacts can also excavate subsurface material which can then be directly studied. Excitingly, impact experiments on real-world solar system bodies \citep[e.g.,][]{ahearn2005_deepimpact1,schultz2010_lcross,saiki2017_hayabusa2impact} as well as studies of natural impact events occurring in real time (see Section~\ref{section:active_asteroids_354P_Scheila}) have also recently become possible.
The velocity dispersion in the asteroid belt is $\sim$5 km s$^{-1}$, resulting in intermittent asteroid-asteroid collisions that are highly erosive or, if sufficiently energetic, destructive.  Collisions also occur at the smallest scales, leading to steady erosion by micrometeoroids.  Although small and large impacts are part of a continuum, it is useful to describe these extremes separately.
\\

%Collisions  loft dust at velocities exceeding an object's escape velocity \citep[e.g.,][]{ishiguro2011_scheila2}, creating a observable, but transient, dust cloud.  The rates at which impacts are expected to occur in specific regions of the solar system are difficult to estimate, given that observational limitations limit our ability to constrain the abundance and distribution of potential impactors at the size scale (meter-sized) that are of particular interest.  Nonetheless, current large-scale wide-field surveys have been able to detect multiple such events in progress over the last decade.

\subsubsection{Micrometeoroid Erosion}
\label{section:micrometeorites}

Micrometeoroid impacts set an absolute lower limit to the rate of steady mass loss from a body.  Few impacting  particles can penetrate deeply because of the steep size and energy distributions of projectiles, and so the erosion is fastest at the surface.  On the Moon, where the gravity is non-negligible ($g_{\leftmoon} = 1.6$~m~s$^{-2}$),  fragments produced by micrometeoroid impact are recaptured, forming a regolith that is overturned or ``gardened'' by continued impacts.  Measurements of the lunar gardening rate show that the top 1 m of the regolith is over-turned in about 10$^9$ yr while the top 1 cm is mixed in only 10$^6$ yr \citep{Arvidson75,Heiken91,Horz97}.  To set a crude upper limit to asteroid erosion, we assume that the material that is gardened on the Moon would instead be launched above escape speed and lost from a km-sized (low gravity) asteroid, and that the impact rates on the Moon and in the asteroid belt are comparable. We take the lunar surface value,  $dr/dt = 10^{-8}$ m yr$^{-1}$, as a representative rate of loss.   With $\rho = 10^3$ kg m$^{-3}$, the micrometeorite erosion mass flux is $\mathcal{L} = \rho dr/dt \sim 10^{-5}$ kg m$^{-2}$ yr$^{-1}$.
\\

Sandblasting by micrometeorites, at a rate given by $\dot{M_{\mu}  = 4\pi r_n^2 \mathcal{L}}$ for a spherical body of radius $r_n$,  is evidently very slow.  Expressing $r_n$ in km, we obtain
\begin{equation}
    {\dot{M_{\mu}}} = 4\times10^{-6} r_n^2~~\textrm{kg~s}^{-1}.
    \label{micrometeorite}
\end{equation}
A  $r_n$$=$1~km body would lose $\sim$4$\times$$10^{-6}$~kg~s$^{-1}$, and could do so indefinitely. Only asteroid belt boulders are limited by micrometeoroid impact.  For example, a $r_n$$=$1~m boulder would have a  lifetime against micrometeoroid erosion, $M/\dot{M_{\mu}} \sim$3$\times$$10^7$ yr. Other than for 0.25 km radius Bennu, whose $\sim$ 10$^{-7}$ kg s$^{-1}$  mass loss rate \citep{Hergenrother20} is $\sim$$0.4\dot{M_{\mu}}$ by Equation \ref{micrometeorite}, all remote detections of asteroid, comet, and Centaur activity have thus far exceeded $\dot M_{\mu}$ by 3 to 10 orders of magnitude.   Micrometeoroid impact erosion is an unlikely source of observable asteroid activity.
\\

\subsubsection{Macroscopic Impacts}
\label{section:macroscopic}

Most of the mass and available impact energy in the asteroid population is carried by the largest projectiles, but large bodies and impacts are rare and their effects are  transient.  Nevertheless, individual large impactors are easily capable of generating  observable   quantities of dust.  Several impact events have been detected in the main asteroid belt within the last decade. The best example of impact destruction of an asteroid is provided by the 100 m scale 354P/LINEAR \citep[Section~\ref{section:active_asteroids_354P_Scheila};][]{Jewitt10,Snod10,Kim17b}.  P/2016 G1 may also be a collisionally disrupted asteroid of similar scale \citep{moreno2016_p2016g1}. Similar collisions must also occur in the Kuiper belt but have yet to be observed. \\

%Compared to fully destructive collisions, collisions resulting in less than total destruction, so-called ``cratering collisions'', should be more common. The minimum size of cratering impact that gives rise to an observable signature is a function of many poorly known parameters \citep[e.g.,][]{mcloughlin2015_outburstphotometry}.  For example, the mass and size distribution of the ejecta determine the peak brightness and persistence time (because  particles are cleared  by solar radiation pressure, which acts in proportion to the reciprocal particle size). Given better determinations of the rate of such collisions, we should be able to map the spatial distribution of small (meter scale) impactors through the asteroid belt.

\begin{figure}[htb!]
\begin{center}
\includegraphics[width=2.5in]{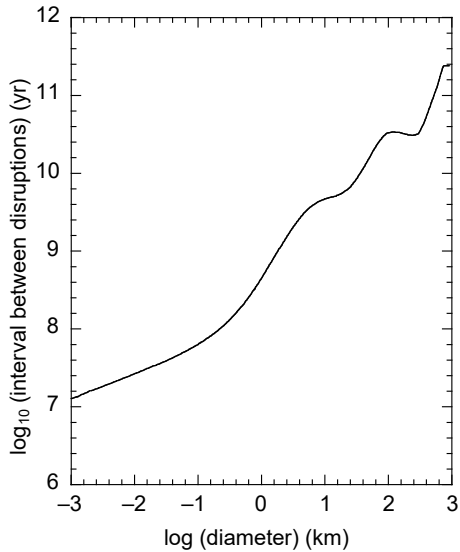}
\caption{Interval between collisional disruptions of main-belt asteroids as a function of their diameter in the range 1 m to 1000 km. Adapted from \citet{Bottke05}.
}
\label{bott1}
\end{center}
\end{figure}

\begin{figure}[htb!]
\begin{center}
\includegraphics[width=2.5in]{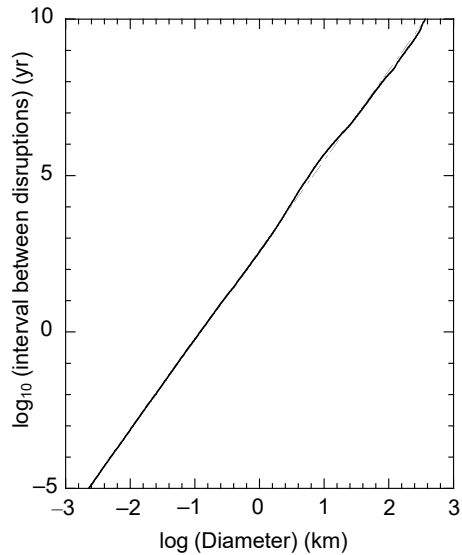}
\caption{Interval between collisional disruptions as a function of asteroid diameter, integrated over the asteroid size distribution. Adapted from \citet{Bottke05} 
}
\label{bott2}
\end{center}
\end{figure}

Figure~\ref{bott1} shows the timescale for the impact destruction of a given main-belt asteroid as a function of its diameter, $D$, where we see that a $D=1$~km asteroid has a collisional lifetime $\sim$0.4~Gyr, decreasing to $\sim$60 Myr for $D=0.1$~km.  Asteroids larger than $D\sim10$~km, on average, should survive against collisions over the 4.5~Gyr age of the solar system.  The likelihood of witnessing a destructive collision on any {\it specific} asteroid in a human lifetime is therefore negligibly small.  However, 
%there are many asteroids of a given diameter, and 
the likelihood of observing a collisional destruction \textit{somewhere} in the asteroid belt, when integrated over the asteroid size distribution, is much higher (Figure~\ref{bott2}).  Specifically, $\tau_{CD}$ (yr), the interval between successive destructions averaged over the whole main belt is well represented by 

\begin{equation}
\log_{10} \tau_{CD} = 2.62 + 2.88 \log_{10} D.
\label{tauCD}
\end{equation}

\noindent for $D$ in km (see Figure~\ref{bott2}). Equation~\ref{tauCD} gives $\tau_{CD} \sim400$~yr for $D=1$~km, falling to $\tau_{CD}\sim1$~yr for $D =0.1$~km and $\tau_{CD}\sim30$~s (!) for $D$ = 1 m. Independent modeling of survey data suggests that up to 10 catastrophic disruption events brighter than $V=18.5$~mag may be discoverable each year by current and future asteroid surveys \citep{Denneau15}.
\\

Compared to catastrophic collisions, collisions resulting in less than complete destruction, so-called ``cratering collisions'', should be more common. The minimum size of cratering impact that gives rise to an observable signature is poorly modeled, being a function of many uncertain parameters \citep[e.g.,][]{mcloughlin2015_outburstphotometry}.  For example, the mass and size distribution of ejecta determine the peak brightness and persistence time (because particles are cleared by solar radiation pressure, which acts in inverse proportion to the particle size). Given better collision rate determinations, we should be able to map the spatial distribution of small (m-scale) impactors through the asteroid belt.
\\

A cratering impact between a 20 m to 40 m scale projectile and the 113 km diameter asteroid (596) Scheila (see section \ref{section:active_asteroids_354P_Scheila}) was a one in $\sim$10$^4$ year event for this object but, given that there are of order 200 asteroids of similar or larger size, should occur once every $\sim50$ years somewhere in the asteroid belt \citep{Jewitt12}.  As monitoring of the sky improves, we should soon be able to detect cratering collisions throughout the asteroid belt in great abundance (see Section~\ref{section:future}).
\\

%\citet{housen2011_craterejecta} - equations for scaling laws for impacts in gravity and strength regimes; impact effects depend on impactor radius, velocity, and mass density, and material properties (e.g., in particular, composition and porosity) of the target body

%Amount of ejecta

%Fading rates of impact ejecta

%Impact rate estimates \citep{denneau2015}

\subsection{Rotational Destabilization}
\label{section:rotation_general}

Rotational instability causes mass loss when the centrifugal forces exceed the sum of gravitational and cohesive forces acting towards the center of mass. 
Like impact disruptions (Section~\ref{section:impacts_general}), rotational disruptions can place important constraints on a body's interior properties \citep[e.g.,][]{hirabayashi2015_rotationalshedding}.
%Suspected examples of solar system bodies that have exhibited mass loss events due to rotation include (6478) Gault \citep[e.g.,][]{devogele2021_gault}, (62412) 2000 SY$_{178}$ \citep{sheppard2015_sy178},
For a strengthless $a\times b\times c$ ellipsoid, with $a = b \le c$, bulk density $\rho$, and rotation about a minor (minimum energy, maximum moment of inertia) axis, the critical, size-independent rotation period is
\begin{equation}
    P_c =  \left(\frac{3\pi}{G \rho}\right)^{1/2} \left(\frac{c}{a}\right)
    \label{equation:tc}
    \end{equation}
\noindent where $G = 6.67\times10^{-11}$ N kg$^{-2}$ m$^2$ is the gravitational constant.  For a sphere, $c/a$ = 1, with $\rho = 10^3$ kg m$^{-3}$, we find $P_c =$ 3.3 hours. This rises to $P_c$ = 4.0 hours for a body with  $c/a = 1.2$ \citep[the modal axis ratio of small asteroids from][]{Szabo08},  with the instability occurring at the tips of the ellipsoid. Several active asteroids rotate with periods comparable to $P_c$ (c.f.~Table \ref{AAs}), raising the possibility that rotational instability plays a role in the ejection of material from those objects.
\\

\begin{figure}[htb!]
\begin{center}
\includegraphics[width=8.0cm]{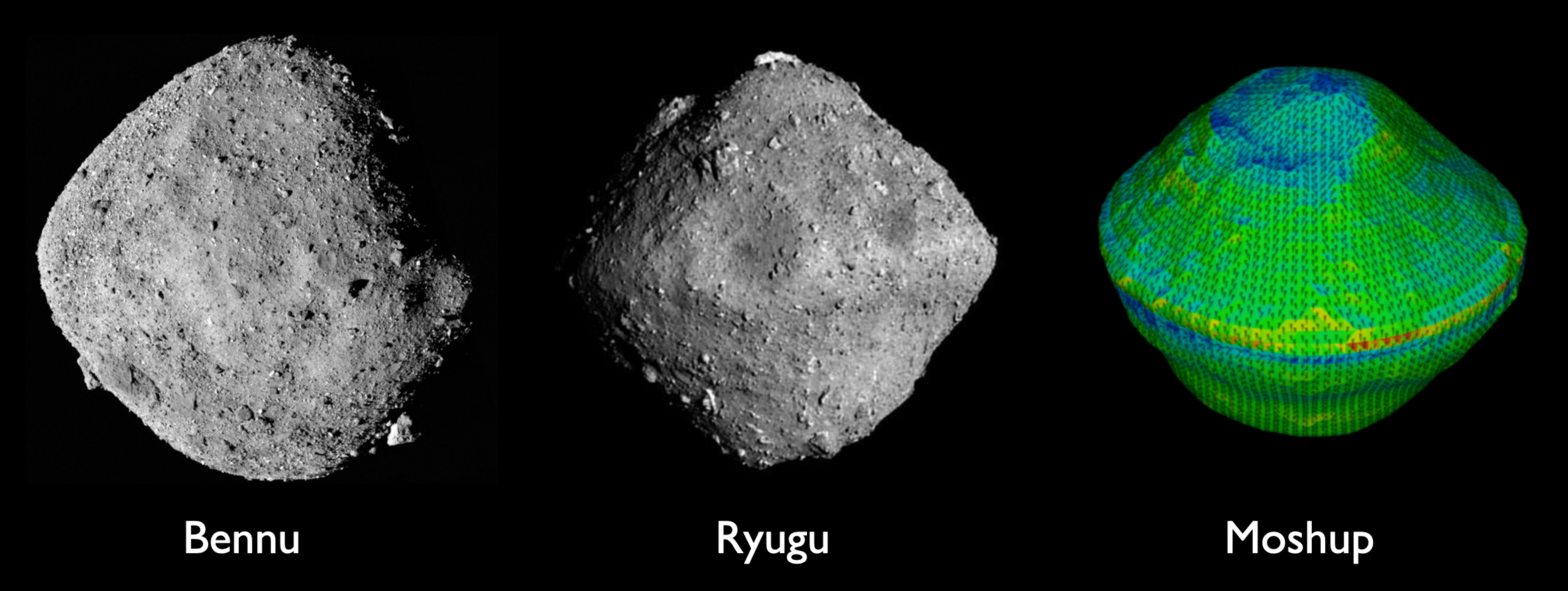}
\caption{Three examples of rotationally-shaped asteroids with equatorial skirts.  % caused by equator-ward movement of material.  
(101955) Bennu, (162173) Ryugu and (66391) Moshup have diameters 500 m, 900 m and 1300 m, respectively.  Bennu and Ryugu were
%from NASA and JAXA spacecraft, respectively, 
imaged by spacecraft,
while the Moshup image was derived from radar %scattering
observations.
Image credits: NASA/Goddard/Univ.\ of Arizona (Bennu); JAXA, Univ.\ of Tokyo, and collaborators (Ryugu); \citet{ostro2006_1999kw4} (Moshup).
%, where radar observations were also used to determine that (65803) Didymos (the target of NASA's Double Asteroid Redirection Test mission, or DART) has a similar top shape with an equatorial bulge \citep{naidu2020_didymos}.
}
\label{figure:muffins}
\end{center}
\end{figure}

A complicating factor in the application of Equation~\ref{equation:tc} is that even very modest values of material cohesion (used here to mean any combination of tensile strength and shear strength) can hold together a rotating asteroid.  To see this, we use the rotational energy density in a  uniform sphere as an estimate of the average rotational stress, $S_0$,  obtaining
\begin{equation}
    S_0 \sim  \rho V_{eq}^2,
\end{equation}
\noindent where $V_{eq} = 2\pi r/P_c$ is the equatorial velocity.  With the above parameters and $r$ = 1 km, we find $S_0 \sim$ 280 N m$^{-2}$, which is easily overcome by material cohesion.  
Most asteroids and comets likely possess a weak, rubble-pile structure in which  gravity, rotation and van der Waals forces all play a role. The magnitude of the latter is $\sim S_0$ \citep{Scheeres10,Sanchez14,Sanchez21}, meaning that even weak cohesion can nevertheless substantially control a body's mass loss and shape.  Bodies with a weakly cohesive regolith are subject to mass movement and, eventually, mass shedding as particulate material escapes from the equator.  Evidence for this process is compelling in ``top-shaped'' asteroids (Figure~\ref{figure:muffins}), whose shapes are formed by skirts of material that has migrated from higher latitudes towards the equator \citep{harris2009_asteroidshapesspins}.  
\\

\begin{figure}[htb!]
\begin{center}
\includegraphics[width=7.0cm]{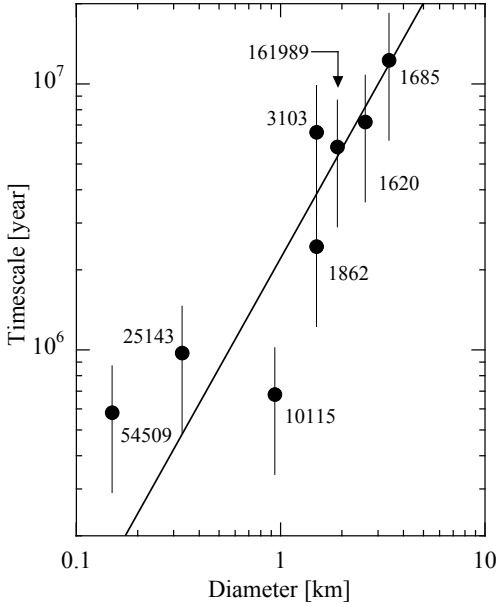}
\caption{Empirical YORP timescales for small near-Earth asteroids (marked by their numerical designations) as a function of diameter, using published measurements of spin-period changes. The plotted objects have semimajor axes $a$ = 1.0 to 1.5 au, and the plotted line is given by Equation~\ref{tauY2}.}
\label{figure:YORP}
\end{center}
\end{figure}

The spin of an inert body can change because of radiative torques \citep[e.g., the Yarkovsky–O'Keefe–Radzievskii–Paddack, or YORP, effect;][]{bottke2006_yarkovsky}, torques from outgassing if volatiles are present \citep{Jewitt21}, and impacts \citep{Marzari11}.  Eight asteroids in the published literature display evidence for spin-up, namely (1620) Geographos \citep[which interestingly is also a suspected meteor stream parent;][also see Section~\ref{section:activity_detection_debris}]{ryabova2002_geographos_meteors}, (1862) Apollo, (3103) Eger, (25143) Itokawa, (54509) YORP, and (161989) Cacus,   (1685) Toro, and (10115) 1992 SK \citep{Rozitis13,Lowry14,Durech18,Durech21}.  Figure~\ref{figure:YORP} shows characteristic spin-up timescales for these asteroids computed using $\tau = \omega/\dot{\omega}$, where $\omega$ is the measured angular frequency, as a function of diameter, where the 10 m scale asteroid 2012 TC$_{4}$ shows evidence for multiple periods and spin-up \citep{Lee21} but is not plotted here. Uncertainties for $\tau_Y$ are dependent on uncertainties for the albedo, size, shape, density and thermal properties of the measured asteroids. For illustration, we show error bars indicating $\pm$50\% uncertainties. 
\\

The timescale for spin-up by YORP torque, $\tau_Y$, is expected to vary as $\tau_Y = \Gamma  a_{au}^2 D^2$, where $\Gamma$ is a constant, $a_{au}$ is the semimajor axis in au, and $D$ is the effective diameter.  The $a_{au}^2$ dependence results from the fading of sunlight described by the inverse square law.  The $D^2$ dependence arises because the timescale is given by the ratio of the moment of inertia (which varies as $D^5$) to the torque (given by the product of the surface area and the moment arm for the torque, together $\propto D^3$).  The constant, $\Gamma$,  must be determined empirically, since it is dependent on many unknown details of each asteroid.  Under the assumption that the spin-up of objects in Figure~\ref{figure:YORP} is due to YORP, we compute a least-squares fit for the YORP timescale given by

\begin{equation}
    \tau_Y = (1.4\pm0.3)\times10^6 a_{au}^2 D^{1.4\pm0.3}
     \label{tauY2}
\end{equation}

\noindent which is shown in the figure as a solid line.  The best-fit  index, $g = 1.4\pm0.3$, is consistent  with the expected value of $g$ = 2 to within two standard deviations\footnote{The small difference, if real, could result from observational selection against the detection of small asteroids with short spin-up timescales. Such objects are more likely to have been spun-up to destruction and will not survive to be measured}. If we instead force a $D^2$ dependence, a  fit to the data gives $\tau_Y = (0.7\pm0.1)\times10^6 a_{au}^2 D^{2} $.  Both fits (as well as Figure~\ref{figure:YORP}) indicate that a 1 km diameter object located at $a_{ au}$ = 3 au should have a characteristic spin-up time  from $\tau_Y \sim 6 \textrm{ to } 13$ Myr, with a considerable scatter due to intrinsic differences (in shape, roughness, thermal diffusivity, spin vector magnitude and direction) between the asteroids.  Scaled to $a_{au}$ = 3 au we take $\tau_Y = 10$ Myr as a middle value. This is much shorter than the $\sim$0.4 Gyr collisional lifetime of a 1 km diameter main-belt asteroid \citep{Bottke05}, showing the potential importance of rotational breakup in weak asteroids.  If sublimating volatiles are present, equally rapid spin-up can be achieved by tiny (e.g.~$10^{-3}$ kg s$^{-1}$) mass loss rates, provided they are sustained over long times \citep{Jewitt21}.    
\\

A major unknown in Figure~\ref{figure:YORP} and Equation~\ref{tauY2} concerns the long-term stability of the applied torque.  Models show that even very small changes in surface topography can result in significant changes in the magnitude and even direction of the YORP torque \citep[e.g.,][]{Statler09}.  Topographical changes can result from YORP itself (leading to feedback  in which the change in shape caused by the equatorward movement of material can alter the body shape and YORP torque that caused the movement), or from cratering by impacts.  As a result, instead of driving inexorably towards breakup, asteroids more likely migrate erratically towards it.  For example, the ``top-shaped'' objects in Figure~\ref{figure:muffins} are not necessarily all rotating near break-up at the present time; the periods of Bennu, Ryugu, and (66391) Moshup are 4.3, 7.6 and 2.8 hours, respectively.  This scatter reflects the temporal instability of the YORP torque and, perhaps, the past loss of angular momentum and subsequent de-spinning by mass redistribution or shedding in these objects.
\\

The stochastic effects of impacts can rival torques due to YORP for bodies in the size range of interest here \citep{Wiegert15}. In impact spin-up, the dominant effects are from the small number of the largest projectiles \citep{Holsapple19}.  In the limiting case, an asteroid can be impact-disrupted, forming an unstable disk from which the asteroid re-accretes mass and angular momentum.  Reassembly after impact is  still capable of producing top-shaped bodies like Bennu and Ryugu \citep{Michel20}.
\\

\subsection{Radiation Pressure Sweeping}

Solar radiation can facilitate mass loss in two ways: via direct radiation pressure and thermal radiation pressure.  
In the first case, a dust grain detached from the surface of a spherical asteroid of mass $M_n$ and radius $r_n$  feels both a gravitational attraction, $g = G M_n/r_n^2$, towards the asteroid and an acceleration away from the Sun, $\alpha$, caused by solar radiation pressure.  We write $\alpha = \beta g_{\odot}$, where $g_{\odot} = G M_{\odot}/r_H^2$ is the gravitational acceleration to the Sun, whose mass is $M_{\odot} = 2\times 10^{30}$ kg, and $\beta$ is the dimensionless radiation pressure efficiency of a particle of radius $a$. For dielectric spheres, $\beta \sim a_{\mu{\rm m}}^{-1}$, where $a_{\mu{\rm m}}$ is the particle radius in units of microns.  Whereas $g$ is independent of $r_H$, $\alpha$ varies as $r_H^{-2}$, meaning that small asteroids close to the Sun are most susceptible to dust loss from radiation pressure sweeping.    For a spherical asteroid of density $\rho$,  $\alpha/g > 1$ is satisfied when 
\begin{equation}
a_{\mu{\rm m}} < \frac{3 M_{\odot}}{4\pi\rho r_n r_H^2}
\end{equation}
\citep{Jewitt12}.  Equivalently, 
\begin{equation}
    a_{\mu{\rm m}} \lesssim 10 \left(\frac{1~\textrm{km}}{r_n}\right) \left(\frac{1~\textrm{au}}{r_H}\right)^2.
\end{equation}
At 3~au, a km-sized asteroid can only lose sub-$\mu$m grains but on Phaethon ($r_n = 3$~km) at perihelion ($r_H =0.14$~au), for example, particles up to $a_{\mu{\rm m}} \sim 170$~$\mu$m can be swept away.  This is only possible, however, for particles which are both detached from the asteroid by another process (i.e., not subject to cohesive forces binding them to the surface) and near the illuminated limb (otherwise the radiation pressure accelerates the particles downwards into the surface).
Meanwhile, sunlight absorbed and thermally re-radiated has the same effect as direct radiation pressure, but with the advantage that thermal radiation is always directed outwards and so dust can be expelled from the entire dayside of the asteroid \citep{Bach21}.   
\\

Neither direct nor thermal radiation pressure can detach small dust particles held to an asteroid surface by contact forces.  Once these are broken, however, by sublimation drag, thermal fracture, or other processes, radiation pressure acceleration provides a mechanism to remove that dust.  As such, while there are currently no known active objects for which radiation pressure sweeping is considered a dominant driver of activity, it is possible that either direct or thermal radiation pressure may serve to enhance the dust ejection efficacy of other mechanisms.
\\

\subsection{Electrostatic ejection}
Solid surfaces exposed to solar UV photons become photoionized, with the loss of electrons resulting in a net positive sunlit surface charge \citep[e.g.,][]{Criswell73}.  Meanwhile, photoelectrons re-impacting the surface concentrate in shadowed regions. As a result, spatial gradients in illumination of a surface, for example at the terminator or between the sunlit side of a particle and its shadow, produce local electric fields.  These fields are capable of moving regolith dust particles, provided that inter-particle cohesive forces can be overcome.  This has been known since the Lunar Surveyor missions of the mid-1960s, when unexpected horizon glow was detected after sunset and interpreted as forward scattering from $\sim$10 $\mu$m sized regolith particles at heights of $h \lesssim$ 1 m above the lunar surface \citep{Criswell77}.  To reach $h$  requires an ejection speed $V = (2 g_{\leftmoon} h)^{1/2} \sim 1.8$ m s$^{-1}$, where $g_{\leftmoon}$ = 1.6 m s$^{-2}$ is lunar surface gravity.   In turn, $V$ is comparable to the gravitational escape speed from a km-sized body, meaning that electrostatic ejection is  a possible mechanism behind activity in asteroids \citep{Jewitt12}.  Spacecraft-imaged dust ``ponds'' settled in local gravitational potential minima on  the 17 km diameter asteroid (433) Eros  provide additional evidence for the mobility of surface dust, likely influenced by electrostatic forces \citep{Colwell05}.  
\\

The physics of electrostatic launch is complicated and still under investigation, particularly concerning the question of how electrostatic repulsion  can overcome attractive van der Waals contact forces that bind small particles together.  One recent   finding is that dust ejection can occur because of the development of small scale but extremely large (10$^5$ to 10$^6$ V m$^{-1}$)  electric fields  in cavities between grains in a porous regolith.  This is the so-called ``patched charge'' model \citep{wang16,Hood22}.  Experiments show that  particles up to 10 to 20 $\mu$m can be ejected by these strong fields at m~s$^{-1}$ speeds.  
\\

Unlike on the Moon, where electrostatically launched  particles are pulled back by gravity, dust is easily lost from small asteroids, and surface charging drives a net loss of surface fines.   \citet{Criswell73} estimated the overturning or ``churn rate'' of lunar dust due to electrostatic levitation to be $C$ = 0.1 kg m$^{-2}$ yr$^{-1}$.   We  estimate  an upper limit to electrostatic dust loss from a small asteroid by assuming that all this churned material will be lost, at a rate  $\dot{M} \sim 4\pi r_n^2 C$. Substituting $r_n$ = 1 km gives $\dot{M} \sim 0.04$ kg s$^{-1}$, a value that is  small compared to the $\sim$ 1 kg s$^{-1}$ rates inferred in the active asteroids, but large enough to be telescopically detectable under ideal circumstances. However, this estimate is an upper limit both because some particles will fall back to the surface, and because the lunar data show that electrostatic churning is spatially localized near shadow boundaries, not global.  \\

The fundamental problem for electrostatic ejection from asteroids, however, is that   $C \gg \mathcal{L}$, where $\mathcal{L}$ is the rate of production of fresh particles by micrometeoroid impact (Section~\ref{section:micrometeorites}). With no adequate source of replenishment, detachable surface dust on a small asteroid would be quickly depleted.  An additional concern is that electrostatic charging is a very general process, dependent only on the presence of ionizing radiation and dust.  If electrostatic losses are important, why would only some km-sized asteroids  be active rather than all of them?  For these reasons, we must conclude that electrostatic loss is unlikely to be a major contributor to activity on most  solar system bodies.
\\

\subsection{Thermal fracture\label{section:thermal_fracture}}

Thermal fatigue, defined as weakening of a material through the application of a cyclic temperature-related stress, has been suggested as a driver of the particle loss from Phaethon and Bennu (Sections~\ref{section:phaethon} and \ref{section:bennu}). The fractional expansion resulting from a temperature change $\Delta T$ is $\alpha \Delta T$, where an expansivity of $\alpha \sim 10^{-5}$ K$^{-1}$ is typical \citep{Konietzky92}.  The  resulting strain is $S \sim  \alpha Y \Delta T$, where $Y$ is Young's modulus and a multiplier of order unity, called Poisson's Ratio, is ignored.  $Y$ values for rock are very large  (e.g.~$10^{10}$ N m$^{-2}$ to  $10^{11}$ N m$^{-2}$) meaning that huge stresses of $S \sim 10^5 \Delta T$  to $10^6 \Delta T$ (N m$^{-2}$) can result from modest temperature excursions. For example, diurnal temperature variations $\Delta T \sim$ 10$^2$ K, common on airless bodies, can generate estimated stresses of $S \sim 10^7$ to 10$^8$ N m$^{-2}$. Since rocks are especially weak in tension (e.g.~the tensile strength of basalt is only $\sim4\times10^6$ N m$^{-2}$), thermal fracture is expected to be an important erosive process in space, even at asteroid belt distances and temperatures \citep{Molaro15}.  
\\

In a homogeneous material, the temperature gradients and resulting thermal stresses are largest on length scales comparable to the thermal skin depth, $\ell_D \sim (\kappa P)^{1/2}$, where $\kappa$ is the thermal diffusivity and $P$ is the timescale for the insolation.  By substitution, the skin depth for a solid rock asteroid ($\kappa = 10^{-6}$ m$^2$ s$^{-1}$) with a rotational period of $P$ = 5 hours is  $\ell_D \sim$ 10 cm. Boulders of size $\ell \gg \ell_D$ can be progressively fractured into smaller rocks but particles with $\ell \ll \ell_D$ will be nearly isothermal and thus less susceptible to cracking.  As a result, the breakdown of rocks by thermal fracture should progressively modify the size distribution of particles in an asteroid regolith relative to the power-law distribution produced by impact fragmentation.  A separate effect arises because  many rocks are  built from mm-sized mineral grains having different compositions and expansivities. Differential expansion between the grains naturally causes weakening and disintegration down to mm scales.    On Earth,  the effects of thermal fracture are amplified by the intervention of liquid water followed by freeze-thaw cracking \citep{Eppes16}.  On asteroids, where there is no liquid water, thermal fracture must be less effective. Still, boulders on asteroids, for example, are often surrounded by skirts of debris likely produced by thermal cracking from the day-night temperature cycle.    
\\

Separately, the thermal unbinding of water from hydrated minerals (e.g., hydrous phyllosilicates like serpentine, brucite, muscovite, and talc) causes shrinkage and is another process capable of cracking rocks and producing dust (as in dusty, dry mud lake beds on Earth).  The activation energies for dehydration of some minerals correspond to temperatures of $T$$\sim$10$^3$~K \citep{Bose94}, as reached by (3200) Phaethon and other small perihelion objects.  While Phaethon shows no evidence for hydrated minerals, it is not clear if this because they have already been lost or were never present \citep{Takir20}. Already suspected in Phaethon, thermal destruction  may  play an important role in the reported depletion of low-perihelion asteroids \citep{Granvik16}.
\\

\begin{figure}[htb!]
\begin{center}
\includegraphics[width=5.0cm]{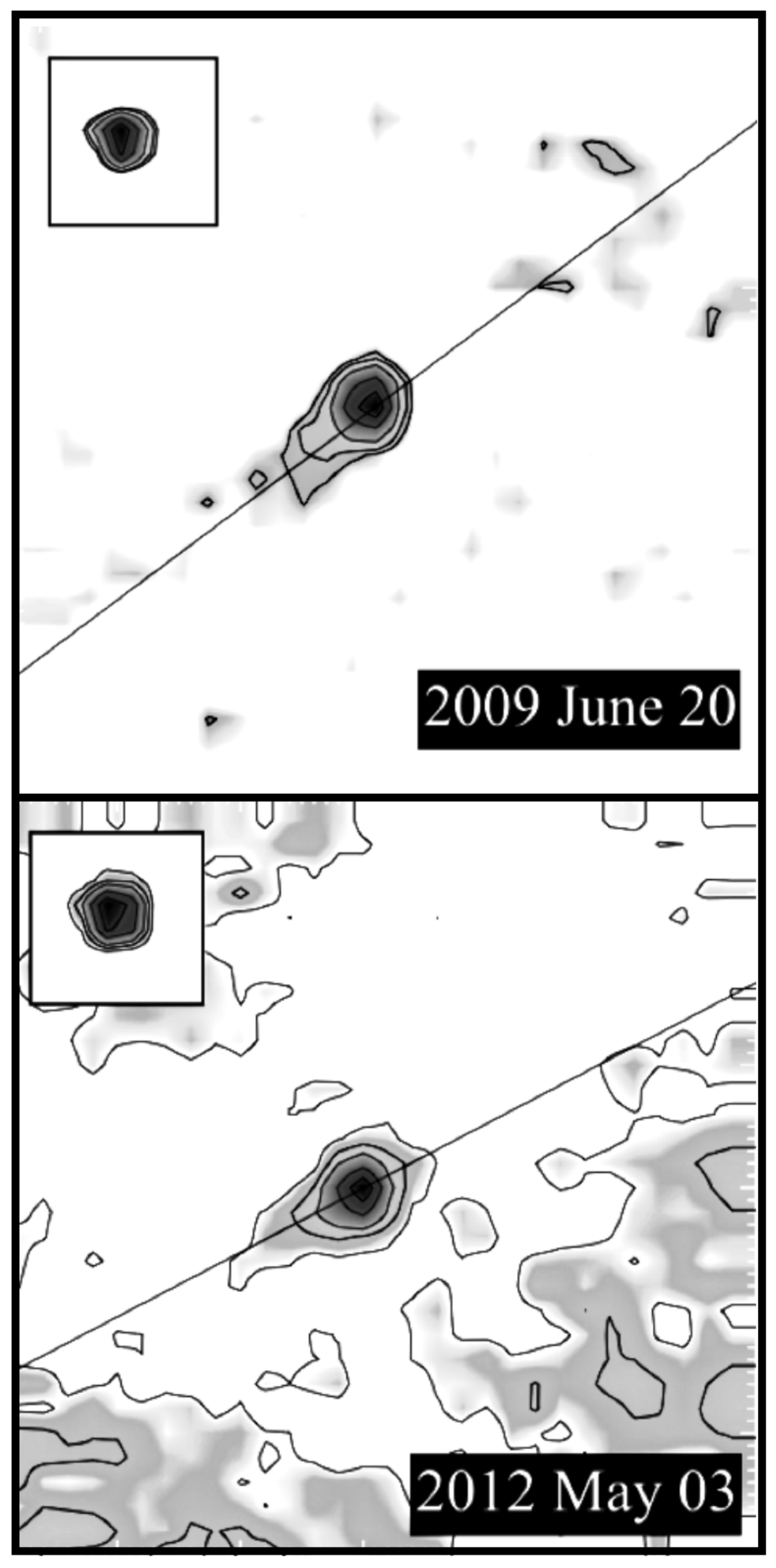}
\caption{Contoured STEREO images of (3200) Phaethon against the bright background of the solar corona at perihelion in 2009 and 2012, from \citet{Jew13} showing its faint tail. Inset boxes showing contoured point sources are $\sim$8\arcmin~square, corresponding to 350,000 km at Phaethon. Straight lines show the projected Sun-Phaethon line.}
\label{figure:3200}
\end{center}
\end{figure}

In order to produce measurable activity as seen in the active asteroids, however, particles produced by thermal fracture must achieve escape velocity.  Order of magnitude considerations show that m~s$^{-1}$ speeds can be generated in 10 $\mu$m-sized fragments \citep{Jewitt12}, sufficient to escape from the gravitational fields of 1 km asteroids, although larger particles will be too slow to escape.  In such cases, additional forces, notably radiation pressure sweeping, might combine with thermal fracture to accelerate fresh fragments above the escape speed.  Without a removal mechanism, the surface layers of an asteroid would eventually clog with small particles that could not be further fractured.
\\

%Phyllosilicate dehydration (mentioned in Bennu activity paper)

\subsection{Interplanetary Field Enhancements}
\label{section:IFEs}

Several NEAs (e.g.~(2201) Oljato and (138175) 2000 EE$_{104}$) are associated with distinct and repeated disturbances in the magnetic field carried by the solar wind, called Interplanetary Field Enhancements (IFEs).  These disturbances, detected by spacecraft crossing the orbits of the asteroids, can last from minutes to hours and have profiles distinct from magnetic structures emanating from active regions in the Sun \citep{Russell84, Lai17}. 
\\

The suggested mechanism underlying IFEs is the impact ejection of nano-dust, which quickly becomes photoelectrically charged  and then loads the passing solar wind, causing a magnetic field disturbance \citep{Russell84, Lai17}.  While nano-dust itself is not optically detectable because the scattering efficiency of nano-particles is negligible, impacts should eject particles with a range of sizes, some of which should be optically detectable.  A targeted search for such particles at (138175) 2000 EE$_{104}$ detected none, however, setting limits to the optical depth $\tau \lesssim 2\times10^{-9}$ \citep{Jewitt20}, comparable to faint JFC trails \citep{Ishiguro09}.   
\\

Furthermore, the  production of sufficient masses of nano-dust by impact is problematic.  \citet{Lai17} estimated that a mass of $M_0$ = 10$^5$ to 10$^6$ kg of nano-dust would be needed to produce a single IFE. Given a range of normal (top-heavy) ejecta power-law size distributions in which most of the mass is contained in the largest particles, nano-dust constitutes only $f \sim 10^{-3.8}$  to $10^{-5.5}$ of the total mass.  Therefore, to supply $M_0$ requires implausible source masses of $M = M_0/f \sim 10^{8.8}$~kg to $10^{11.5}$~kg, rivalling the $\sim10^{11}$ kg mass of (138175) itself   \citep{Jewitt20}.  Moreover, the planet-crossing asteroid population is comparatively rarefied, and the collision rate for such objects is likely too small to supply IFEs at the measured rate.  
\\

While no other explanations have been suggested to date, IFEs appear to be real.  If they are caused by charged dust loading of the solar wind, then a solution for the mass problem and a non-impact mechanism for the production of this dust remain to be found.  We  know from comet 1P/Halley that nano-dust, perhaps produced by the spontaneous fragmentation of larger particles dragged out of the nucleus by gas, can be abundant in comets \citep{Mann17}. Indeed, IFE-like magnetic field disturbances have been detected along the orbit of comet 122P/de Vico \citep{Jones03}. 
\\

\section{DISTRIBUTION OF VOLATILE MATERIAL}
\label{section:volatile_distribution}

In the classical view, icy objects formed beyond the water snow line, which is the heliocentric distance beyond which temperatures are low enough for   ice grains to form and become incorporated into growing planetesimals.  Water interior to that distance remained in vapor form and was incorporated into forming planetesimals at a far lower rate \citep[e.g.,][]{encrenaz2008_water}. Icy bodies  are preserved in the Kuiper belt and Oort cloud reservoirs, from which  all ice-rich comets were initially believed to originate
\citep[e.g.,][]{duncan2008_cometorigins,dones2015_cometreservoirs}. 
%\citep[e.g.,][]{dones2015_cometreservoirs}. 
Meanwhile, the inactive nature of asteroids in the inner solar system was historically interpreted to indicate a lack of ice, either because they did not accrete much icy material to begin with, or because any such material  has been lost in the heat of the Sun over the age of the solar system.
\\

Initial interpretations of the distribution of asteroid taxonomic types in the asteroid belt placed the water snow line in our solar system at $\sim$2.5~au  \citep[e.g.,][]{gradie1982_asteroidbeltcomposition,jones1990_cpdasteroids}, but the migratory nature of the snow line due to changing protoplanetary disk conditions \citep[e.g.,][]{martin2012_snowline} means that icy material could have accreted throughout the asteroid belt.  
Moreover,  icy objects from the outer solar system, well beyond any plausible location of the snow line, could have been emplaced in the inner solar system after their accretion.  Evidence for this has been discovered in the isotopic compositions of meteorites, which fall into two groups identified with accretion in spatially (and, perhaps, temporally) distinct hot and cold reservoirs \citep{Warren11}.  Other observational studies  have revealed the otherwise unexpected presence of objects in the main asteroid belt with D-type taxonomic classifications  more commonly associated with outer solar system objects \citep[e.g.,][]{demeo2014_dtypes,hasegawa2021_tnos_asteroidbelt}. Numerical models also suggest that small bodies from the outer solar system could have been scattered inward as a consequence of giant planet migration early in the solar system's history \citep{levison2009_tnocontamination,walsh2011_grandtack}, or perhaps even as a consequence of giant planet formation itself \citep{Raymond17}.
\\

%inward scattering of small bodies from the outer solar system onto now-stable orbits in the present-day

%, \citet{levison2009_tnocontamination} and \citet{} have suggested that during periods of giant planet migration early in the solar system's history, small solar system objects from the outer solar system could have been scattered inward onto now-stable orbits in the present-day main asteroid belt, while \citet{Raymond17} suggest that such inward scattering of icy outer solar system bodies could be an inevitable consequence of giant planet formation itself, with no need to invoke planetary migration at all.
%Moreover, numerical models of the solar system's early dynamical evolution as well as newer observations of the distribution of asteroid compositional types in the asteroid belt also demonstrate that icy objects from the outer solar system, well beyond any plausible location of the snow line, could have been emplaced in the inner solar system after their accretion \citep[e.g.,][]{walsh2011_grandtack,demeo2014_astbeltmapping,Raymond17}.
%Meanwhile, on the observational side, several D-type asteroids, which is a taxonomic classification typically associated with outer solar system objects like Jupiter Trojans and cometary nuclei, have been unexpectedly found in the inner and middle regions of the main asteroid belt \citep{demeo2014_dtypes} 

%\citet{hasegawa2021_tnos_asteroidbelt} reported the identification of two ultra-red objects in the main asteroid belt which they suggest could have originated in the outer solar system.
%\\

In addition to finding that some inner solar system objects may contain  more ice than typically assumed, recent work has also demonstrated that some outer solar system objects may also contain  less ice than typically assumed.  Dynamical models of early solar system evolution, both including and excluding planetary migration, predict that some objects originally formed in the terrestrial planet region could have been ejected from the inner solar system, resulting in a small fraction of planetesimals that are predominantly rocky currently residing in the Oort Cloud \citep{weissman1997_1996PW,walsh2011_grandtack,izidoro2013_earthwaterorigin,Shannon15}.
When perturbed into the inner solar system, these objects would then be identifiable as possessing orbits characteristic of long-period comets, while also being nearly or completely inert \citep[e.g.,][]{weissman1997_1996PW,Meech16,piro2021_manx_A2018V3}.
\\
%%THIS PARA SEEMS A BIT OUT OF PLACE

\section{\textbf{DYNAMICS}}
\label{section:dynamics}

\begin{figure*}[htb!]
\begin{center}
\includegraphics[width=16cm]{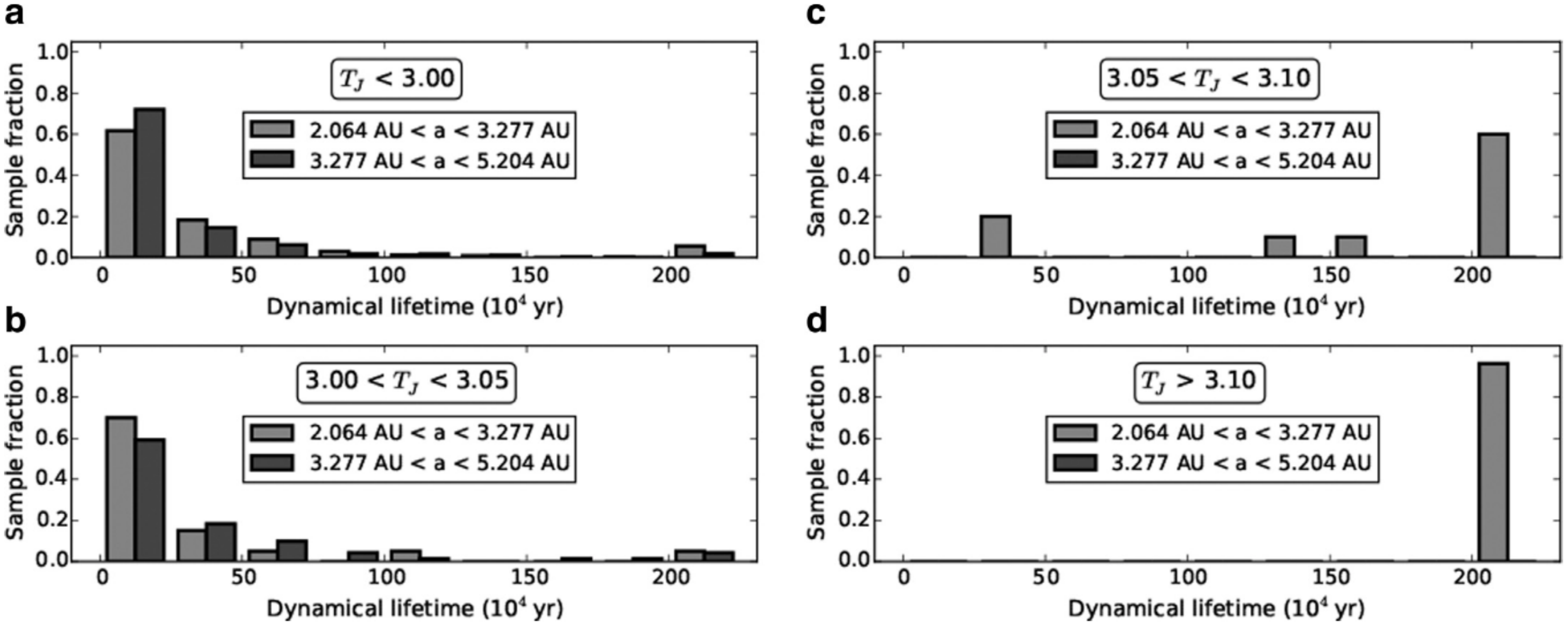}
\end{center}
\caption{Histograms of dynamical lifetimes for known comets with $a\leq5.204$~au with $T_{J,s}$ values of (a) $T_{J,s}<3.00$, (b) $3.00<T_{J,s}<3.05$, (c) $3.05<T_{J,s}<3.10$ and (d) $T_{J,s}>3.10$, where light and dark gray bars indicate the fraction of comets with $2.064~{\rm au}<a<3.277~{\rm au}$ and $a>3.277~{\rm au}$, respectively, that are lost due to ejection or planetary/solar impact within a particular time interval. From \citet{hsieh2016_tisserand}.}
\label{figure:dynamical_lifetimes}
%\end{center}
\end{figure*}

As mentioned in Section~\ref{section:intro}, the canonical $T_J=3$ boundary between asteroids and comets is strictly valid only under the assumptions of the circular restricted three-body approximation (in which the only gravitational perturbers are Jupiter and the Sun, and Jupiter's orbit is exactly circular). A more nuanced understanding of $T_J$ and small body dynamics in general, however, is necessary for properly interpreting the implications of studies that rely on assumptions about locations of origin based on this parameter.
\\

The real solar system contains other massive bodies, and non-gravitational perturbations from asymmetric outgassing from active bodies or radiative effects can also alter an object's motion in ways that are not considered in the circular restricted three-body approximation.  As a result, real asteroids and comets are not separated by a sharp dynamical boundary at $T_J=3$, but instead by a less distinct boundary zone spanning a range of values from $T_J\sim3.05$ to $T_J\sim3.10$, with numerical integrations showing that dynamical lifetime distributions of comets with $3.00<T_J<3.05$ and comets with $T_J<3$ are effectively indistinguishable, while comets with $3.05<T_J<3.10$ have a mix of short dynamical lifetimes ($t<2$~Myr), characteristic of objects with $T_J<3$, and longer dynamical lifetimes ($t>200$~Myr) characteristic of objects with $T_J>3.10$
\citep[Figure~\ref{figure:dynamical_lifetimes};][]{hsieh2016_tisserand}. These results are consistent with the comet-asteroid thresholds of $T_J=3.05$ and $T_J=3.08$ used by \citet{tancredi2014_asteroidcometclassification} and \citet{JHA15}, respectively.
\\

Dynamical classification systems based on objects' current orbital parameters can of course be made more complex to try to better capture the dynamical nuances of ``true'' asteroids and comets \citep[e.g.,][]{tancredi2014_asteroidcometclassification}, but even these can fail since the idea that an impenetrable boundary in $T_J$ space exists at all between asteroids and comets is  fundamentally flawed.  Numerical integrations have shown that small bodies can evolve from one side of the $T_J$ boundary to the other, effectively changing their dynamical identities as asteroids or comets as determined by $T_J$.  In a dynamical study of 58 JFCs on near-Earth orbits, \citet{fernandez2015_jfcinterlopers} found several comets that avoided very close encounters with Jupiter and remained dynamically stable on much longer timescales than other JFCs in the study, and hypothesized that they could have originated in the main asteroid belt.  Most of these objects also exhibit relatively weak activity for their sizes compared to other JFCs, and some show other asteroid-like physical properties \citep[e.g., 249P/LINEAR;][]{fernandez2017_249p}, further supporting possible main-belt origins for those objects.
\\

\begin{figure}[htb!]
\begin{center}
\includegraphics[width=6.5cm]{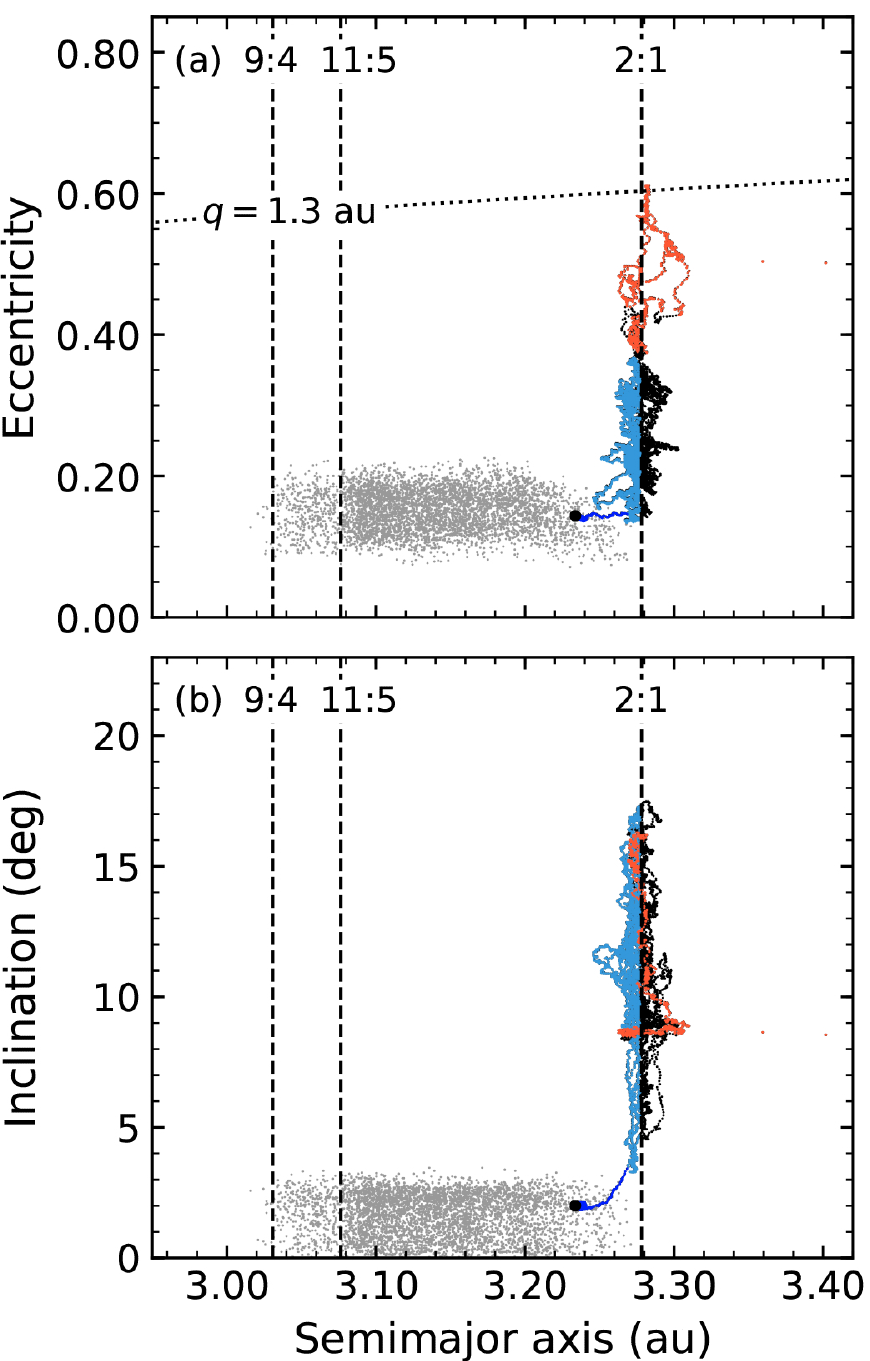}
\end{center}
\caption{Plots of the forward dynamical evolution of Themis family asteroid (12360) Unilandes in (a) semimajor axis vs.\ eccentricity space, and (b) semimajor axis vs.\ inclination space.  Large black circles in each panel mark the current orbital elements of the asteroid, gray dots mark the current orbital elements of Themis family asteroids, and other colored dots mark intermediate orbital elements in the forward integrations of the object.  Of the latter, blue dots mark intermediate orbital elements meeting criteria for being classified main-belt-like, and red dots mark orbital elements that meet criteria (including $T_J<3$) for being classified as JFC-like. From \citet{hsieh2020_themisjfcs}.}
\label{figure:unilandes_evolution}
%\end{center}
\end{figure}

While the orbits of many active asteroids appear stable on Gyr timescales (e.g.~\cite{Haghighipour09}), numerical integrations performed by \citet{hsieh2020_themisjfcs} subsequently revealed a pathway by which transitions inferred by \citet{fernandez2015_jfcinterlopers} could occur on  timescales of $<$100~Myr.  In that pathway (see Figure~\ref{figure:unilandes_evolution}), the 2:1 mean-motion resonance (MMR) with Jupiter can excite the eccentricities of main-belt asteroids (in that work, members of the Themis asteroid family) close to the resonance, lowering their perihelion distances and increasing their aphelion distances.  These orbital changes increase the potential for significant interactions with planets other than Jupiter, invalidating the approximation from which the theoretical $T_J=3$ asteroid-comet boundary is derived, ultimately allowing a small but non-negligible number of objects to evolve from high-$T_J$ orbits to low-$T_J$ orbits.
\\

Statistical evidence (Section \ref{section:inactive_comets}) indicates that main-belt contamination dominates the population of bodies with $2.8 \le T_J \le$ 3.0.  A spectroscopic survey by \citet{licandro2008_acos1} found that 7 of 41 observed objects with $T_J < 3$ showed absorption bands suggestive of S-type asteroids \citep[known for their distinctly un-comet-like silicate-containing compositions; see][for a detailed discussion about asteroid taxonomic classification]{demeo2015_astbeltcomposition_ast4}.  All seven had $T_J >$ 2.8.  \citet{geem2021_acos} proposed that asteroids and comets might be distinguished by their polarization vs.\ phase angle relations, and gave three examples.  One of these, (331471) 1984 QY$_{1}$, had a small polarization amplitude deemed more characteristic of an S-type asteroid than of a comet.  Future  measurements might be able to establish a polarization versus $T_J$ relation that would help distinguish displaced main-belt asteroids from JFCs.
\\
%%ACO not defined yet?

Meanwhile, dynamical evolution of small bodies in the opposite direction, from cometary orbits to asteroidal orbits, also appears to be possible. although unlikely.  \citet{hsieh2016_tisserand} found that a small number of synthetic test particles with JFC-like initial orbital elements reached fully main-belt-like orbits
%(semimajor axis internal to the 2:1 MMR with Jupiter, perihelion and aphelion distances that do not cross the orbits of either Mars or Jupiter, and $T_J>3$)
in 2-Myr-long integrations, suggesting that JFCs could occasionally become interlopers in the main asteroid belt.  Such an origin has been suggested for  P/2021 A5 (PANSTARRS) \citep{moreno2021_2019a4_2021a5}.
%The identification of two ultra-red (but apparently inactive) main-belt asteroids by \citet{hasegawa2021_tnos_asteroidbelt} may indicate that some inactive objects from the outer solar system may also be present in the asteroid belt.
That said, the JFC interlopers found by \citet{hsieh2016_tisserand} are dynamically distinct from the active asteroids in having higher mean eccentricities and inclinations. Efforts to use active asteroids to study primordial volatile material in the asteroid belt (see Section~\ref{section:mbcs}) might be able to mitigate the impact of such interlopers by focusing only on the low-inclination, low-eccentricity portion of the main belt population.
\\

Even if interactions with all other gravitating solar system bodies are taken into account, non-gravitational accelerations from asymmetric outgassing can cause further deviations from gravity-only dynamical evolution.
%, further eroding the diagnostic power of the $T_J$ parameter.
For example, 2P/Encke is a highly active comet with
%whose orbit is distinct from those of JFCs in having 
$T_J=3.025$, whose unusual orbit \citet{Fernandez02} found could be produced from a typical JFC orbit by non-gravitational accelerations, but only if they were sustained for 10$^5$ years. This physically implausible conclusion (2P does not contain enough ice to remain active that long) was also reached by \citet{Levison06}.  
An astrometric study of 18 active asteroids demonstrated that only two (313P and 324P) exhibited strong evidence of non-gravitational acceleration \citep{Hui17}. Nevertheless, the possibility that stronger asymmetric mass loss in the past could have produced their current asteroid-like orbits cannot be ruled out.
%, although their results also suggest 2P would also need to have been in a dormant state during much of its dynamical evolution in order to preserve its volatility to the present day.
\\

In summary, as with other binary criteria for distinguishing comets and asteroids (see Sections~\ref{section:activity_detection}, \ref{section:activity_mechanisms}, and \ref{section:volatile_distribution}),  dynamical classification  is also not always clear-cut.  Orbit evolution is an intrinsically chaotic, and therefore non-deterministic, process.  Backward and forward numerical integrations can indicate what range of past and future dynamical evolution is possible (with dynamical clones typically used for this purpose), but cannot  track an object's exact dynamical origin or fate beyond a certain length of time.
Conclusions about a specific object's origin are unavoidably statistical in nature.
\\

%and even undergo sudden changes, especially in the cases of close approaches to major planets other than Jupiter.  

%Therefore, real small solar system bodies can occasionally transition from one dynamical classification to another.  

%Just as Centaurs are understood to evolve onto Jupiter-family comet orbits \citep[e.g.,][]{tiscareno2003_centaurdynamics,disisto2007_centaurs,sarid2019_29p}, it appears to be possible for objects to evolve from unambiguously comet-like orbits to asteroid-like orbits, and vice versa.

%\citep[e.g., P/2019 LD2;][]{hsieh2021_p2019ld2,steckloff2020_p2019ld2}

%Possible comet contamination of MBAs and vice versa \citep{hsieh2016_tisserand,hsieh2020_themisjfcs}

%Different dynamical classes of MBCs? \citep{hsieh2016_tisserand}

%\section{\textbf{ACTUAL TRANSITION OBJECTS (OR A BETTER SECTION TITLE)?}}
%\label{transition}

%we do see some objects transitioning onto comet orbits, e.g., 2019 LD2 \citep{steckloff2020_p2019ld2}

%An example of a citation \citep{AHearn1995}.   \citep{kleyna2019_gault}
%Also see \citet{hsieh2006_mbcs} and \citet{jewitt2015_actvasts_ast4}.

%See Gault \citep{kleyna2019_gault}

%Need a broad intro to explain why they're interesting.

\section{\textbf{ACTIVE ASTEROIDS}}
\label{section:AA}

\subsection{Overview}
\label{section:AA_overview}

\begin{figure*}[htb!]
\begin{center}
\includegraphics[width=5.5in]{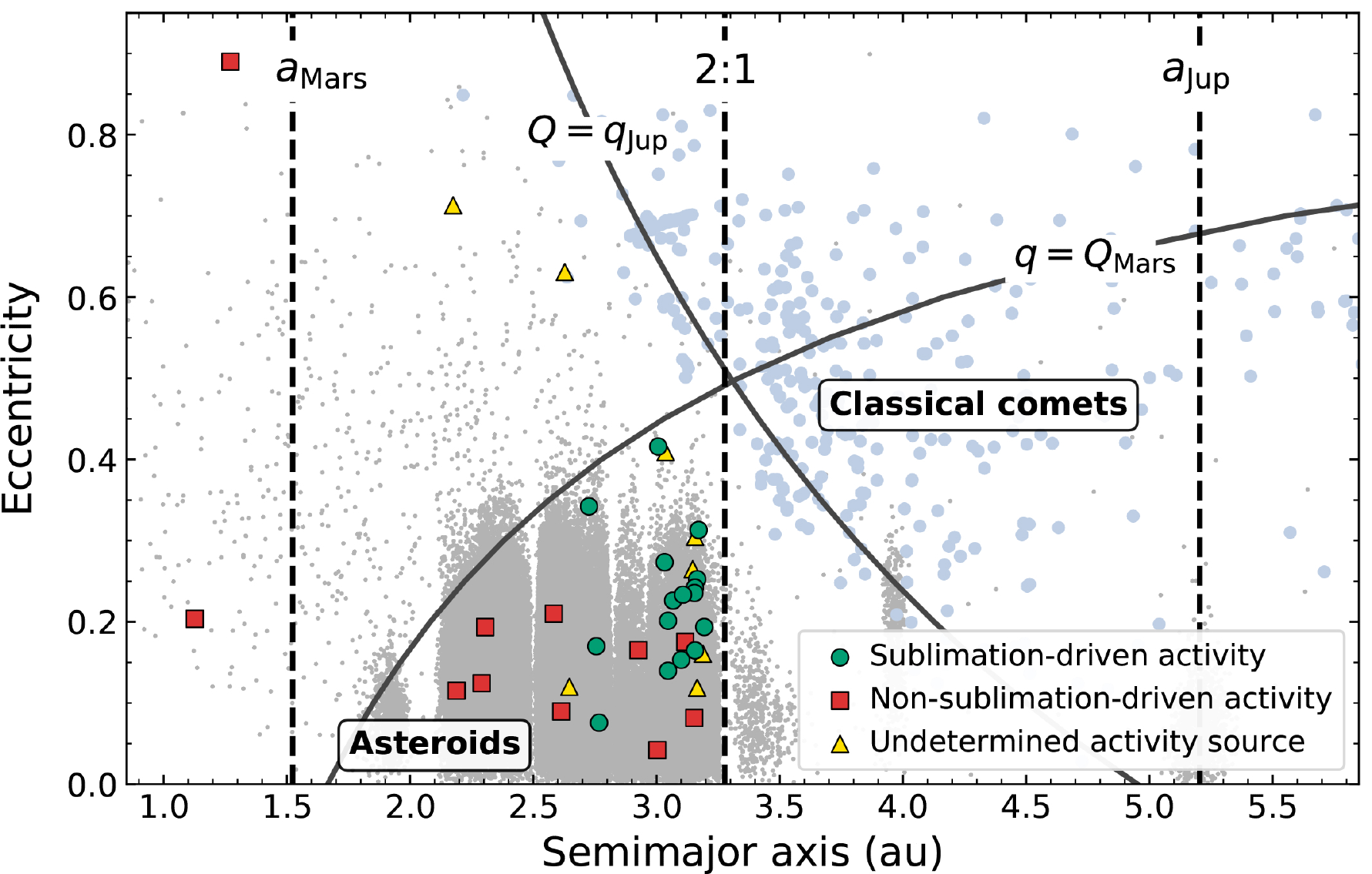}
\end{center}
\caption{Semimajor axis vs.\ orbital eccentricity plot for active asteroids determined to exhibit likely sublimation-driven activity (open green circles), activity driven by processes other than sublimation (open red squares), and activity of currently undetermined origin (open yellow triangles), where classical comets are marked as filled light blue circles and asteroids as small gray dots.  Curved arcs show the loci of points having aphelion and perihelion distances equal to the aphelion of Mars and the perihelion of Jupiter, respectively, as marked.  Objects above these lines cross the orbits of those planets.  Vertical dashed lines mark the orbital radii of Mars and Jupiter and the location of the 2:1 mean-motion resonance with Jupiter.  } 
\label{figure:aeplot}
%\end{center}
\end{figure*}

Active asteroids have dynamical properties characteristic of asteroids and the physical appearances of comets. They are one of the more recent groups of objects to be recognized as blurring the lines between asteroids and comets, although many individual objects in this category were known long before the term ``active asteroid'' was first used.  A wide range of potential mechanisms for the observed comet-like mass loss of active asteroids has been identified, including sublimation (Section~\ref{section:ice_sublimation}), impacts (Section~\ref{section:impacts_general}), rotational instability (Section~\ref{section:rotation_general}), thermal fracture (Section~\ref{section:thermal_fracture}), and combinations of these processes, making these objects valuable for the opportunities they provide to study these processes in the real world.  Reviews addressing different aspects of activity in asteroids include  \citet{Jew04}, \citet{Jewitt12}, \citet{JHA15}, and \citet{Kasuga19}.  Table~\ref{AAs} lists the active asteroids known as of 2022 January 1, while their distribution in semi-major axis versus eccentricity space is shown in Figure~\ref{figure:aeplot}.
\\

%For additional discussion, the reader is referred to previous reviews relevant to this subject, including \citet{Jew04}, \citet{Bertini11}, \citet{Jew12}, \citet{JHA15}, \citet{hsieh2017_continuumobjects}, \citet{Snodgrass17} and \citet{Kasuga19}.

\subsection{\textbf{Main-Belt Comets}}
\label{section:mbcs}

The main-belt comets (MBCs) are a noteworthy subset of the active asteroids, comprising objects for which sublimation appears to plays a role in producing the observed mass loss  \citep{hsieh2006_mbcs}. The nature of their activity and largely  stable orbits in the main asteroid belt indicate that near-surface volatile material has managed to remain preserved in the inner solar system until the present day. They have attracted significant scientific interest since their discovery due to the opportunities they provide to place constraints on the volatile content of  inner solar system bodies and their possible connections to the primordial delivery of water and other volatile material to the early Earth  \citep[e.g.,][]{hsieh2014_mbcsiausproc}.
\\

The MBCs discovered to date are too weakly active (production rates $\lesssim$1 kg s$^{-1}$ compared with 10$^2$ to 10$^3$ kg s$^{-1}$ for typical comets) for gas to be spectroscopically detected given the sensitivity limits of current ground-based telescopes \citep{Jewitt12}.  Instead, the most distinctive observational property of the MBCs is repeated near-perihelion activity in different orbits, indicating that their activity is thermally modulated (see Section~\ref{section:ice_sublimation}).  
Of the many processes considered in Section~\ref{section:activity_mechanisms}, only temperature-dependent sublimation can easily account for repeated near-perihelion activity.  
Independent evidence for outgassing comes from the detection of non-gravitational acceleration (caused by recoil from the anisotropic mass loss) in 313P/Gibbs and 324P/La Sagra \citep{Hui17}.  Only gas carries enough momentum to explain the observed non-gravitational accelerations.  Evidence for sublimation as the driver of activity in MBCs is thus indirect but compelling.  Reviews specifically focusing on MBCs include \citet{bertini11} and \citet{Snodgrass17}.  MBCs identified through the repetition of their activity in more than one orbit or through other means (e.g., dust modeling indicating prolonged dust emission events characteristic of sublimation-driven activity) are shown as green symbols in Figure~\ref{figure:aeplot}.  \\

The ``duty cycle'', $f_{dc}$, defined as the fraction of the total time during which an MBC is active, is an important quantity because it is related to the abundance of ice in the asteroid belt. Values of $f_{dc} \ll$ 1 are automatically required in order for the MBCs to have survived against devolatilization \citep{hsieh2006_mbcs} or destruction by spin-up \citep{jewitt2019_p2017s5}.   The most stringent estimates, $f_{dc} \lesssim 3\times10^{-5}$, are obtained from surveys conducted in search of asteroid activity \citep{Gilbert10,Waszczak13}. Such small values  identify the MBCs as the tip of the proverbial iceberg; each detected MBC represents $f_d^{-1} \gtrsim$  30,000 similarly icy asteroids that were not detected as such because they exist in the dormant state.  It is entirely possible, for example, that \textit{all} outer-belt asteroids contain ice and that the MBCs are notable simply because we observe them in an unusual outgassing state. Additional characteristics \citep[e.g., fast rotation, small nucleus sizes and therefore low escape velocities, moderate orbital eccentricities, small obliquities, membership in young asteroid families, and so on;][]{hsieh2004_133p,hsieh2018_activeastfamilies,hsieh2009_htp,kim2018_mbcalignment,schorghofer2016_asteroidice,novakovic2022_433p}, may aggregate to help produce observable activity.  
\\
  
%\subsection{DYNAMICS}
%\label{section:dynamics}

\setlength{\tabcolsep}{4.5pt}
\begin{table*}[htb!]
\caption{Currently Known Active Asteroids}
\centering
\smallskip
%\footnotesize
\small
\begin{tabular}{lrrrrr@{.}lr@{.}lr@{.}lccl}
\hline\hline
\multicolumn{1}{c}{Object}
 & \multicolumn{1}{c}{$a$$^a$}
 & \multicolumn{1}{c}{$e$$^b$}
 & \multicolumn{1}{c}{$i$$^c$}
 & \multicolumn{1}{c}{$T_J$$^d$}
 & \multicolumn{2}{c}{$H_V$$^e$}
 & \multicolumn{2}{c}{$r_n$$^f$}
 & \multicolumn{2}{c}{$P_{\rm rot}$$^g$}
 & \multicolumn{1}{c}{Mech.$^h$}
 & \multicolumn{1}{c}{Excl.$^i$}
 & \multicolumn{1}{c}{Ref.$^j$}
 \\
\hline
(1) Ceres                        & 2.766 & 0.078 & 10.588 & 3.310 &  3&53 & 469&7 &  9&07 & S & --- & [1] \\
(493) Griseldis                  & 3.116 & 0.176 & 15.179 & 3.140 & 10&97 &  20&78 & 51&94 & I & S & [2] \\
(596) Scheila                    & 2.929 & 0.163 & 14.657 & 3.209 &  8&93 &  79&86 & 15&85 & I & S & [3] \\
(2201) Oljato                    & 2.174 & 0.713 &  2.522 & 3.299 & 15&25 &   0&90 & $>$26& & --- & --- & [4]  \\
(3200) Phaethon                  & 1.271 & 0.890 & 22.257 & 4.510 & 14&32 &   3&13 & 3&60 & RTP & --- & [5] \\
(6478) Gault                     & 2.305 & 0.193 & 22.812 & 3.461 & 14&81 & 2&8 &  2&49 & R & S & [6] \\
(62412) 2000 SY$_{178}$          & 3.159 & 0.079 &  4.738 & 3.195 & 13&74 &  5&19 &  3&33 & R & --- & [7] \\
(101955) Bennu                   & 1.126 & 0.204 &  6.035 & 5.525 & 20&21 &  0&24 &  4&29 & --- & --- & [8] \\
107P/(4015) Wilson-Harrington    & 2.625 & 0.632 &  2.799 & 3.082 & 16&02 &  3&46 & 7&15 & --- & --- & [9] \\
133P/(7968) Elst-Pizarro         & 3.165 & 0.157 &  1.389 & 3.184 & 15&49 &  1&9 & 3&47 & SR & --- & [10] \\
176P/(118401) LINEAR             & 3.194 & 0.193 &  0.235 & 3.167 & 15&10 &  2&0 & 22&23 & S & R & [11]  \\
233P/La Sagra (P/2005 JR$_{71}$) & 3.033 & 0.411 & 11.279 & 3.081 & 16&6 &  1&5  & \multicolumn{2}{c}{---} & --- & --- & [12]  \\
238P/Read (P/2005 U1)            & 3.162 & 0.253 &  1.266 & 3.153 & 19&05 &  0&4 & \multicolumn{2}{c}{---} & S & --- & [13]  \\
259P/Garradd (P/2008 R1)         & 2.727 & 0.342 & 15.899 & 3.217 & 19&71 &  0&30  & \multicolumn{2}{c}{---} & S & --- & [14]  \\
288P/(300163) 2006 VW$_{139}$    & 3.051 & 0.201 &  3.239 & 3.203 & \multicolumn{2}{c}{17.8,18.2} & \multicolumn{2}{c}{0.9,0.6} & \multicolumn{2}{c}{---} & S & --- & [15]  \\
311P/PANSTARRS (P/2013 P5)       & 2.189 & 0.116 &  4.968 & 3.660 & 19&14 & 0&2 & \multicolumn{2}{c}{>5.4} & R & S & [16]  \\
313P/Gibbs (P/2014 S4)           & 3.154 & 0.242 & 10.967 & 3.133 & 17&1 & 1&0  & \multicolumn{2}{c}{---} & S & --- & [17]  \\
324P/La Sagra (P/2010 R2)        & 3.098 & 0.154 & 21.400 & 3.099 & 18&4  &  0&55 & \multicolumn{2}{c}{---} & S & --- & [18]  \\
331P/Gibbs (P/2012 F5)           & 3.005 & 0.042 &  9.739 & 3.228 & 17&33 &  1&77 & 3&24 & R & S & [19]  \\
354P/LINEAR (P/2010 A2)          & 2.290 & 0.125 &  5.256 & 3.583 & \multicolumn{2}{c}{---} & 0&06  & 11&36 & I & S & [20]  \\
358P/PANSTARRS (P/2012 T1)       & 3.155 & 0.236 & 11.058 & 3.134 & 19&5  &   0&32  & \multicolumn{2}{c}{---} & S & --- & [21]  \\
426P/PANSTARRS (P/2019 A7)            & 3.188 & 0.161 & 17.773 & 3.103 & 17&1 & 1&2 & \multicolumn{2}{c}{---} & --- & --- & [22]  \\
427P/ATLAS (P/2017 S5)           & 3.171 & 0.313 & 11.849 & 3.092 & 18&91 &  0&45  & 1&4 & SR & --- & [23]  \\
432P/PANSTARRS (P/2021 N4)                        & 3.045 & 0.244 & 10.067 & 3.170 & $>$18&2 & $<$0&7 & \multicolumn{2}{c}{---} & --- & --- & [24]  \\
433P/(248370) 2005 QN$_{173}$    & 3.067 & 0.226 &  0.067 & 3.192 & 16&32 &  1&6  & \multicolumn{2}{c}{---} & SR & --- & [25] \\
P/2013 R3 (Catalina-PANSTARRS)   & 3.033 & 0.273 &  0.899 & 3.184 & \multicolumn{2}{c}{---} & \multicolumn{2}{c}{$\sim$0.2 ($\times$4)} & \multicolumn{2}{c}{---} & SR & --- & [26]  \\
P/2015 X6 (PANSTARRS)            & 2.755 & 0.170 &  4.558 & 3.318 & $>$18&2 & $<$0&7 & \multicolumn{2}{c}{---} & S & --- & [27]  \\
P/2016 G1 (PANSTARRS)            & 2.583 & 0.210 & 10.968 & 3.367 & \multicolumn{2}{c}{---} & $<$0&4 & \multicolumn{2}{c}{---} & I & S & [28]  \\
P/2016 J1-A/B (PANSTARRS)        & 3.172 & 0.228 & 14.330 & 3.113 &  \multicolumn{2}{c}{---}  &   \multicolumn{2}{c}{$<$0.4,$<$0.9}  & \multicolumn{2}{c}{---} & S & --- & [29]  \\
P/2017 S9 (PANSTARRS)            & 3.156 & 0.305 & 14.138 & 3.087 & $>$17&8 & $<$0&8 & \multicolumn{2}{c}{---} & --- & --- & [30] \\
P/2018 P3 (PANSTARRS)            & 3.007 & 0.416 &  8.909 & 3.096 & $>$18&6 & $<$0&6 & \multicolumn{2}{c}{---} & S & --- & [31]  \\
P/2019 A3 (PANSTARRS)            & 3.147 & 0.265 & 15.367 & 3.099 &  $>$19&3 & $<$0&4 & \multicolumn{2}{c}{---} & --- & --- & [32]  \\
P/2019 A4 (PANSTARRS)            & 2.614 & 0.090 & 13.319 & 3.365 & \multicolumn{2}{c}{---} & 0&17 & \multicolumn{2}{c}{---} & --- & S & [33]  \\
P/2020 O1 (Lemmon-PANSTARRS)     & 2.647 & 0.120 &  5.223 & 3.376 & $>$17&7 & $<$0&9  & \multicolumn{2}{c}{---} & SR & --- & [34]  \\
P/2021 A5 (PANSTARRS)            & 3.047 & 0.140 & 18.188 & 3.147 & \multicolumn{2}{c}{---} &   0&15 & \multicolumn{2}{c}{---} & S & --- & [35]  \\
P/2021 L4 (PANSTARRS)            & 3.165 & 0.119 & 16.963 & 3.125 & $>$15&8 & $<$2&2 & \multicolumn{2}{c}{---} & --- & --- & [36]  \\
P/2021 R8 (Sheppard)             & 3.019 & 0.294 &  2.203 & 3.179 & \multicolumn{2}{c}{---} & \multicolumn{2}{c}{---} & \multicolumn{2}{c}{---} & --- & --- & [37] \\
\hline
\multicolumn{14}{l}{$^a$ Semimajor axis, in au. $~~^b$ Eccentricity. $~~^c$ Inclination, in degrees. $~~^d$ Tisserand parameter with respect to Jupiter.} \\
\multicolumn{14}{l}{$^e$ Measured $V$-band absolute magnitude, or estimated upper limit based on apparent brightness when active, if available.} \\
\multicolumn{14}{l}{$^f$ Effective nucleus radius (or radii), in km, as estimated from absolute magnitude or via other means as described in references.} \\
\multicolumn{14}{l}{$^g$ Rotation period, in hr, if available.} \\
\multicolumn{14}{l}{$^h$ Mechanisms supported by available evidence as contributing to the observed activity, if any (S: sublimation; I: impact;} \\
\multicolumn{14}{l}{$~~~$  R: rotational destabilization; T: thermal fatigue/fracturing; P: phyllosilicate dehydration)} \\
\multicolumn{14}{l}{$^i$ Mechanisms excluded by available evidence as contributing to the observed activity, if any (S: sublimation; R: rotational} \\
\multicolumn{14}{l}{$~~~$ destabilization), where any mechanisms not explicitly excluded should be considered potential contributors to activity} \\
\multicolumn{14}{l}{$^j$ References: 
[1] \citet{kuppers2014_ceres}; \citet{park2016_ceres}; % Ceres
[2] \citet{tholen2015_griseldis}; % Griseldis
[3] \citet{ishiguro2011_scheila2}; % Scheila
} \\
\multicolumn{14}{l}{$~~~$
[4] \citet{Russell84,tedesco2004_imps,warner2019_lcdb_v3}; % Oljato
[5] \citet{JewLi10,Ansdell14}; % Phaethon
} \\
\multicolumn{14}{l}{$~~~$
[6] \citet{devogele2021_gault}; % Gault
[7] \citet{sheppard2015_sy178}; % SY178
[8] \citet{lauretta2019_bennuactivity}; % Bennu
[9] \citet{fernandez1997_107p}; % 107P
} \\
\multicolumn{14}{l}{$~~~$
     \citet{licandro2009_107p,urakawa2011_107p}; % 107P
[10] \citet{hsieh2004_133p,hsieh2009_albedos,hsieh2010_133p}; % 133P
[11] \citet{hsieh2009_albedos,hsieh2011_176p}; % 176P
} \\
\multicolumn{14}{l}{$~~~$
[12] \citet{mainzer2010_233p}; % 233P
[13] \citet{hsieh2011_238p}; % 238P
[14] \citet{maclennan2012_259p,hsieh2021_259p}; % 259P
} \\
\multicolumn{14}{l}{$~~~$
[15] \citet{agarwal2020_288p}; % 288P
[16] \citet{jewitt2018_311p}; % 311P
[17] \citet{hsieh2015_313p}; % 313P
[18] \citet{hsieh2014_324p}; % 324P
} \\
\multicolumn{14}{l}{$~~~$
[19] \citet{drahus2015_331p}; % 331P
[20] \citet{jewitt2010_p2010a2,Snod10,Kim17b,Kim17a}; % 354P
} \\
\multicolumn{14}{l}{$~~~$
[21] \citet{hsieh2013_p2012t1,hsieh2018_358p}; % 358P
%[22] \citet{ramanjooloo2019_p2019a7}; % 426P
[22] \citet{rudenko2021_p2019a7_cbet}; % 426P
[23] \citet{jewitt2019_p2017s5}; % 427P
[24] \citet{wainscoat2021_p2021l4}; % 432P
} \\
\multicolumn{14}{l}{$~~~$
[25] \citet{hsieh2021_433P,novakovic2022_433p}; % 433P
[26] \citet{jewitt2014_p2013r3}; % P/2013 R3
[27] \citet{moreno2016_p2015x6}; % P/2015 X6
} \\
\multicolumn{14}{l}{$~~~$
[28] \citet{moreno2016_p2016g1,hainaut2019_p2016g1}; % P/2016 G1
[29] \citet{moreno2017_p2016j1,hui2017_p2016j1}; % P/2016 J1
[30] \citet{weryk2017_p2017s9}; % P/2017 S9
} \\
\multicolumn{14}{l}{$~~~$
[31] \citet{weryk2018_p2018p3,kim2019_p2018p3,kim2022_p2018p3}; % P/2018 P3
[32] \citet{weryk2019_p2019a3}; % P/2019 A3
[33] \citet{moreno2021_2019a4_2021a5}; % P/2019 A4
[34] \citet{weryk2020_p2020o1} % P/2020 O1
} \\
\multicolumn{14}{l}{$~~~$
     \citet{kim2022_p2020o1}; % P/2020 O1
[35] \citet{moreno2021_2019a4_2021a5}; % P/2021 A5
[36] \citet{wainscoat2021_p2021l4}; % P/2021 L4
[37] \citet{tholen2021_p2021r8} % P/2021 R8
} \\
\multicolumn{14}{l}{$~~~$}\\
\end{tabular}
\label{AAs}
\end{table*}

\begin{figure*}[htb!]
\begin{center}
\includegraphics[width=5.5in]{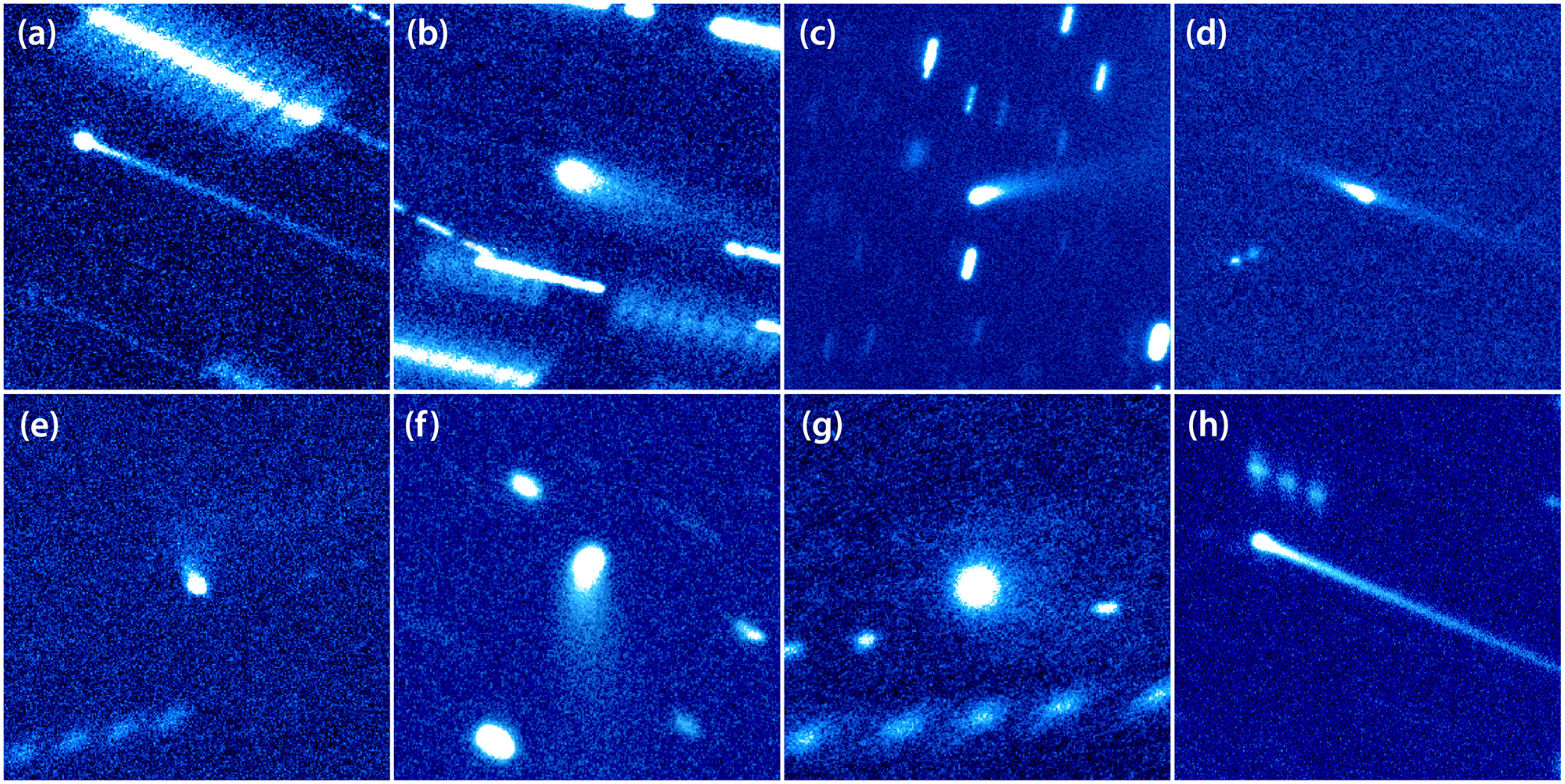}
\caption{Gallery of $1'\times1'$ images of MBCs confirmed to exhibit recurrent activity: (a) 133P/Elst-Pizarro, (b) 238P/Read, (c) 259P/Garradd, (d) 288P/(300163) 2006 VW$_{139}$, (e) 313P/Gibbs, (f) 324P/La Sagra, (g) 358P/PANSTARRS, and (h) 433P/(248370) 2005 QN$_{173}$.  Images adapted from data published in \citet{hsieh2004_133p,hsieh2009_238p,hsieh2012_288p,hsieh2012_324p,hsieh2013_p2012t1,hsieh2015_313p,hsieh2021_433P,hsieh2021_259p}.}
\label{figure:recurrent_mbcs}
\end{center}
\end{figure*}

%\begin{figure*}[htb!]
%\begin{center}
%\includegraphics[width=6.0in]{fig_sublim_rate_heliodist.pdf}
%\caption{From \citet{hsieh2015_ps1mbcs}}
%\label{figure:thermal_plot}
%\end{center}
%\end{figure*}

%\subsubsection{OTHER}
%\label{section:other}
%Examples that don't fit other categories.
%Maybe electrostatics, thermal fracture, phyllosilicate dehydration (mentioned in Bennu activity paper), IFEs here 

\subsection{Active Asteroid Examples}
\label{section:active_asteroids_examples}

\subsubsection{Sublimation: 133P/(7968) Elst-Pizarro}
\label{section:active_asteroids_133p}

Comet 133P/(7968) Elst-Pizarro, originally discovered as asteroid 1979 OW$_7$, was first found to exhibit comet-like activity on 1996 August 7, shortly after passing perihelion on 1996 April 18 \citep{elst1996_133p}.  At the time, 133P was the first and only main-belt asteroid observed to exhibit comet-like activity, leading to uncertainty about whether the observed mass loss, later determined to have occurred over the course of several weeks or months \citep{boehnhardt1996_133p}, was due to sublimation of volatile material similar to other comets, or to a series of impacts \citep{toth2000_133p}.
Follow-up observations in 2002 revealed 133P to be active  \citep{hsieh2004_133p}, again near perihelion, heavily favoring the hypothesis that its activity was due to  the sublimation of exposed volatile material, instead of a highly improbable succession of impacts.  Since then, 133P has continued to exhibit activity at each of its subsequent perihelion passages \citep[e.g.,][]{hsieh2010_133p,jewitt2014_133p}, firmly establishing the recurrent and likely sublimation-driven nature of its activity.
\\

Calculations described in Section~\ref{section:ice_sublimation} show that sublimation rate differences between perihelion and aphelion for main belt asteroids can easily explain the preferential appearance of detectable activity  only near perihelion, while obliquity-related and surface-shadowing effects may also play a role.  To date, seven other active asteroids (Figure~\ref{figure:recurrent_mbcs}) have been found to exhibit recurrent activity near perihelion \citep[Figure~\ref{figure:thermal_plot}; see][]{hsieh2021_433P}, indicating that they are likely MBCs exhibiting sublimation-driven mass loss, and corroborating the hypothesis that their activity is primarily modulated by temperature and not seasonal effects.  %Furthermore, suspected MBC candidates that have not yet been confirmed to exhibit recurrent activity nonetheless also display activity near perihelion
%Since the initial discovery and detailed characterization of 133P's activity, we have also since seen most suspected main-belt comets exhibit activity preferentially near perihelion (Figure~\ref{figure:thermal_plot}), indicating that their activity is primarily dependent on temperature.
\\

Interestingly, 133P's relatively fast rotation rate of $P_{\rm rot}=3.471$~hr \citep{hsieh2004_133p} suggests that rotational instability could assist in the ejection of dust particles initially lofted by sublimation.  Combined with the hypothesis that 133P's current activity may have required an initial collisional trigger to excavate a sub-surface ice reservoir (see Section~\ref{section:ice_sublimation}), 133P presents an excellent illustrative case of an active object for which multiple mechanisms may be responsible for observed activity.
\\

\subsubsection{Impact: 354P/LINEAR and (596) Scheila}
\label{section:active_asteroids_354P_Scheila}

\begin{figure}[htb!]
\begin{center}
\includegraphics[width=7.0cm]{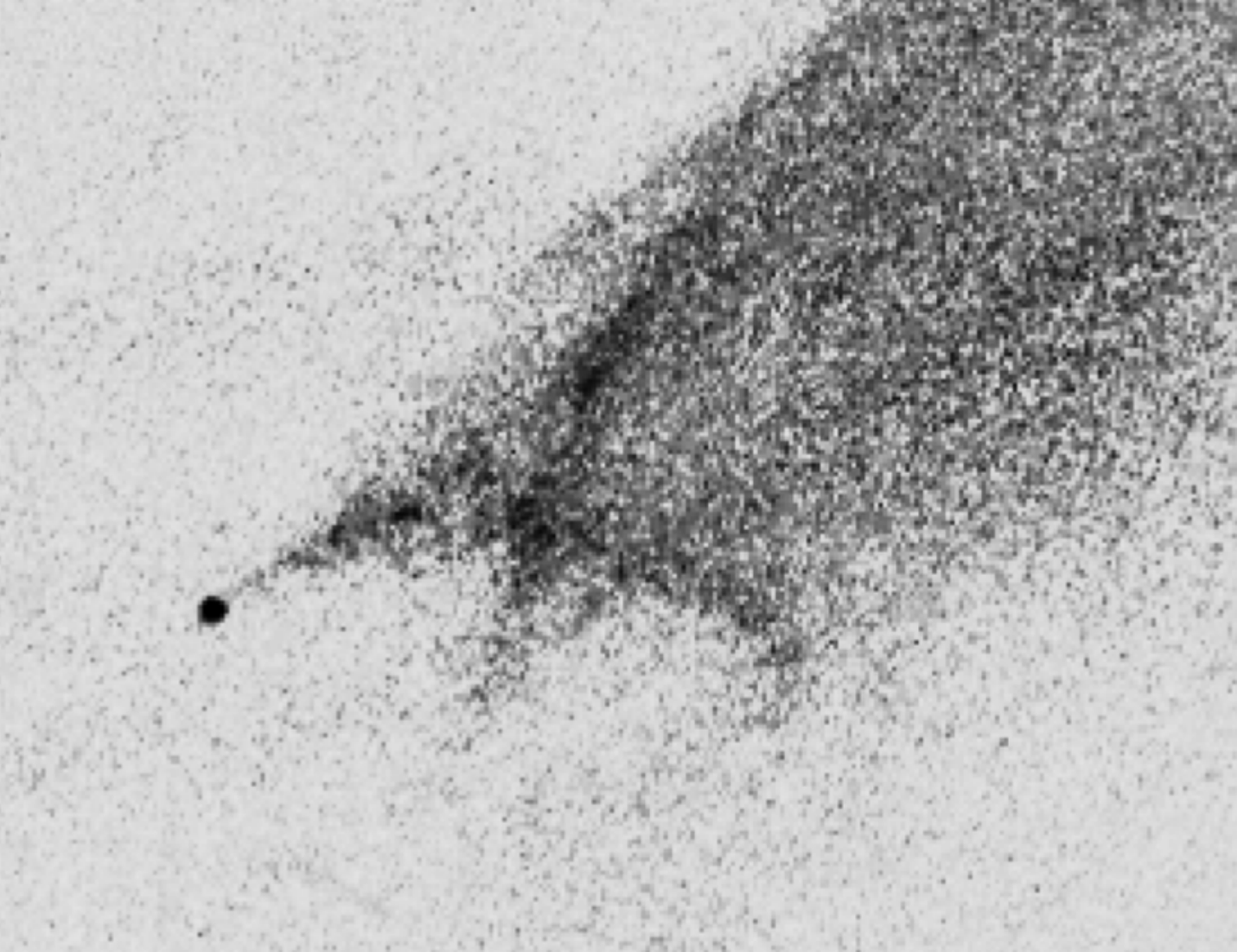}
\includegraphics[width=7.0cm]{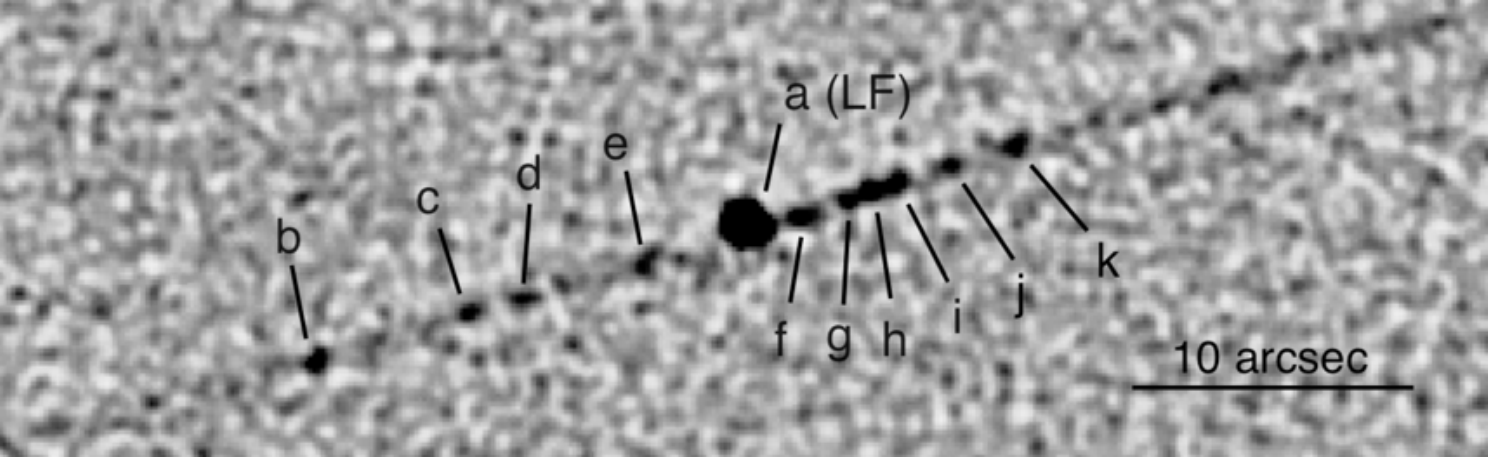}
\caption{(top) Head structure of P/2010 A2 on UT 2010 January 25, from \citet{jewitt2010_p2010a2}. (bottom) fragments in the particle trail on UT 2017 January 27 and 28, from \citet{Kim17b}.}
\label{P2010A2}
\end{center}
\end{figure}

Comet 354P/LINEAR (formerly P/2010 A2) and (596) Scheila were the first objects discovered immediately following impact events.
The largest component in 354P's debris cloud is $\sim$100 m in scale, and is accompanied by a co-moving cluster of smaller bodies embedded in a particle trail \citep{jewitt2010_p2010a2,Snod10,Jew13}. The object was disrupted by a collision nine months before it was discovered, with the delay in the discovery in part due to having an initially small solar elongation \citep{JSL11}. 
While rotational destabilization was also initially considered as a possible alternate explanation for 354P's activity \citep{jewitt2010_p2010a2}, the main nucleus was later found to have a double-peaked rotational period of $P_{\rm rot}=11.36\pm0.02$~hr, ruling out fast rotation as a potential cause of the observed activity \citep{Kim17b}.
\\

%\cite{jewitt2010_p2010a2}, \cite{snod10}, \cite{JSL11}, \cite{Hainaut12}, \cite{JIA13}, \cite{AJW13}, \cite{Kim17a}, \cite{Kim17b} \\

Meanwhile, the 113 km diameter asteroid 596 Scheila ejected $\sim10^8$ kg of dust due to an apparent cratering impact in late 2010 \citep{jewitt2011_scheila}, with that dust clearing in a few weeks. The impulsive nature of the brightening and the monotonic fading thereafter are consistent with the impact of a 20 m t0 40~m-scale projectile.
%The escape velocity from Scheila is $\sim$60 m s$^{-1}$, causing all slower material to fall back to the surface.  
The unusual dust cloud morphology exhibited by Scheila can be reproduced by  an impact ejecta cone and downrange plume produced by an impact at a $\sim45\degr$ angle to the object's surface \citep{ishiguro2011_scheila2}.  \citet{ishiguro2011_scheila1} and \citet{Bodewits14} also reported post-impact changes in the optical lightcurve while \citet{Hasegawa21} reported a reddening in the 1.0 to 2.5 $\mu$m wavelength region, but, curiously, no change from 0.4 to 1.0 $\mu$m or from 2.5 to 4.0 $\mu$m. If real, the observed lightcurve modifications may have resulted from fallback of material ejected at speeds less than Scheila's $\sim$60 m s$^{-1}$ escape velocity, while the reported near-infrared color change,  might suggest burial of a surface previously de-reddened by space weathering.
\\

The discoveries of these two objects, both in 2010, represent a key milestone in our understanding of the comet-asteroid continuum, as they definitively demonstrated for the first time that impact events could  be detected in real time, and also that observations of comet-like activity do not always mean that sublimation of volatile material is taking place.  This realization helped to pave the way for other non-sublimation-related interpretations of future activity detections (Sections~\ref{section:activity_mechanisms}, \ref{section:active_asteroids_p2013r3}, and \ref{section:active_asteroids_rotational_disruptions}).
\\

\subsubsection{Rotational Breakup: P/2013 R3 (Catalina- \\
PANSTARRS)}
\label{section:active_asteroids_p2013r3}

\begin{figure}[htb!]
\begin{center}
\includegraphics[width=8cm]{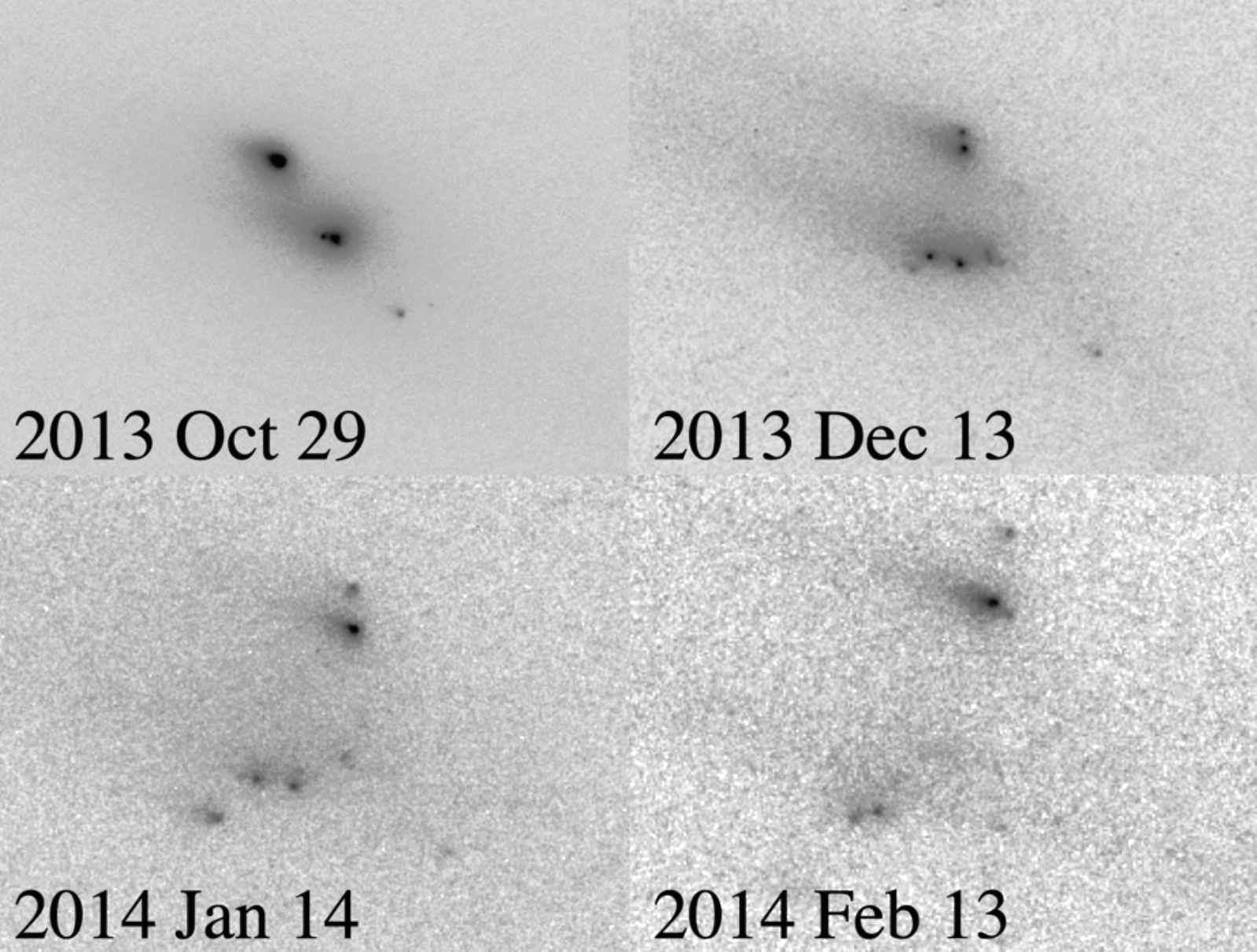}
\caption{Fragments of P/2013 R3 on four dates, as marked.  Each panel shows a region approximately 20,000 km wide.  The anti-solar and negative projected orbit vectors are at position angles 67\degr~and 246\degr, respectively. North to the top, East to the left.   From \citet{Jew17}. }
\label{figure:P2013R3}
\end{center}
\end{figure}

Perhaps the most dramatic example of rotational destabilization found to date is P/2013 R3 (Catalina-PANSTARRS), which was observed in 2013 in the midst of disintegrating into a collection of $\sim10^2$ m scale bodies with a velocity dispersion of $\Delta V \sim$ 0.3 m s$^{-1}$ \citep[Figure~\ref{figure:P2013R3};][]{jewitt2014_p2013r3}.   The parent body size is estimated at a few $\times10^2$ m. Bodies this small are susceptible to spin-up by radiation torques on timescales $\lesssim1$~Myr, leading to the suspicion that P/2013 R3 was a rotationally disrupted asteroid.  Sublimating water ice, if present, could also spin up the body on a very short timescale \citep{Jewitt21}, and in fact was inferred to be present by the dust ejection behavior of individual components of the object as it disintegrated \citep{jewitt2014_p2013r3}.  
%Estimates of the cohesive strength of P/2013 R3 are $S \sim 10^2$ N m$^{-2}$, comparable to the cohesion expected of a sand pile.  
Modeling of the break-up process by \citet{hirabayashi2014_p2013r3} suggested that the parent body had a rotational period of 0.48$-$1.9~hr, implying that it was likely spinning well beyond the critical disruption limit for a rubble pile prior to its breakup, suggesting that the object had some degree of cohesive strength on the order of 40$-$210~Pa, comparable to the cohesion expected of a sand pile.  Similar to 133P (Section~\ref{section:active_asteroids_133p}), P/2013 R3 presents an interesting case where multiple activity mechanisms are operating, with rotational instability likely being the primary cause of the disintegration of the parent body, but with continued decay and dust release then driven by sublimation of interior ices exposed by the disruption event.
\\

%\subsubsection{P/2013 R3: Rotational Breakup}
\subsubsection{Rotational Instability: (6478) Gault, 331P/Gibbs\\
and 311P/PANSTARRS}
\label{section:active_asteroids_rotational_disruptions}

Non-catastrophic mass loss events observed for main-belt asteroid (6478) Gault and comets 311P/PANSTARRS and 331P/Gibbs were all believed to be produced by rotational destabilization, yet were simultaneously remarkably diverse compared to one another (as well as compared to the catastrophic rotational disruption of P/2013 R3; Section~\ref{section:active_asteroids_p2013r3}).  The diversity in active behavior displayed by these three objects demonstrates the complexity of rotational destabilization as an activity mechanism, with potentially very different outcomes depending on the circumstances of each object and event. 
\\

\begin{figure}[htb!]
\begin{center}
\includegraphics[width=8cm]{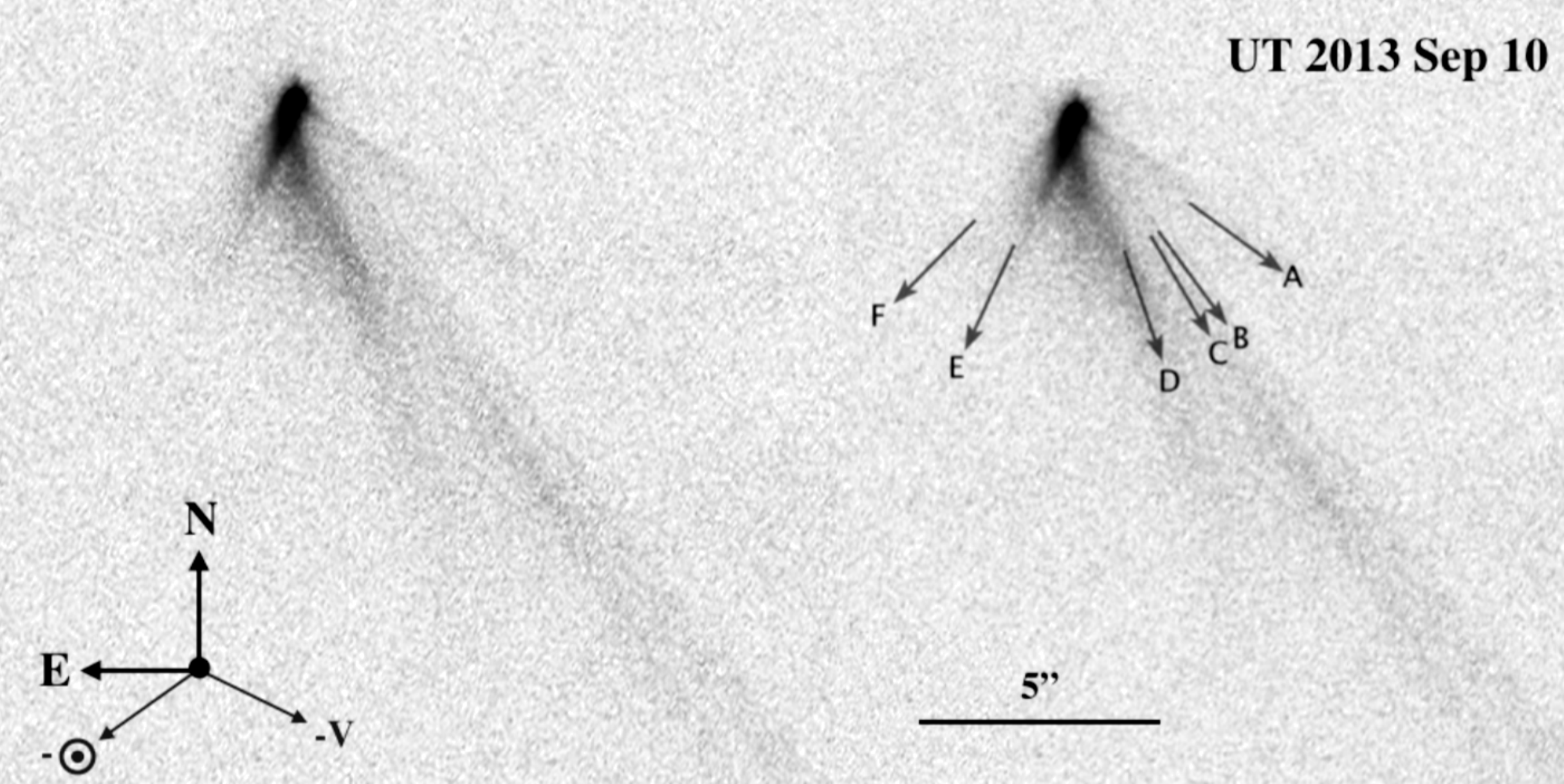}
\end{center}
\caption{Multi-tailed active asteroid 311P.  The first six of nine tails are labeled. From \citet{jewitt2013_311p}.}
\label{figure:311p}
%\end{center}
\end{figure}

\begin{figure}[htb!]
\begin{center}
\includegraphics[width=8cm]{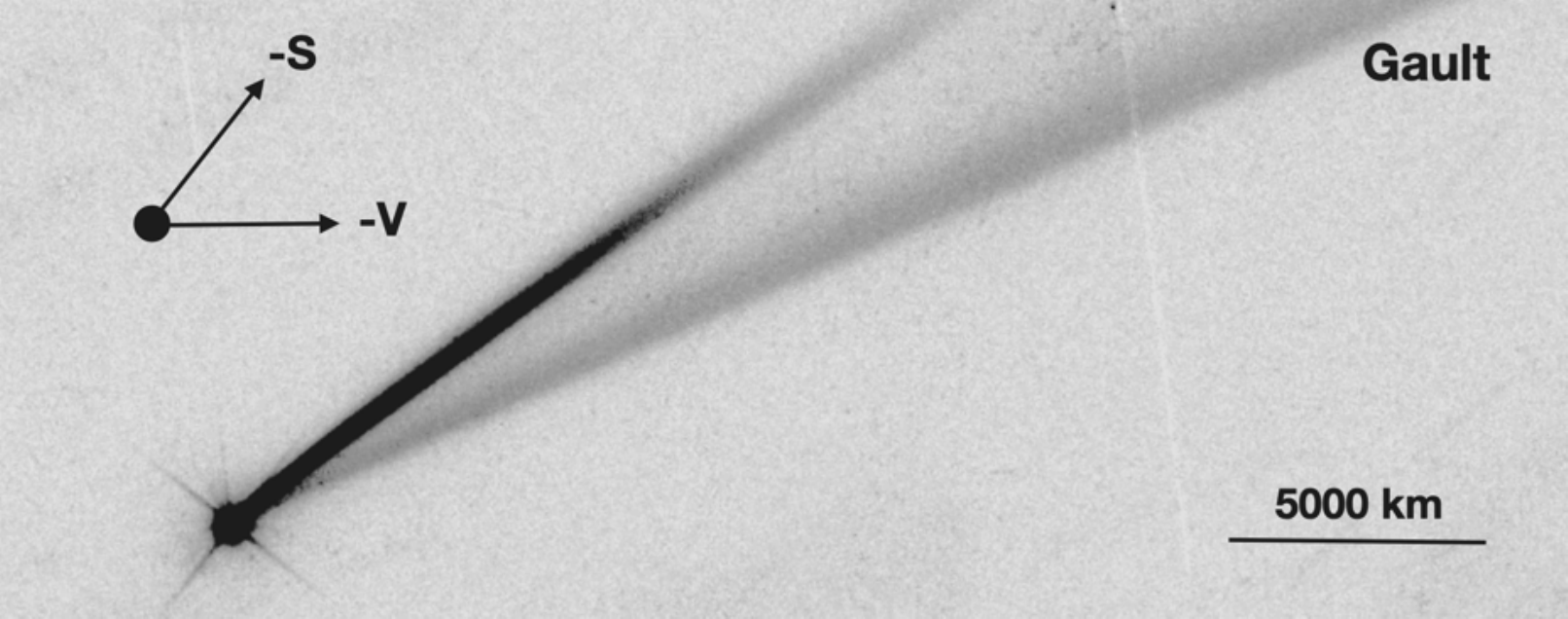}
\caption{Gault on UT 2019 February 5, showing two of the three tails ejected in 2019. From \citet{kleyna2019_gault}. }
\label{figure:Gault}
\end{center}
\end{figure}

\begin{figure*}[htb!]
\begin{center}
\includegraphics[width=13cm]{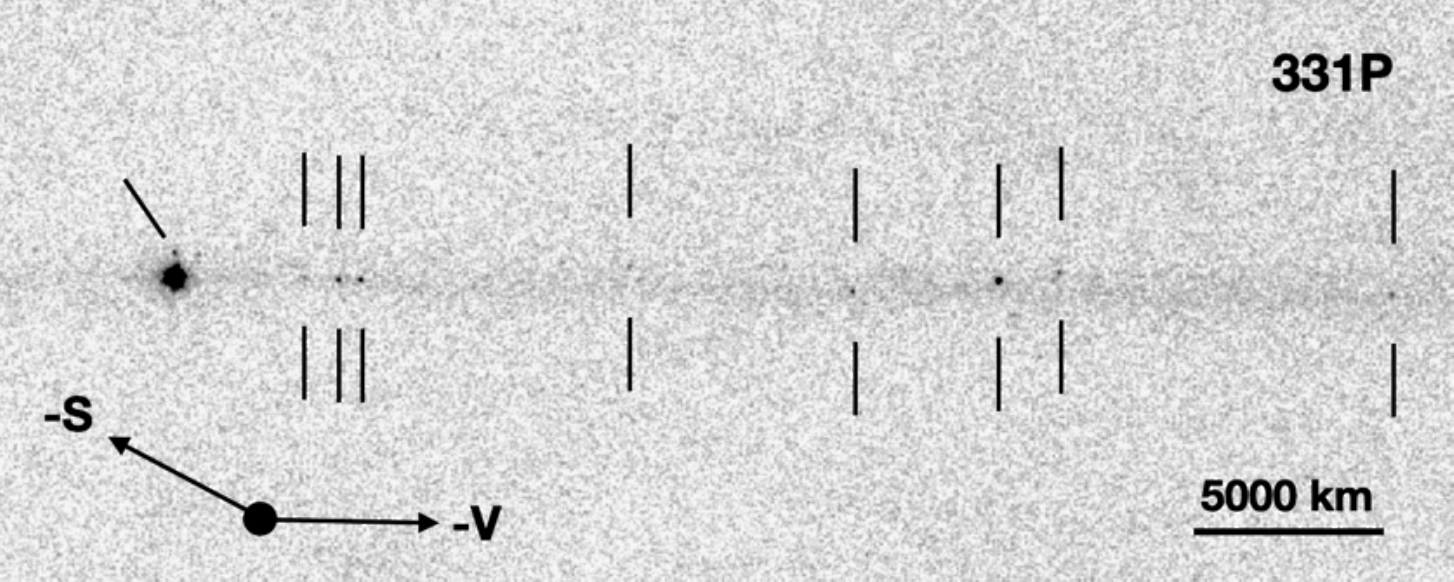}
\caption{Some of the 19 fragments of 331P on UT 2013 December 25. The fragments, many of which are very faint and only confirmed by their reappearance in images from different dates, are marked by solid black lines. The 1.6$\pm$0.2 km diameter primary is the bright object to the left. Suspected contact binary 331P-A is third from the right. From \citet{Jewitt21b}. }
\label{figure:331P}
\end{center}
\end{figure*}

Comet 311P (formerly P/2013 P5) displayed a distinctive set of  discrete dust tails (Figure~\ref{figure:311p}), formed by submillimeter-sized debris ejected in nine bursts over nine months, with no evidence of periodicity in the mass-loss events and no evidence for macroscopic fragments in the ejecta \citep{jewitt2013_311p}.  The directions of the tails are determined by the ejection epoch and the subsequent action of radiation pressure, creating a size gradient along each tail.  311P's activity likely resulted from the loss of loose surface material which avalanched to the equator in response to rapid rotation. A reliable period has not been determined for 311P.  Instead, photometry of its nucleus shows complex structure, suggestive of an eclipsing, possibly contact, binary \citep{jewitt2018_311p}.
\\

Meanwhile, when discovered to be active in 2019, Gault exhibited two bright tails \citep[Figure~\ref{figure:Gault}; e.g.,][]{kleyna2019_gault}, while a third, fainter tail emerged in later follow-up imaging \citep{jewitt2019_gault}. Intriguingly, searches of archival data revealed multiple additional past mass loss events over six years \citep{Chandler19}.  Like for 311P, the episodic nature of Gault's dust ejection events, as well as a lack of evidence that other activity mechanisms such as sublimation could be driving the observed activity, indicated that its activity was likely to be driven by rotational destabilization.  This hypothesis was later supported by observations of the nucleus after activity had subsided, revealing an extremely short rotational period of $P_{\rm rot}=2.5$~hr with a lightcurve amplitude of 0.06~mag \citep{devogele2021_gault,Luu21}.  Together, these observations suggest that Gault is a rapidly spinning asteroid held continuously at the brink of rotational disruption, perhaps by the YORP effect, that periodically undergoes mass loss events that temporarily restabilize it.  Unlike the eight MBCs showing confirmed recurrent activity near perihelion (Section~\ref{section:mbcs}), activity in Gault is unrelated to its   heliocentric distance, evidently because sublimation  does not play a role in its activity.
\\

331P/Gibbs ejected only a small fraction  of the parent body mass ($\Delta M/M \sim 0.01$), visible as a set of fragments embedded in a long debris trail \citep[Figure~\ref{figure:331P};][]{Jewitt21b}.  The 0.8 km radius (assuming an albedo of 0.05) primary body again shows clear evidence of rapid rotation, with period $P_{\rm rot}=3.26$~hr. At this period, Equation~\ref{equation:tc} gives $\rho = 1600$~kg~m$^{-3}$ as the minimum density required for a cohesionless spherical body to resist mass shedding at its equator.
The 19 measurable fragments ejected from 331P travel away from the source at incredibly low speeds, $\sim 0.1$~m~s$^{-1}$. The fragments show no sign of further fading or disintegration over a three year window of observation, consistent with having wholly refractory compositions.  The largest, the 100 m scale 331P-A (visible as the second brightest object in Figure \ref{figure:331P}), shows a large-amplitude rotational lightcurve with morphological characteristics consistent with a contact binary structure.  They are embedded in a debris trail, also characterized by a small velocity dispersion.  Despite its designation, ``comet'' 331P is probably a non-volatile, rotationally unstable asteroid.
\\

%\subsubsection{Gault: Rotational Mass-Shedding}
%\subsubsection{331P: Mass Shedding} 

%Whereas P/2013 R3 completely disintegrated, 
%\\

%\subsubsection{311P: Rotational Mass-Shedding}

\subsubsection{Thermal Breakdown: (3200) Phaethon}
\label{section:phaethon}

The Geminid meteoroid stream consists of millimeter-sized and larger debris with a total mass $M_G \sim$ (2 to 7)$\times10^{13}$ kg \citep{Blaauw17}.  The stream is $\tau \sim10^3$ years old, as judged by the spread of orbital elements of the Geminid meteors \citep{Gustafson89,Williams93} and, to be supplied in steady state, would require production at the rate $M_G/\tau \sim$700 to 2000 kg s$^{-1}$.  Such large rates would be comparable to those of bright comets, but no such bright comet source of the Geminids exists.  Instead, the Geminids appear to be supplied by the 5 to 6 km diameter B-type asteroid (3200) Phaethon.
\\

Curiously, dark sky observations have consistently shown no visible evidence for on-going mass loss from Phaethon at the $\sim$10$^{-2}$ kg s$^{-1}$ level \citep{Hsieh2005}, or even from the vantage point of a near-Earth (0.07 au) flyby in 2017, where limits $\le 10^{-3}$ kg s$^{-1}$ were set \citep{Jew18,Jew19}. However, observations near perihelion at $r_H = 0.14$~au (Figure~\ref{figure:3200}) show photometric behavior and a resolved tail consistent with the release of $\mu$m-sized particles at about 3 kg s$^{-1}$ \citep{JewLi10,LiJew13,Jew13,HuiLi17}.  This rate is still two to three orders of magnitude too small to supply the Geminids in steady state.  Furthermore, the $\mu$m-sized particles detected near perihelion are so strongly affected by solar radiation pressure  that they cannot be confined to the Geminid stream.    Either the particle size distribution is shallow, such that unseen large particles dominate the ejected mass or, more likely,  the source of the Geminids is episodic, or even catastrophic in nature, and unrelated to the activity at perihelion.  \\

Several mechanisms have been suggested to drive Phaethon's observed repeated perihelion activity.  The  subsolar temperature on Phaethon is $\sim$1000 K, suggesting that stresses induced by thermal expansion and desiccation shrinkage could eject fragments \citep{JewLi10,Jewitt12}.    Phaethon's $P_{\rm rot}=3.6$~hr rotation period \citep{Ansdell14} suggests that centrifugal effects might also contribute to mass loss \citep{Nakano20}, as does its likely top-shape \citep{Hanus16,Taylor19}. Another potential consequence of Phaethon's proximity to the Sun at perihelion is the liberation of volatile materials from minerals exposed on the surface.  In this regard, \cite{Masiero21} suggested that escaping sodium might create a tail, while  \cite{Hui22} used an upper  limit to the degree of forward scattering from Phaethon dust to assert  that the tail is produced by sodium emission, not by dust.  The required sodium production rate is $\sim$10$^{24}$ s$^{-1}$ \citep{Hui22}.  When averaged over the dayside hemisphere of Phaethon this corresponds to a flux $\sim4\times10^{12}$ cm$^{-2}$ s$^{-1}$, four orders of magnitude larger than the peak sodium flux from Mercury ($\sim$10$^8$ cm$^{-2}$ s$^{-1}$; \cite{Schmidt13}).   After scaling by the inverse square law (a factor of 8) to account for the smaller heliocentric distance, the reported Phaethon flux is still 5,000 times larger than the sodium flux from Mercury, a difference which is unexplained. A direct (spectroscopic) detection or limit to the presence of sodium at Phaethon is sorely needed to make progress.
%In addition, the sodium D lines are 99\% blocked by the nominal filter bandpass of the STEREO spacecraft cameras used to study Phaethon at perihelion, so that detection would require a dramatic change in the bandpass since the spacecraft was launched.  \\
Finally, \citet{kimura2022_phaethon} suggested that Phaethon's activity could be produced by the electrostatic lofting of dust particles with the aid of mobile alkali atoms at high temperatures, which the authors notably claim would be able to not only explain the appearance of Phaethon's dust tail but also the estimated mass of the Geminid meteoroid stream.
  \\
  
Phaethon's dynamical lifetime is $\lesssim 10^8$ years, implying delivery from elsewhere in the solar system \citep{deLeon10}.  A cometary origin is unlikely, given that the orbit is strongly decoupled from Jupiter, and has the very large Tisserand parameter $T_J$ = 4.5. However, the specific source region, presumably somewhere in the main asteroid belt, is unknown. Suggested sources include %a fragment of 
the large B-type asteroid 2 Pallas \citep{deLeon10} and, more likely, a collisional family in the inner asteroid belt \citep{MacLennan21}.   The km-sized asteroid 2005 UD appears dynamically related to Phaethon \citep{Ohtsuka06} and is coincidentally also a B-type object \citep{JH06}, \citep[although recent observations show dissimilarity in the near infrared;][]{Kareta21}.  A dynamical relation to another kilometer-sized asteroid, 1999 YC, has also been proposed but is less certain, where 1999 YC is a neutral C-type spectrally distinct from the blue B-types Phaethon and 2005 UD \citep{KJ08}.  \\

Phaethon, 2005 UD and the Geminid stream are  collectively known as the ``Phaethon-Geminid Complex'' \citep{Ohtsuka06}. Unfortunately, the genetic relationships between these bodies remain completely unknown, as does the relevance of the current mass loss from Phaethon at perihelion \citep{Kasuga19}.  We hope for clarification of the nature of Phaethon's surface  by the upcoming DESTINY+ flyby mission (see Section~\ref{section:future}), although the 33 km s$^{-1}$ encounter velocity and planned 500 km distance of closest approach will limit the data obtained.   
%A segment of Phaethon's dust trail has been (faintly) recorded from the Parker Solar Probe \citep{Battams20}. %DJ - so what?  \\
%The Geminids must be produced on a timescale ($\sim$10$^3$ years) much shorter than any timescale plausibly associated with the separation of Phaethon from 2005 UD.  
\\

%\begin{figure}[htb!]
%\begin{center}
%\includegraphics[width=6.0cm]{figures/3200.pdf}
%\caption{Contoured STEREO images of (3200) Phaethon at perihelion in 2009 and 2012, from \citet{Jew13} showing its faint tail. The inset boxes are about 8\arcmin~square, corresponding to 350,000 km at (3200) Phaethon. Straight lines show the projected Sun-Phaethon line.}
%\label{figure:3200}
%\end{center}
%\end{figure}

%Curiously, dark sky observations have consistently failed to show evidence for on-going mass loss from Phaethon at the $\sim10^{-2}$ kg s$^{-1}$ level \citep{Hsieh2005}, or even from the vantage point of a near-Earth (0.07 au) flyby in 2017, where limits $\le 10^{-3}$ kg s$^{-1}$ were set \citep{Jew18,Jew19}. However, observations near perihelion at $r_H = 0.14$ au (Figure~\ref{figure:3200}) show photometric behavior and a resolved tail consistent with the release of $\mu$m-sized particles at about 3 kg s$^{-1}$ \citep{JewLi10,LiJew13,Jew13,HuiLi17}.  This rate is still two to three orders of magnitude too small to supply the Geminids in steady state.  Furthermore, the $\mu$m-sized particles detected near perihelion are strongly affected by solar radiation pressure so that they cannot contribute to the Geminid stream.  Larger particles emitted at perihelion would not be detected in the available data.  Either the particle size distribution is shallow, such that unseen large particles dominate the mass , or  the source of the Geminids must be episodic, or even catastrophic in nature. \\

\subsubsection{Unknown: (101955) Bennu}
\label{section:bennu}

Repeated instances of particle ejection from Bennu were observed from the OSIRIS-REx  spacecraft \citep[Figure~\ref{figure:bennu};][]{lauretta2019_bennuactivity}.   
Particles ejected from Bennu, estimated to be cm-sized, were emitted erratically, with one large burst containing up to 200 particles.
%(Figure~\ref{figure:bennu}).  
Particle speeds were estimated to range from 0.1~m~s$^{-1}$ to 3~m~s$^{-1}$, with $\sim$70\% of the particles being too slow to escape and falling back to the surface. The peak observed ejection velocity, 3 m s$^{-1}$, exceeded the 0.2 m s$^{-1}$ escape velocity from Bennu by an order of magnitude.  Dissimilar size distributions were measured in three outburst events, with differential power laws indices varying from $-3.3$ to $-1.2$ \citep{Hergenrother20}.  The total ejected mass, however, was very small.  \citet{Hergenrother20} estimated a mass loss of $\sim10$ kg per orbit, corresponding to a mean mass loss rate of $\dot{M} \sim10^{-7}$ kg s$^{-1}$ over the 1.2 year orbit period.  This is $\sim$$10^{6}$ to $10^{7}$ times smaller than the sensitivity limits achieved through ground-based observations of other active asteroids. 
\\

The chance, in-situ discovery of Bennu's activity  raises the possibility that many asteroids lose mass at rates that are individually beneath the ground-based detection threshold.  To consider the possible magnitude of low-level dust production across the asteroid belt, we scale the surface area of Bennu ($\sim$1 km$^2$) to the total surface area of all asteroids having escape velocities  less than  the peak ejection velocities of Bennu ejecta. The latter corresponds to a radius of $r_n <$ 3 km. This total area, calculated by integrating over the asteroid size distribution,  is $\sim$10$^8$ km$^2$.  Therefore, if all  $r_n\le3$ km asteroids eject dust at the same rate per unit area as Bennu, the total mass loss would be $10^8 \dot{M} \sim$10 kg s$^{-1}$.  While this is a very crude calculation, it serves to show that Bennu-like production of dust in the asteroid belt is, at best, a very minor contributor to the $10^3$ to $10^4$ kg s$^{-1}$ production rate needed to sustain the Zodiacal cloud \citep{Nesvorny11}.  We also note that the Hayabusa2 spacecraft \citep{tsuda2019_hayabusa2} did not detect any Bennu-like loss of particles from (162173) Ryugu during its encounter with the asteroid in 2018-2019, despite the fact that Ryugu is also taxonomically classified as a primitive asteroid \citep{Watanabe19} and was also observed in situ by a visiting spacecraft.  We do not know if this reflects an intrinsic difference between the two asteroids, or merely a difference in the sensitivity to small particles between the two spacecraft.
\\

The mechanism driving Bennu's mass loss is currently undetermined, despite the availability of abundant spacecraft data. The primary constraints are the sizes, numbers and speeds of the ejected fragments, as well as an apparent preference for launch in the local afternoon.  Of the mechanisms already proposed for the active asteroids, electrostatic lofting, micrometeorite impact \citep{Bottke20}, thermal fracture and dehydration cracking \citep{molaro20} have all been considered in the case of Bennu.  
\\

\section{\textbf{INACTIVE COMETS}}
\label{section:inactive_comets}

%\subsection{Asteroids on Cometary Orbits}

The average dynamical lifetimes of short-period comets exceed their volatile lifetimes. Therefore, the fate of old comets, if they do not disintegrate completely, is to leave behind inert, asteroid-like  bodies travelling in comet-like orbits.  These are the ACOs and Damocloids mentioned in Section~\ref{section:intro}.  
Proper identification of such objects is essential for setting constraints on the physical lifetimes of comets \citep[e.g.,][]{Jew04} and models of cometary source regions \citep[e.g.,][]{wang2014_htcorigins,brasser2015_jfcpopulation}.
ACOs and Damocloids are presumed to be comets for which activity has weakened below detectable limits or completely ceased due to the depletion of surface volatiles or mantling \citep[e.g.,][]{podolak1985_cometmantles,prialnik1988_cometmantles,jewitt1996_dormantcomets}, although other processes like hydrostatic compression have also been proposed to contribute to activity quenching \citep{Gundlach16}.
%ACOs can be further divided into JFC-like ACOs, whose orbits have $2\le T_J\le 3$, and Damocloids, which have $T_J<2$. 
%Numerous asteroids on cometary orbits (``ACOs''; defined as being point-like with Tisserand parameter values of $2 \le T_J \le 3$) are known, but the fraction of these bodies which are inactive comets (as opposed to asteroids scattered inwards from the main-belt) is a subject of research. 
%where a spectroscopic study revealed the population of JFC-like ACOs to largely consist of a mix of D- and X-type objects, and the population of Damocloid ACOs to be dominated by D-type objects \citep{Licandro18}.
\\

%%%%%%%%%%%%%%%%%%%%%%%%%%
%%%%%%%%%%%%%%%%%%%%%%%%%%
%%%%%%%%%%%%%%%%%%%%%%%%%%

\setlength{\tabcolsep}{4.5pt}
\begin{table}[htb!]
\caption{Empirical Asteroid:Comet Ratio vs.~$T_J$ $^1$}
\centering
\smallskip
%\footnotesize
%\small
\begin{tabular}{rrrcrr@{.}lr@{.}lr@{.}lcl}
\hline\hline
\multicolumn{1}{c}{$T_J$}
 & \multicolumn{1}{c}{$N_a^2$}
 & \multicolumn{1}{c}{$N_c^3$}
 & \multicolumn{1}{c}{$\mathcal{R}^4$}
 \\
\hline
$<$1        & 28 & 372 & 0.08$\pm$0.01  \\
$<$2        & 453 & 108 & 0.24$\pm$0.02  \\
%1.0 - 2.0    & 80 & 88   & 0.9$\pm$0.1 \\
2.0 - 2.2   & 55 & 7 & 7.9$\pm$3.2 \\
2.2 - 2.4   & 101  & 23 & 4.4$\pm$1.0\\
2.4 - 2.6   & 202  & 38 & 5.3$\pm$0.9  \\
2.6 - 2.8   & 553  & 75 & 7.4$\pm$0.9 \\
2.8 - 3.0   & 2727  & 89 & 30.6$\pm$3.3 \\
$>$3        & 315331 & 12 & 2.6$\times10^4$ \\
\hline
\multicolumn{4}{l}{$^1$ Perihelion $q < 2$~au only} \\
\multicolumn{4}{l}{$^2$ Number of asteroids} \\
\multicolumn{4}{l}{$^3$ Number of comets} \\
\multicolumn{4}{l}{$^4$ Ratio $\mathcal{R} = N_a/N_c$} \\

\end{tabular}
\label{table:Ratio}
\end{table}

%%%%%%%%%%%%%%%%%%%%%%%%%%
%%%%%%%%%%%%%%%%%%%%%%%%%%
%%%%%%%%%%%%%%%%%%%%%%%%%%

Table~\ref{table:Ratio} lists the numbers of asteroids, $N_a$,  and comets, $N_c$, compiled from the JPL Horizons orbital element database (\url{https://ssd.jpl.nasa.gov/horizons/}), together with their ratio, $\mathcal{R} = N_a/N_s$, all  as functions of the binned Tisserand parameter, $T_J$.   The listed uncertainty on $\mathcal{R}$ has been calculated assuming Poisson counting statistics.  In this table, asteroids and comets are distinguished only by whether or not visible activity has ever been reported. Objects with large perihelia, even those known from detailed observations to be active comets, are too cold and weakly active for outgassing to be easily measured.  Therefore, in order to make the empirical asteroid/comet distinction meaningful, we only list objects with perihelia $q\le2.0$~au, where activity is relatively easy to detect.   We also exclude from consideration the sungrazing comets, which generally have poorly determined orbits.  The table provides a first-order estimate of the asteroid/comet number ratio as a function of a simple dynamical parameter.    
\\

As expected, the Table shows that almost all objects with $T_J > 3$ are asteroids ($\mathcal{R}\gg 1$), while most of those with $T_J <$ 2 are comets ($\mathcal{R} < 1$).  In between, $\mathcal{R}$ is roughly constant across the range assigned to the JFCs (namely 2 $\le T_J \le$ 3) except for a higher value ($\mathcal{R}$ = 31) in the range   $T_J >$ 2.8 to 3.0.  This is presumably because of contamination of the 2.8 $\le T_J \le$ 3.0 sample by  scattered main-belt asteroids
%; although $T_J$ is a constant of the motion, it does in fact change modestly in response to multi-planet scattering 
(see Section~\ref{section:dynamics}). An independent study of objects with $q < 1.3$~au also finds that $T_J \sim 2.8$ marks a change in the spectral characteristics, with featureless C and X type spectra more common at  $T_J < 2.8$ than for larger $T_J$ \citep{Simion21}. Across the range $2.0 \le T_J \le 2.8$, the  asteroid to comet ratio settles to a steady mean value $\mathcal{R} = 6.2\pm$0.8, meaning that, in this $T_J$ range, asteroids outnumber comets by $\sim$6:1.  
%, c.f.~\cite{Licandro18}).
\\

%%%%
In the absence of main-belt contamination, $\mathcal{R}$ would give a measure of the ratio of the dynamical to the active cometary lifetimes. The measured $\mathcal{R} \sim 6$ would result if comets are active for only $\sim$1/6th of their dynamical lifetimes.  However, this is a lower limit because main-belt contamination is presumably non-zero.  Furthermore, $\mathcal{R}$ provides no clue as to whether comets simply stop outgassing (because they exhaust near-surface volatiles), or cycle between active and inactive states \citep[perhaps because of cyclic mantle build-up and blow-off correlated with changes in the perihelion distance and peak surface temperature;][]{Rickman90}.
\\

Dynamical simulations provide an independent perspective.  The mean dynamical  lifetime of JFCs is $\tau_d \sim4\times10^5$ years \citep{Levison94}.  Repeated scattering by the terrestrial planets causes the mean inclination of the JFCs to increase with time. In order to reproduce the modest inclinations of the JFCs (the mean and median values are 14.3\degr~and 12.2\degr, respectively), these authors were forced to assume that the JFC physical lifetime is short compared to $\tau_d$.  Specifically, they inferred that the ratio of inactive to active JFCs should be in the range of 5 to 20, barely consistent with the measured  $\mathcal{R} \sim 6$ for $2.0 \le T_J \le 2.8$.   
\\

\begin{figure}[htb!]
\begin{center}
\includegraphics[width=7.5cm]{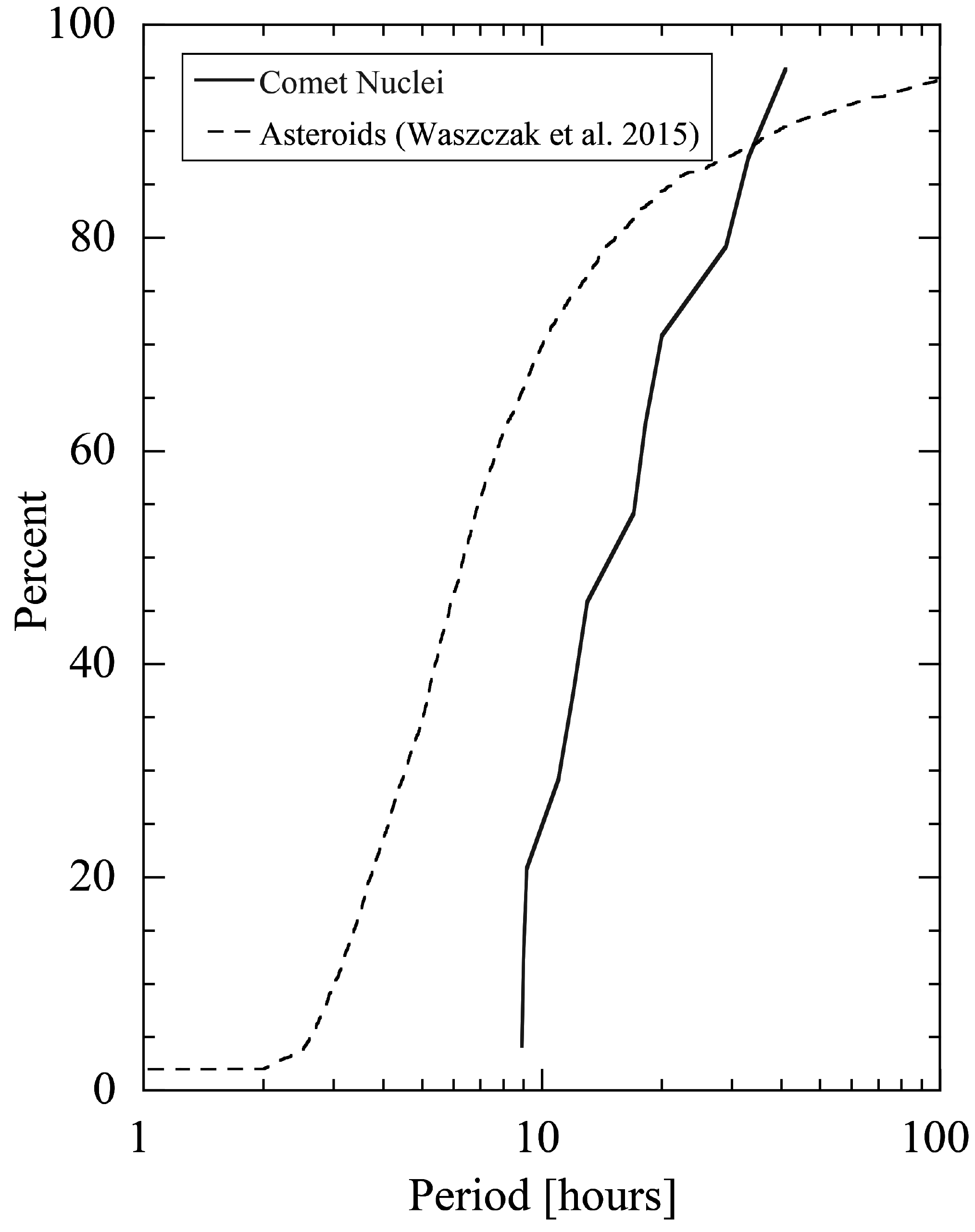}
\caption{Comparison of the cumulative distributions of the rotation periods of JFCs (solid  line) and asteroids (dashed line) in the 1 to 10 km size range. The much longer mean period of the JFCs is evident.  From \citet{Jewitt21}.}
\label{figure:periods}
\end{center}
\end{figure}

Another distinctive feature of JFC nuclei that might be helpful in the statistical identification of such objects when dormant is  their rotation period distribution.  The measured rotation periods of comets are highly biased towards long periods (median 15 hour) compared to the period distribution in small NEAs (median $\sim$6 hour) \citep[Figure~\ref{figure:periods}; also see][]{Binzel92,Jewitt21}.  This difference, which likely reflects the lower bulk density of comets, where the critical period for rotational instability scales as $\rho^{-{1/2}}$ (Equation \ref{equation:tc}), would suggest that few  ACOs are dead comets.  Another pointer is the truncated size distribution and distinct lack of sub-km cometary nuclei compared to the high abundance of sub-km asteroids.
\\
%%%%

%\subsection{$T_J <$ 2; DAMOCLOIDS}
%\label{section:damocloids}

\begin{figure}[htb!]
\begin{center}
\includegraphics[width=7.0cm]{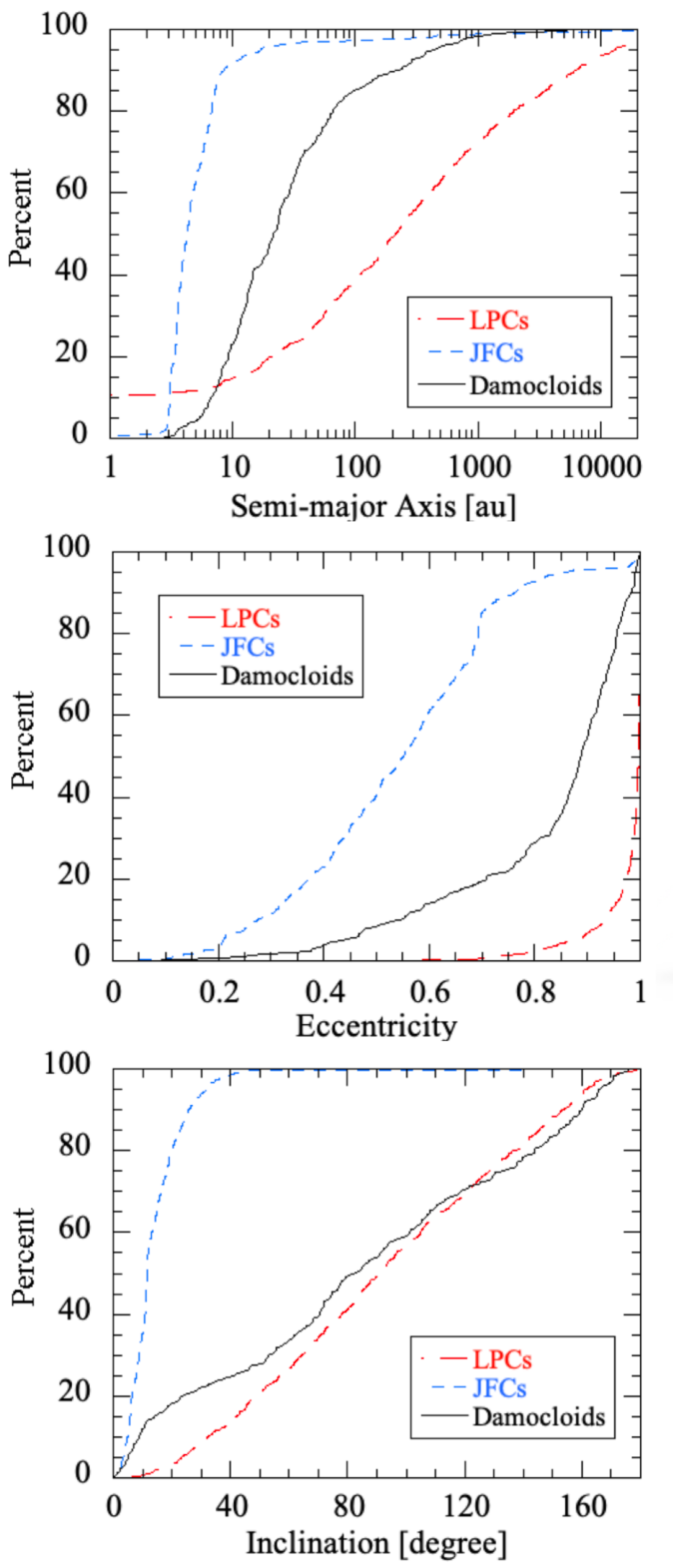}
\caption{Comparison of the cumulative distributions of the orbital elements of Damocloids (solid black lines), JFCs (dashed blue lines), and LPCs (long-dashed red lines); (top) semi-major axis, (middle) eccentricity and (bottom) inclination. Compiled from JPL Horizons.}
\label{figure:damo_plot}
\end{center}
\end{figure}

Table~\ref{table:Ratio} shows that Damocloids \citep[named after the prototype object (5335) Damocles;][]{Jew05} are comparatively rare compared to comets with $T_J < 2$, accounting for only $\sim$20\% of  the total ($\mathcal{R}$ = 0.2) having $q < 2$~au. The ratio becomes even more extreme for smaller values of $T_J$.  The Damocloid population grows larger when the perihelion distance constraint in the table is relaxed, but this might just be because activity is harder to detect when $q \ge 2$ au.  The table shows that, whereas apparent asteroids outnumber apparent comets in the range $2 \le T_J \le 3$, the opposite is true for $T_J < 2$.  
Currently 325 Damocloids are known, almost all of them reddish D-types \citep{Licandro18}.  
\\
  
The orbital elements of Damocloids are compared with those of JFCs and LPCs in Figure~\ref{figure:damo_plot}.  In all three panels, the Damocloids are intermediate between the JFC and LPC populations. While the inclination distributions of Damocloids and LPCs are similar, albeit with a preponderance of prograde orbits in the former (median inclinations of $i=73\degr$ for Damocloids vs.\ $i=90\degr$ for the LPCs), the semimajor axis and eccentricity distributions are distinctly different.  Therefore, we can reject the possibility that the Damocloids are exclusively dead or dormant LPC nuclei, and additional sources must be considered.  \citet{Wang12} suggest a combination of sources both in the Oort cloud and in the scattered disk component of the Kuiper belt.  Numerical simulations show that transfer to and from the Scattered Disk population in the Kuiper belt can occur.  
\\

Several objects initially classified as Damocloids (e.g., C/2002 CE$_{10}$ (LINEAR) and C/2002 VQ$_{94}$ (LINEAR)) have subsequently been found to show activity.  In some cases, the detection of coma simply reflects the acquisition of more sensitive measurements, but, in others, it appears that coma is intermittently present.  These objects provide firm evidence that Damocloids and comets are compositionally as well as dynamically related.   
\\

\begin{figure}[hbp!]
\begin{center}
\includegraphics[width=8.0cm]{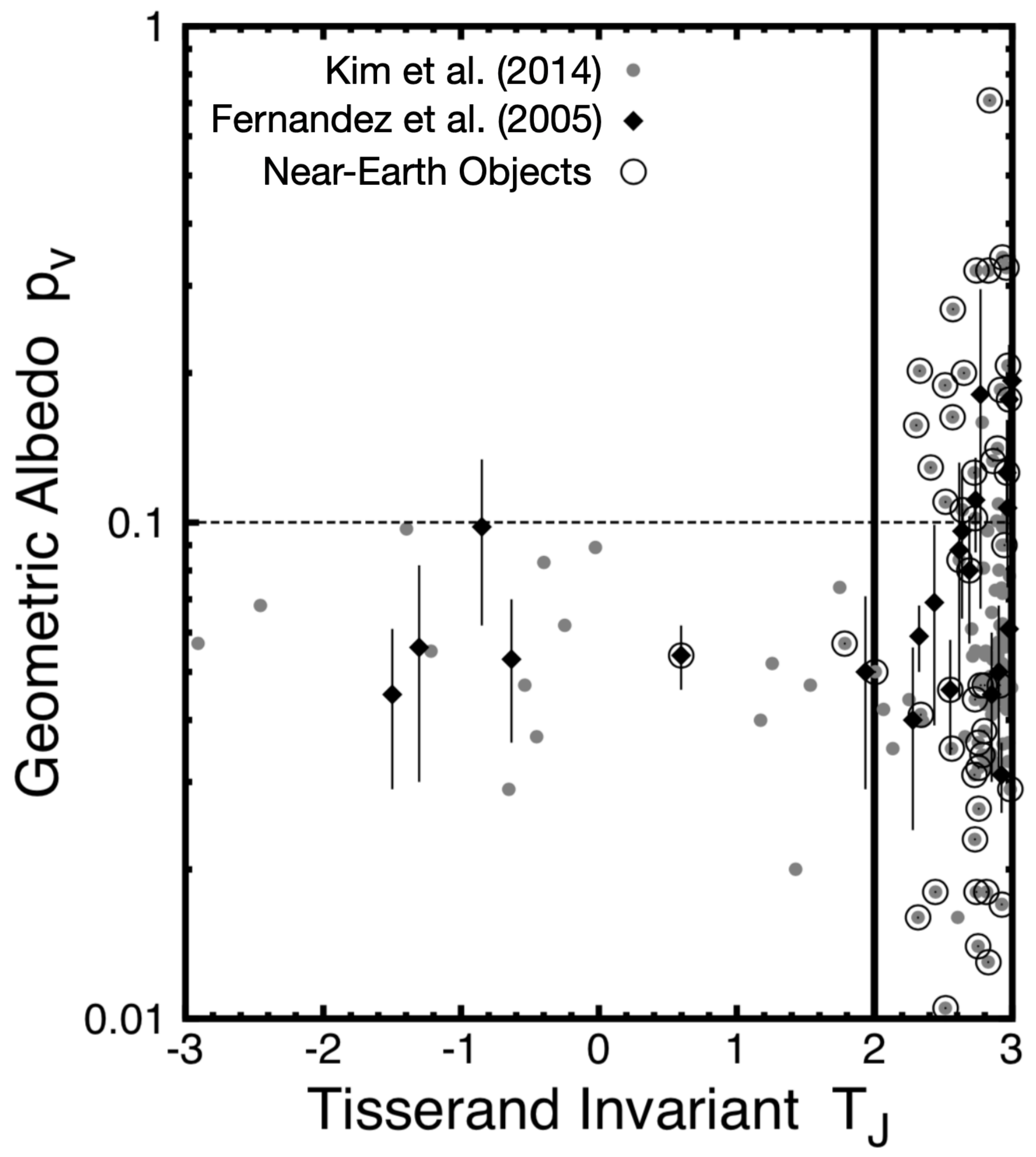}
\caption{Albedo vs.\ Tisserand parameter.  All objects with $T_J <$ 2 have low (comet-like) albedos. From \citet{Kim14}.}
\label{ACO}
\end{center}
\end{figure}

An observational connection between  dynamical and physical properties was established by \citet{Fernandez01}, and confirmed by the addition of new measurements by \citet{Fernandez05} and \citet{Kim14} (see Figure~\ref{ACO}).  Specifically, geometric albedos, $p_V$, are observed to vary systematically with $T_J$, such that bodies with $T_J<2$ have comet-like values $p_V=0.05\pm0.02$, regardless of whether or not they show evidence for cometary activity, whereas bodies with $2 \le T_J \le 3$ show a wide range of albedos, particularly for $T_J \gtrsim 2.5$.  \citet{Kim14} emphasized the existence of high albedo ($p_V > 0.1$) objects in this population, similar to what is seen in the bona-fide asteroids ($T_J\ge3$).  The albedo trends in Figure~\ref{ACO} support the conclusions from Table~\ref{table:Ratio}, to the effect that Damocloids are mostly inactive comets while the ACO population consists of a mixture of (dark) inactive comets and (on average, brighter) bona-fide asteroids, though dominated by the latter. Models indicate that objects  can be scattered out of the protoplanetary disk by the giant planets from  a wide range of initial orbits, even those interior to Jupiter \citep{Hahn99}. The systematically low (i.e., comet-like) albedos of Damocloids  (Figure~\ref{ACO}) indicate that such asteroidal bodies are uncommon.
\\

LPCs showing weak evidence for activity are sometimes referred to as ``Manx'' objects, an allusion to the short-tailed mutant cats of the same name.  It is not clear that the Manx objects are more than semantically different from weakly active LPCs; they could simply be low activity LPCs in which surface volatiles have been depleted in previous orbits.  However,  one  object, C/2014 S3, shows a spectral continuum downturn near 1~$\mu$m suggestive of that seen in S-type asteroids \citep{Meech16}.  This might suggest the presence of more highly metamorphosed solids like those found in the S-type asteroids, with a possible origin nearer the Sun at higher temperatures than typical for the nuclei of LPCs.  Unfortunately, it is not known whether the albedo of C/2014 S3 is like that of an S-type asteroid and we possess no spectrum suited to a detailed comparison.  The efficiency with which objects formed inside the snow-line can be captured into the Oort cloud is uncertain, given our limited knowledge of the orbital architecture of the planets in the young solar system.  \citet{Shannon15} estimated that $\sim8\times10^9$ asteroids could be so emplaced, which would comprise only $\sim$1\% of the Oort cloud comet population ($\sim 10^{12}$ total objects).   More albedo determinations and 1 $\mu$m continuum spectra of long-period objects are needed to make progress in this area.
\\

%\begin{figure*}[htb]
%\begin{center}
%\includegraphics[width=7.0cm]{size_plot.pdf}
%\caption{example of full-width plot.}
%\label{fullwidth}
%\end{center}
%\end{figure*}

%An observational connection between  dynamical and physical properties was established by \citet{Fernandez01},  and confirmed by the addition of new measurements by \citet{Fernandez05} and \citet{Kim14} (c.f.~Fig. \ref{ACO}).  The geometric albedo, $p_V$, is observed to vary systematically with $T_J$, such that bodies with $T_J<2$ have comet-like values $p_V=0.05\pm0.02$, regardless of whether or not they show evidence for cometary activity.  This matches the conclusion from Table~\ref{table:Ratio}, to the effect that the long-period ($T_J <$ 2) objects are mostly comets, and also supports the idea that the Damocloids are the inactive nuclei of comets.  For bodies with $2 \le T_J \le 3$ the evidence is less clear-cut, showing a wide range of albedos particularly for $T_J \gtrsim$ 2.5.  \citet{Kim14}, in particular, noted the existence of high albedo ($p_V > 0.1$ objects in this population, similar to what is seen in the bona-fide asteroids ($T_J\ge3$), and consistent with asteroidal contamination.

%\cite{Gundlach16}

\begin{figure*}[htb]
\begin{center}
\includegraphics[width=15.0cm]{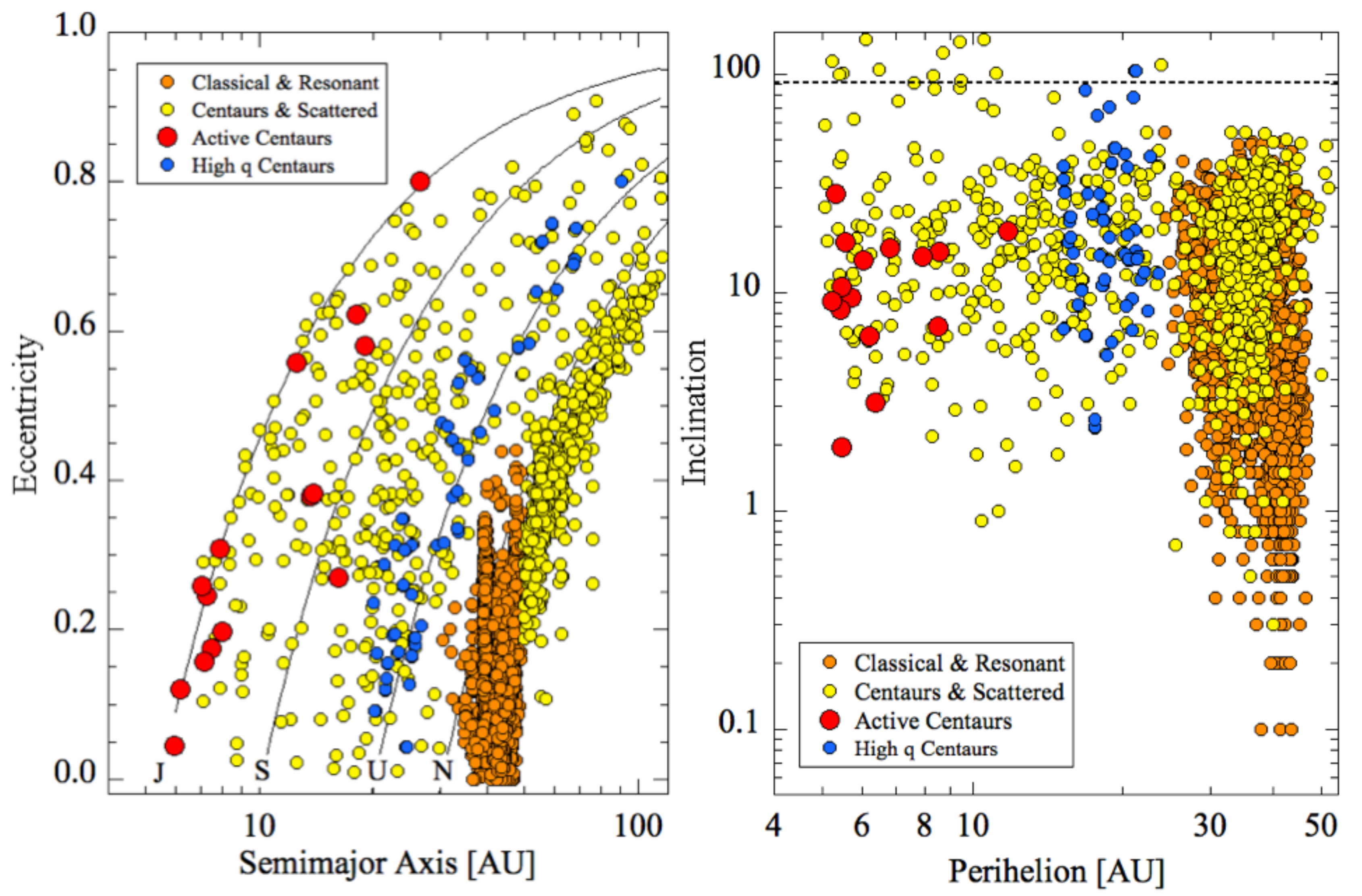}
\caption{Orbital elements of outer solar system populations including the active (red) and inactive Centaurs.  Semimajor axes of the major planets are marked J, S, U, and N.  Diagonal arcs in the left panel show the loci of orbits having fixed perihelia, $q = a(1-e)$, equal to the semi-major axes of the giant planets. The dotted horizontal line in the right panel is at $i = 90\degr$; objects above it are retrograde. The active Centaurs cluster within $q \lesssim 10$ au.  Adapted from \citet{Li20}}
\label{centaur_orbits}
\end{center}
\end{figure*}

\section{\textbf{CENTAURS}}
\label{section:centaurs}

Centaurs are generally considered to be an intermediate population of small bodies that are in dynamical transition from the trans-Neptunian population to the JFCs (see \cite{dones2015_cometreservoirs} and the chapter by Fraser et al.) and thus are scientifically interesting as JFC precursors.
Several working definitions of this population exist.  Here, we refer to objects with both perihelia and semi-major axes in between the orbits of Jupiter (5.2 au) and Neptune (30 au) which are not in mean-motion resonances with the planets (e.g., excluding Jupiter and Neptune Trojans).  The defining object in this class is 2060 Chiron ($a=13.6$~au, $q=8.5$~au), discovered in 1977 \citep{kowal78}, although comet 29P/Schwassmann-Wachmann 1, which has only recently come to be regarded as a Centaur, was actually discovered five decades before Chiron.  
About 430 Centaurs are known as of late 2021. %Chiron at first exhibited an asteroid-like appearance but later showed evidence for variable mass loss, unconnected to the heliocentric distance.  
\\

The Centaurs interact strongly with the giant planets and are consequently dynamically short-lived compared to the age of the solar system \citep{Horner04,Volk08}. Centaur dynamical half-lives, $\tau_{1/2}$, are spread over a wide range, becoming shorter as the perihelion distance decreases, which is a reflection of the large mass and gravitational influence of Jupiter. \citet{Horner04} bracketed the range of half-lives (expressed in Myr) as
\begin{equation}
    0.39 e^{0.135 q} \lesssim \tau_{1/2} \lesssim 0.06 e^{0.275 Q}
    \label{equation:Horner}
\end{equation}
\noindent where $q$ and $Q$ are the perihelion and aphelion distances. Substituting $q$ = 5.2 au and $Q$ = 30 au, for example, gives 0.8 $\lesssim \tau_{1/2} \lesssim 230$ Myr for the full range of dynamical ages.  The Centaur source, at first unknown, is now firmly recognized as the Kuiper belt, specifically  the scattered disk component of the belt from which objects are destabilized by  interactions with Neptune at perihelion \citep{Volk08}.  The fate of the Centaurs, other than a few that collide with a giant planet, is either to be ejected from the solar system to interstellar space, or injected into the terrestrial planet region where surface ices sublimate and they are re-labeled  as JFCs \citep{Levison94,sarid2019_29p}.  Unfortunately, the current orbit of any particular Centaur provides only a weak guide to its dynamical past because of the role of chaos induced by gas giant planet scattering.  While there is a statistical flow of Centaurs from their source beyond 30 au inwards, the instantaneous orbit of an individual Centaur may be diffusing inwards or outwards at any given time.  
\\

About 31 of the known Centaurs have a cometary designation, corresponding to about 7\% of the total.  This is a lower limit to the true incidence of Centaur activity because most such objects have not been studied in detail sufficient to reveal activity even if it is present.  Furthermore, most Centaur activity is transient, so that observations over a long period can only increase the estimated active fraction. Chiron itself  shows intermittent  mass loss, unconnected to its heliocentric distance.  
\\

\begin{figure}[htb!]
\begin{center}
\includegraphics[width=7.0cm]{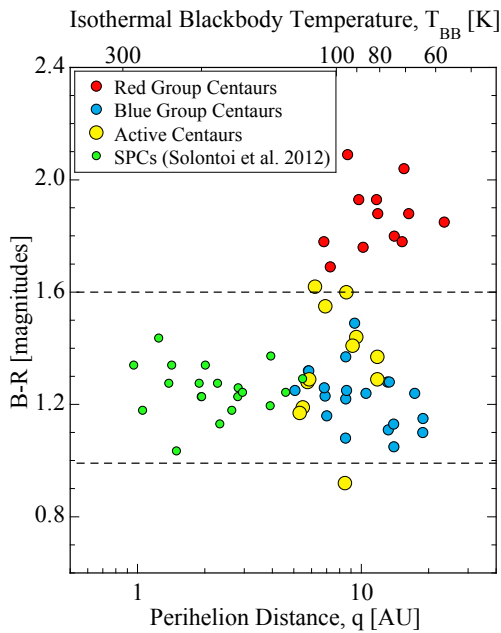}
\caption{$B-R$ optical color as a function of perihelion distance for short-period comets (SPCs; synonymous here with JFCs) and active and inactive Centaurs, as marked.  Horizontal dashed lines at $B-R = 0.99$ and $B-R = 1.6$ mark the colors of the Sun and the nominal beginning of the ultra-red objects, respectively.  Adapted from \citet{Jewitt15}.}
\label{centaur_colors}
\end{center}
\end{figure}

Two properties distinguish active Centaurs from others \citep{Jewitt09, Jewitt15}. First, their perihelia are preferentially small, typically $q \lesssim8$ to 10 au (Figure~\ref{centaur_orbits}, left panel).  While observational selection does favor the detection of activity at small $q$,  a sensitive, high resolution (Hubble Space Telescope based) search for activity in 53 Centaurs with perihelia $q >$ 15 au found none \citep{Li20}.  Second, while Centaurs as a whole show a bimodal distribution of optical colors \citep{Peixinho12}, this bimodality vanishes for objects with $q \lesssim$ 8 to 10 au (c.f.~Figure~\ref{centaur_orbits}).  The loss of bimodality occurs because the Centaurs lose ultrared matter, defined to have color index B-R $>$ 1.6\footnote{The color index B-R is defined as 2.5 times the logarithm of the ratio of the flux density measured in the R filter (center wavelength $\sim$6500\AA) to that in the B filter (4500\AA). Larger values indicate redder colors}, and thought to consist of highly irradiated organics.  The onset of activity and the disappearance of the ultrared matter appear to be correlated, in the sense that bodies which are active are rarely very red. Active centaur (523676) 2013 UL$_{10}$ with $B-R = 1.8\pm0.1$ has been reported as a possible exception \citep{Mazzotta18}, although the colors are contradicted by  \citet{Tegler16}, who measured $B-R = 1.62\pm0.03$ (plotted in Figure~\ref{centaur_colors}).   This could mean that the ultrared matter, whose precise composition is unknown, is thermodynamically unstable at temperatures $\gtrsim$124 K.  Alternatively, a thin layer of ultrared matter could be ejected or simply buried by fallback debris excavated from beneath, as soon as the activity begins \citep{Jewitt02}.  The smaller average perihelia of the active Centaurs correlate, as expected from Equation~\ref{equation:Horner}, with shorter mean dynamical lifetimes \citep{Melita12,Fernandez18}. 
\\

What triggers Centaur activity near 8 to 10 au, where the isothermal blackbody temperature of a sphere is about 88 to 98 K and even peak sub-solar temperatures are 124 to 138 K? This is far too cold for water ice to significantly sublimate.  While water is not volatile enough,  exposed super-volatile ices (e.g.~CO) are too volatile, in the sense that they would easily drive activity at much larger distances than observed.  For example, CO ice would sublimate strongly even at Kuiper belt distances and temperatures, where no activity is observed, and even CO$_2$ ice is in the strong-sublimation regime at distances $\sim$10 au.  Burial of super-volatile ices beneath a refractory mantle could, in principle, delay the onset of sublimation from an incoming body but would give no clear reason for a distinct turn-on distance at 8 to 10 au (c.f.~Figure \ref{centaur_orbits}).  Moreover, thermal evolution models suggest that \textit{pure} CO cannot survive long-term in comets \citep{Davidsson21}. Explanations attempting to match activity to the sublimation of less volatile, less abundant materials (such as H$_2$S) struggle to provide a convincing match to the 8 to 10 au critical distance \citep{Poston18}.  
\\

More promisingly, the crystallization timescale for amorphous water ice decreases exponentially with temperature, and first becomes less than  the orbital period at $\sim$10 au, suggesting that crystallization might drive Centaur activity \citep{Jewitt09,Guilbert12}.   Amorphous ice is the natural form of ice formed at low temperatures and gas pressures, so its incorporation in to the icy bodies of the outer solar system would not be surprising. Crystallization also fits the observed bursty nature of Centaur activity, corresponding to short-lived exothermic runaways in which trapped gases are expelled (see chapter by Prialnik and Jewitt).  In addition, Centaur activity may be correlated with recent inward drift of the perihelion distance \citep{Fernandez18}, consistent with the crystallization hypothesis.  Amorphous ice can trap CO, which  has been reported in 29P/Schwassmann-Wachmann 1, (60558) Echeclus and 2060 Chiron \citep[see review by][]{Womack17} but has not been detected in a majority of the Centaurs in which it has been sought \citep{Morbid01,Jewitt08,Drahus17}.  Absence of evidence is not necessarily evidence of absence, however, and the non-detections could simply reflect the difficulty in measuring CO rotational transitions in bodies far from the Earth.  The onset of Centaur activity (Figure~\ref{centaur_orbits}) and the collapse of the bimodal color distribution (Figure~\ref{centaur_colors}) are certainly consistent with an origin in the crystallization of amorphous ice, but it must be said that they do not prove that this is the responsible mechanism.  Establishing proof of the existence of amorphous ice in JFCs and Centaurs would be of great scientific value. It would imply that ice in the precursor Kuiper belt objects has also survived in the amorphous state,  significantly limiting the post-accretion thermal processing of these bodies  (see Davidsson 2021 and the chapter in this book by Prialnik and Jewitt).
\\

\section{FUTURE PROSPECTS AND CONCLUSIONS}
\label{section:future}

%Our state of understanding of the small solar system body continuum has evolved greatly over the past several decades and will undoubtedly continue to advance as new observing facilities begin operations, permitting ever more sensitive searches for activity that may reveal active objects in small body populations in which activity has yet to be found (e.g., the Jupiter Trojans), and with exciting new space missions being planned or proposed (also see chapter by Snodgrass et al.).
%\\

Looking ahead, Rubin Observatory's Legacy Survey of Space and Time (LSST) should generate a massive survey of the time-variable sky over 10 years and is expected to increase the numbers of known small solar system bodies in different populations by an order of magnitude or more \citep{jones2009_lsst,ivezic2019_lsst}.  Its use of a 8.4~m-aperture telescope
%(equivalent to a telescope 6.7~m in diameter in total collecting area) 
means that Rubin Observtory should achieve a sensitivity to faint activity that is unprecedented among wide-field surveys.
%, meaning that the survey has the potential to revolutionize our understanding of where active objects can be found in the solar system.
Meanwhile, relevant upcoming NASA space-based observatories include the {\it Near-Earth Object Surveyor} \citep[hereafter, {\it NEO Surveyor;}][]{mainzer2021_neosm}, which aims to identify near-Earth objects that might present impact threats to Earth, and the Nancy Grace Roman Space Telescope \citep{akeson2019_wfirst}, which will conduct a wide-field survey that is nominally focused on dark energy and exoplanets, but which is expected to observe many solar system objects as well (although mostly at high ecliptic latitudes).
Observing in the infrared, both spacecraft will be sensitive to thermal emission from long-lasting large dust grains ejected by active small bodies, potentially widening the observability window for detection of active events \citep[e.g.,][]{holler2018_wfirst_solarsystem}.
In particular, {\it NEO Surveyor's} focus on discovering NEOs means that it should be able to greatly improve our understanding of the ratio of active and inactive small bodies in the terrestrial planet region.
%comet to asteroid population ratio in the terrestrial planet region.
Taken together, these ever more sensitive searches for activity may reveal active objects in small body populations in which activity has yet to be found (e.g., the Jupiter Trojans) and will also provide a more thorough accounting of the extent of activity in populations already known to contain active objects like the NEOs, main-belt asteroids, and Centaurs, potentially allowing for further classification of active objects in those populations into smaller sub-groups based on physical or dynamical properties \citep[e.g.,][]{hsieh2016_tisserand}. 
%Similarly, NASA's Near-Earth Object Surveyor (hereafter, NEO Surveyor), intended to identify near-Earth objects which might present an impact threat to Earth, is expected to detect $\sim$8 million asteroids and find  $\sim$10$^5$ new NEOs. 
%It will provide sensitivity to thermal emission from dust in trails as well as to small parent bodies themselves, and should provide a better assessment of the comet to asteroid population ratio in the terrestrial planet region.
\\

Continuing efforts to characterize individual active objects will also shed light on their physical nature, provide needed constraints for thermal models, and help to identify physically plausible activity mechanisms for each object.  As most active events are of limited duration,  sometimes lasting just a few days, useful characterization of new active objects discovered by LSST, {\it NEO Surveyor}, or the {\it Roman Space Telescope} will require  rapid-response follow-up with deep imaging and spectroscopy \citep[e.g.,][]{najita2016_lsstobserving,street2018_toms}.
Long-term monitoring of known active objects \citep[e.g.,][]{hsieh2018_238p288p,wierzchos2020_29p}
%\citep[e.g.,][]{hsieh2018_238p288p,wierzchos2020_29p,clements2021_29p}
is also needed to assess both short-term and long-term activity evolution, while detailed characterization (e.g., measurements of phase functions, colors, and rotation states) of the nuclei of active objects during periods of inactivity (e.g., also see chapter by Knight et al.) provide information essential for thermal modeling studies \citep[e.g.][]{schorghofer2020_icepreservation}.
Meanwhile, the recently launched {\it James Webb Space Telescope} and the upcoming generation of extremely large ($\sim$30~m-class) ground-based telescopes should enable searches for gas with unprecedented sensitivity  \citep{crampton2009_tmtsolarsystem,kelley2016_jwstcomets,wong2020_elts}.
Targeted dynamical studies investigating the connections between small body populations (see Section~\ref{section:dynamics}) will also be useful for identifying potential opportunities for close-proximity studies of active objects originally from the main asteroid belt \citep[e.g.,][]{fernandez2017_249p}, much as classical comets give us the opportunity to study  the composition of objects displaced from the outer solar system.
\\

We can also look forward to a number of in-situ space missions both planned and proposed (see chapter by Snodgrass et al.).
The DESTINY+ spacecraft being developed by the Japan Aerospace Exploration Agency (JAXA) will visit Phaethon and has an expected launch date in 2024 \citep{Arai21,ozaki2022_destinyplus}. It will reach a close approach distance of 500 km from the object, briefly permitting imaging of the surface at $\le$5 m pixel$^{-1}$, albeit at a flyby speed of 36 km s$^{-1}$. An impact ionization time-of-flight mass spectrometer will provide  elemental compositions of any dust particles impacting the spacecraft, with a mass resolution of $M/\Delta M \sim$ 100 - 150 \citep{Kruger19}. It should be noted, however, that the total dust mass that will be intercepted is modest: less than that of a single 30 $\mu$m particle \citep{Jew18}.
Meanwhile, the Chinese National Space Agency's {\it ZhengHe} mission aims to visit active asteroid 311P, with an expected launch in 2025.  Missions targeting MBCs have been proposed to both NASA and ESA \citep[e.g.,][]{meech15,jones2018_caroline,snodgrass2018_castalia}, so far without success.  In-situ studies are particularly needed to resolve major issues about the nature of MBC outgassing, because they provide the only means for investigating outgassed material with a mass spectrometer.  
\\

%Other interesting mission concepts exist.
%, ongoing work continues to reveal new and exciting targets for future investigation using spacecraft.  
In terms of other interesting potential mission concepts, active asteroid 331P ejected $\sim$1\% of its mass as recently as 2011, producing a debris trail and a chain of 19 or more fragments that grows in length at $\lesssim$10 cm s$^{-1}$ \citep{Jewitt21b}.  A single ion-driven spacecraft sent to this object could determine the (presumably) excited rotation state of the km-sized primary body, investigate the geology of the detachment scar at high resolution in order to assess the mode of failure, and then travel the length of the debris trail, visiting the $\sim$19 fragments one by one.  The largest fragment, 331P-A, has an effective diameter of $\sim$200~m and shows a lightcurve consistent with a contact binary, presenting a particularly attractive target for in-situ investigation.
Of course, we have already had one unexpected mission to an active asteroid, OSIRIS-REx's visit to Bennu.  Because Bennu's activity was unexpected, only limited relevant observations were possible and the study of the ejected material has been inconclusive.  Nevertheless, the discovery of particles too large and sparse to be detected from Earth suggests that many other asteroids may be weakly active when examined closely, and that this possibility should be taken into account for future small body missions in general.  %As we look in greater and greater detail, the continuum expands to consume the objects at the ends of the distribution.
\\

Interestingly, a byproduct of NASA's planetary defense-oriented Double Asteroid Redirection Test (DART) mission, which aims to deliver a kinetic impactor to the secondary component of the binary asteroid (65803) Didymos in an attempt to deflect its orbit, may be to essentially produce an artificial active asteroid.  The ejecta plume caused by the DART impact will be observed by the Italian Space Agency's LICIACube cubesat spacecraft which will ride on the DART spacecraft and be released shortly before the impact, and later by the European Space Agency's separately-launched Hera spacecraft \citep[e.g.,][]{rossi2022_dart}, and may even be observable from the ground \citep{moreno2022_dart}.  The observational results from this experiment should yield useful insights into the detailed mechanics of real-world asteroid impact events, and also provide invaluable context for interpreting ground-based observations of active asteroids whose activity is produced by natural impacts.
\\

%Our state of understanding of the complexity of the small solar system body continuum will undoubtedly continue to evolve and advance, particularly as new observing facilities begin operations, permitting ever more sensitive searches for activity that may reveal active objects in small body populations in which activity has yet to be found (e.g., the Jupiter Trojans) and also provide a more thorough accounting of the extent of activity in populations already known to contain active objects like the NEOs, main-belt asteroids, and Centaurs. 

%In the coming years, we fully expect that future discoveries will continue to challenge our classical assumptions about the origin and physical nature of solar system objects based on the appearance of activity or their current orbital properties. 
%In doing so, however, such developments will deepen our understanding of the origin, physical nature, and evolution of our solar system itself.
%\\

Our understanding of the asteroid-comet continuum has evolved substantially since the publication of Comets II \citep{Jew04}. We now recognize that many small bodies possess diverse combinations of physical and dynamical properties, and evolutionary histories, that cannot be cleanly associated with either asteroids or comets.  Studies of extant volatile material from the inner portions of the protosolar disk are now possible, as are  direct observations of asteroid collisions and rotational disruptions.  New observations raise fresh questions about the incredible range of processes  that can produce mass loss from small bodies and about the  extent to which small bodies have been shuffled radially in the solar system since its formation. We anticipate even more exciting developments in the coming years.

\vskip .5in
\noindent \textbf{Acknowledgments.} 
We thank Masateru Ishiguro and Fernando Moreno for their helpful reviews of this chapter. HHH acknowledges support from NASA grants 80NSSC17K0723, 80NSSC18K0193, and 80NSSC19K0869.   \\

\bibliographystyle{sss-three.bst}
\bibliography{refs}

\end{document}